\documentclass[twocolumn]{IEEEtran}
\usepackage{booktabs}
\usepackage{authblk}
\usepackage{url}
\usepackage{silence}
\usepackage[font=scriptsize]{caption}
\usepackage{subcaption}

\usepackage[pdftex]{graphicx}
\usepackage{epstopdf}
\usepackage{pifont}
\usepackage[table,xcdraw]{xcolor}
\usepackage{colortbl}
\usepackage{array}
\usepackage{verse}
\usepackage{longtable}
\usepackage{supertabular}
\usepackage{cite}
\usepackage{color}
\usepackage{amsmath}
\usepackage{lettrine}
\usepackage{hyperref}
\usepackage{multirow}
\usepackage{dirtytalk}
\usepackage{graphicx}
\usepackage{soul}
\usepackage{comment}
\usepackage{verbatim}
\usepackage[utf8]{inputenc}
\usepackage{authblk}
\usepackage{tikz}
\def\checkmark{\tikz\fill[scale=0.4](0,.35) -- (.25,0) -- (1,.7) -- (.25,.15) -- cycle;}
\usepackage{amsmath}
\usepackage{pifont}

\newcommand{\xmark}{\text{\ding{55}}}
\graphicspath{ {images/} }

\hypersetup{
    colorlinks=false,
    linkcolor=blue,
    filecolor=magenta,
    urlcolor=cyan,}
    
\definecolor{LightCyan}{rgb}{0.88,1,1}

\newcolumntype{a}{>{\columncolor{LightCyan}}c}

\title{Blockchain and the Future of the Internet: \\A Comprehensive Review}

\author[1]{Fakhar ul Hassan}
\author[2]{Anwaar Ali}
\author[3]{Mohamed Rahouti}
\author[4]{Siddique Latif}
\author[5]{Salil Kanhere}
\author[6]{Jatinder Singh}
\author[7]{Ala Al-Fuqaha}
\author[8]{Umar Janjua}
\author[9]{Adnan Noor Mian}
\author[10]{Junaid Qadir}
\author[11]{Jon Crowcroft}
\affil[1,8,9,10]{Information Technology University (ITU), Punjab, Pakistan} 
\affil[2,6,11]{Computer Laboratory, University of Cambridge, United Kingdom}
\affil[3]{Computer \& Information Science Dept., Fordham University, NY USA} 
\affil[4]{University of Southern Queensland, Australia}
\affil[5]{University of New South Wales, Australia}
\affil[7]{Hamad Bin Khalifa University, Qatar; Western Michigan University, USA}

\begin{document}

\maketitle

\begin{abstract}

Blockchain is challenging the status quo of the central trust infrastructure currently prevalent in the Internet towards a design principle that is underscored by decentralization, transparency, and trusted auditability. In ideal terms, blockchain advocates a decentralized, transparent, and more democratic version of the Internet. Essentially being a trusted and decentralized database, blockchain finds its applications in fields as varied as the energy sector, forestry, fisheries,  mining, material recycling, air pollution monitoring, supply chain management, and their associated operations. In this paper, we present a survey of blockchain-based network applications. Our goal is to cover the evolution of blockchain-based systems that are trying to bring in a renaissance in the existing, mostly centralized, space of network applications. While re-imagining the space with blockchain, we highlight various common challenges, pitfalls, and shortcomings that can occur. Our aim is to make this work as a guiding reference manual for someone interested in shifting towards a blockchain-based solution for one's existing use case or automating one from the ground up.
\end{abstract}

\section{Introduction}
\label{sec: intro}
The paradigm shift entailed by blockchain's premise of decentralization envisages an eventual migration from the end-to-end principle to trust-to-trust principle \cite{ali2017trust}. According to this new design principle, a user should ideally always have complete control over the trust decisions particularly pertaining to user's data that powers a network application such as an online social network. This decentralization aspect forms the basis of the blockchain-based networks. This further paves the path for an era of distributed trust and consensus. This implies that large networks, in a peer-to-peer configuration, will guarantee the integrity of transactions (simply put interactions) among their peers without the involvement of any centrally trusted mediating third party. The provision of verifiable trust guarantees further entails that such networks can be audited in a trusted and transparent manner. This audit ability is useful to enforce the networked systems accountability over malfunctioning or an activity of foul play. Moreover, any application that requires interactions among various stakeholders for its operations in a mutually non-trusting environment (where the stakeholders do not have to or do not want to trust one another) can benefit from blockchain as it creates transparency and trust in interactions among the stakeholders without involving any third party. That is the reason why industries such as transport, energy sector, insurance, finance, and logistics have started to show their interest in blockchain technology to automate their solutions \cite{hamida2017blockchain, swan2015blockchain, mori2016financial, WEFReali3:online}.

It can be observed that although the onset of the Internet revolution heightened the societal collaboration among people, communities, and businesses \cite{peters2016understanding} many of the Internet applications, however, such as email and Domain Name Systems (DNS), largely remain centralized as far as their management and core development are concerned. The centralized governing bodies are usually behind the trust guarantees associated with such online applications. Similarly, the issue of trust in cloud-hosted data storage is another contemporary challenge predicated on the inherent centralized nature of the Internet \cite{ali2017trust}. The clients of such online and cloud-based services, such as cloud storage and computation, usually put their trust in the claims put forward by the third party cloud providers. It raises the pressing need for \emph{verifiability} that the cloud is not tampering with a client's stored data and is always returning correct results in response to the requested computation. A single instance of a data breach in cloud storage or a faulty execution of a requested set of computations can lead to disastrous ramifications for such a business. As it has been seen in a famous data breach that calls the trust in central management of online services such as Facebook (an online social network) into question \cite{facebook_breach:online}. Blockchain, on the other hand, with its premise of immutability, transparency, and peer-to-peer consensus can provide the means for a trusted audit of networked systems while at the same time giving much of the control back to the edges of a network.

%

\begin{figure*}[!t]
\centering
\captionsetup{justification=centering}
\centerline{\includegraphics[height=6cm, width=17cm]{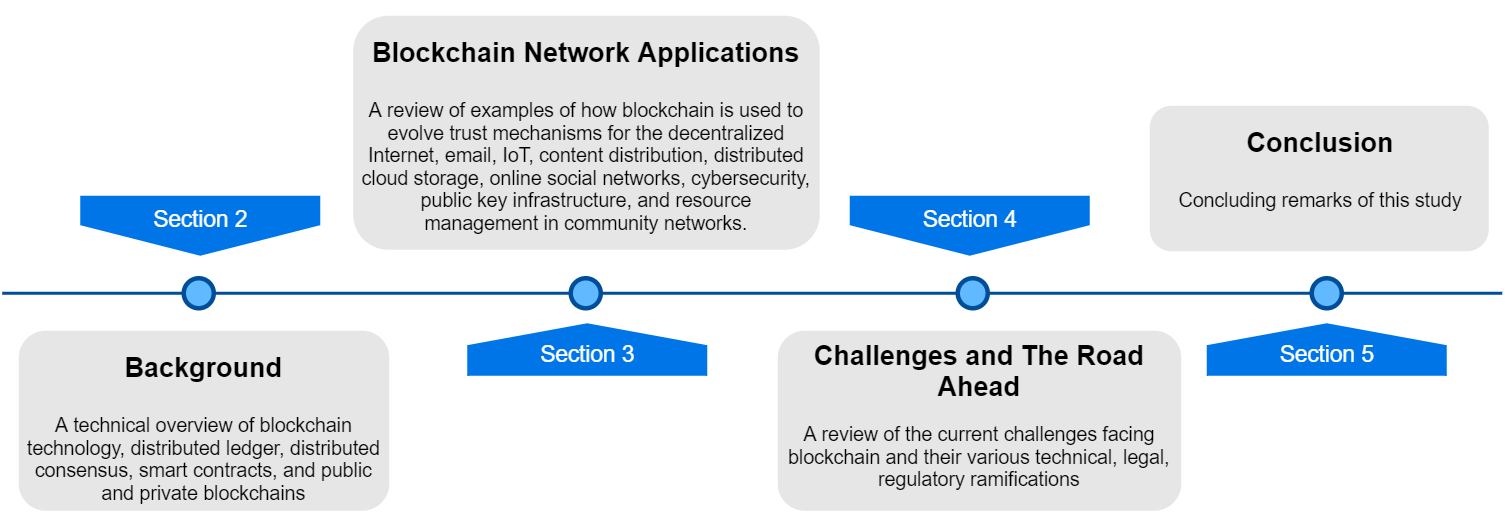}}
\caption{Overview of the paper}
\label{fig:roadmap}
\end{figure*}

\subsection{Contribution of the survey}
\label{subsec: contribution}
In this paper we provide a broad ranging survey of the implications of blockchain on the future of the Internet with a comprehensive take on their legal and regulatory ramifications as well. Instead of limiting ourselves to one particular use case or application (such as the Internet of Things (IoT) \cite{ali2018applications, ferrag2018blockchain}), we cover a wide range of use cases and try to observe the common patterns, differences, and technical limitations so that a more informed decision can be made by someone interested in deploying a use case from ground up or translating one's use case to a blockchain-based solution. We provide a comparison of our paper with other recent blockchain-based surveys in Table \ref{table:comparison}. Apart from encompassing most of the issues covered by recent survey literature, a clear distinguishing feature of this paper is that we also discuss a few of the most important legal and regulatory challenges and ramifications of deploying a blockchain-based solution. This is particularly important given the development of new data protection regulations (such as the advent of the General Data Protection Regulation (GDPR) in Europe), and regular reports of data breaches and government mass surveillance stories coming to light.

\subsection{Structure of the survey}
\label{subsec: structure}

The rest of the paper is organized in three main sections as expressed in Figure \ref{fig:roadmap}. In the section titled \textit{Background} (Section  \ref{sec: background}), we provide the necessary background to understand the big picture of how blockchain works by introducing distributed ledger technology, distributed consensus, smart contracts, and public and private blockchains. In the next section (Section \ref{app}) titled \textit{Blockchain-based Network Applications}, we provide examples of how blockchain can be used to evolve trust mechanisms for the decentralized Internet, email, Internet of Things (IoT), content distribution, distributed cloud storage, online social networks, cybersecurity, public key infrastructure, and resource management in community networks. Thereafter in the section titled \textit{Challenges and The Road Ahead} (Section \ref{sec: challenges}), we discuss the current challenges facing blockchain and their various technical, legal, and regulatory ramifications: in particular, we discuss governance, operational, and regulatory issues, scalability issues, security and privacy concerns, sustainability concerns, anonymity, the use of artificial intelligence (AI) and machine learning (ML), and issues related to usability and key management. Finally the paper is concluded in Section \ref{con}.

\begin{table*}[!ht]
\centering
\scriptsize
\begin{tabular}{ |m{2.2cm} |m{0.5cm}|m{1cm}|m{.7cm}|m{.8cm}|m{.9cm}|m{0.65cm}|m{0.35cm}|m{.7cm}|m{1.0cm}| m{.9cm}|a |}
\hline
\textbf{Papers/Books \newline (Author)}
& \textbf{Year}
& \textbf{\tiny Blockchain \newline Fundamentals} 
& \textbf{\tiny Challenges}
& \textbf{\tiny Smart \newline Contracts}
& \textbf{\tiny Blockchain \newline Applications}
& \textbf{\tiny Future \newline Trends}
& \textbf{IoT}
& \textbf{\tiny Blockchain \newline Types}
& \textbf{\tiny Blockchain \newline Characteristics}
& \textbf{\tiny Consensus \newline Algorithms}
& \textbf{Regulatory \newline Issues}
 \\ \hline

\begin{tabular}[c]{@{}l@{}}Zheng et al.\cite{zheng2016blockchain}\end{tabular}
&\begin{tabular}[c]{@{}l@{}}2016\end{tabular}
&\begin{tabular}[c]{@{}l@{}}\checkmark{} \end{tabular}
&\begin{tabular}[c]{@{}l@{}}\checkmark{} \end{tabular}
&\begin{tabular}[c]{@{}l@{}}\xmark{} \end{tabular}
&\begin{tabular}[c]{@{}l@{}}\checkmark{} \end{tabular}
&\begin{tabular}[c]{@{}l@{}}\checkmark{} \end{tabular}
&\begin{tabular}[c]{@{}l@{}}\xmark{} \end{tabular}
&\begin{tabular}[c]{@{}l@{}}\xmark{} \end{tabular}
&\begin{tabular}[c]{@{}l@{}}\checkmark{} \end{tabular}
&\begin{tabular}[c]{@{}l@{}}\checkmark{} \end{tabular}
&\begin{tabular}[c]{a}\xmark{} \end{tabular}\\\hline

\begin{tabular}[c]{@{}l@{}}Ye et al.\cite{guo2016blockchain}\end{tabular}
&\begin{tabular}[c]{@{}l@{}}2016\end{tabular}
&\begin{tabular}[c]{@{}l@{}}\xmark{} \end{tabular}
&\begin{tabular}[c]{@{}l@{}}\checkmark{} \end{tabular}
&\begin{tabular}[c]{@{}l@{}}\xmark{} \end{tabular}
&\begin{tabular}[c]{@{}l@{}}\checkmark{} \end{tabular}
&\begin{tabular}[c]{@{}l@{}}\xmark{} \end{tabular}
&\begin{tabular}[c]{@{}l@{}}\xmark{} \end{tabular}
&\begin{tabular}[c]{@{}l@{}}\checkmark{} \end{tabular}
&\begin{tabular}[c]{@{}l@{}}\xmark{} \end{tabular}
&\begin{tabular}[c]{@{}l@{}}\xmark{} \end{tabular}
&\begin{tabular}[c]{@{}l@{}}\checkmark{} \end{tabular}\\\hline

\begin{tabular}[c]{@{}l@{}}Yli-Huumo et al.\cite{yli2016current}\end{tabular}
&\begin{tabular}[c]{@{}l@{}}2016\end{tabular}
&\begin{tabular}[c]{@{}l@{}}\xmark{} \end{tabular}
&\begin{tabular}[c]{@{}l@{}}\checkmark{} \end{tabular}
&\begin{tabular}[c]{@{}l@{}}\checkmark{} \end{tabular}
&\begin{tabular}[c]{@{}l@{}}\checkmark{} \end{tabular}
&\begin{tabular}[c]{@{}l@{}}\checkmark{} \end{tabular}
&\begin{tabular}[c]{@{}l@{}}\xmark{} \end{tabular}
&\begin{tabular}[c]{@{}l@{}}\xmark{} \end{tabular}
&\begin{tabular}[c]{@{}l@{}}\xmark{} \end{tabular}
&\begin{tabular}[c]{@{}l@{}}\xmark{} \end{tabular}
&\begin{tabular}[c]{@{}l@{}}\xmark{} \end{tabular}\\\hline


\begin{tabular}[c]{@{}l@{}}Pilkington \cite{pilkington201611}\end{tabular}
&\begin{tabular}[c]{@{}l@{}}2016\end{tabular}
&\begin{tabular}[c]{@{}l@{}}\checkmark{} \end{tabular}
&\begin{tabular}[c]{@{}l@{}}\checkmark{} \end{tabular}
&\begin{tabular}[c]{@{}l@{}}\checkmark{} \end{tabular}
&\begin{tabular}[c]{@{}l@{}}\checkmark{} \end{tabular}
&\begin{tabular}[c]{@{}l@{}}\xmark{} \end{tabular}
&\begin{tabular}[c]{@{}l@{}}\xmark{} \end{tabular}
&\begin{tabular}[c]{@{}l@{}}\xmark{} \end{tabular}
&\begin{tabular}[c]{@{}l@{}}\xmark{} \end{tabular}
&\begin{tabular}[c]{@{}l@{}}\checkmark{} \end{tabular}
&\begin{tabular}[c]{@{}l@{}}\xmark{} \end{tabular}\\\hline
\begin{tabular}[c]{@{}l@{}}Nofer et al.\cite{nofer2017blockchain}\end{tabular}
&\begin{tabular}[c]{@{}l@{}}2017\end{tabular}
&\begin{tabular}[c]{@{}l@{}}\checkmark{} \end{tabular}
&\begin{tabular}[c]{@{}l@{}}\checkmark{} \end{tabular}
&\begin{tabular}[c]{@{}l@{}}\checkmark{} \end{tabular}
&\begin{tabular}[c]{@{}l@{}}\checkmark{} \end{tabular}
&\begin{tabular}[c]{@{}l@{}}\checkmark{} \end{tabular}
&\begin{tabular}[c]{@{}l@{}}\xmark{} \end{tabular}
&\begin{tabular}[c]{@{}l@{}}\xmark{} \end{tabular}
&\begin{tabular}[c]{@{}l@{}}\xmark{} \end{tabular}
&\begin{tabular}[c]{@{}l@{}}\xmark{} \end{tabular}
&\begin{tabular}[c]{@{}l@{}}\xmark{} \end{tabular}\\\hline

\begin{tabular}[c]{@{}l@{}}Zheng et al.\cite{zheng2017overview}\end{tabular}
&\begin{tabular}[c]{@{}l@{}}2017\end{tabular}
&\begin{tabular}[c]{@{}l@{}}\checkmark{} \end{tabular}
&\begin{tabular}[c]{@{}l@{}}\checkmark{} \end{tabular}
&\begin{tabular}[c]{@{}l@{}}\xmark{} \end{tabular}
&\begin{tabular}[c]{@{}l@{}}\xmark{} \end{tabular}
&\begin{tabular}[c]{@{}l@{}}\checkmark{} \end{tabular}
&\begin{tabular}[c]{@{}l@{}}\xmark{} \end{tabular}
&\begin{tabular}[c]{@{}l@{}}\checkmark{} \end{tabular}
&\begin{tabular}[c]{@{}l@{}}\checkmark{} \end{tabular}
&\begin{tabular}[c]{@{}l@{}}\checkmark{} \end{tabular}
&\begin{tabular}[c]{@{}l@{}}\xmark{} \end{tabular}\\\hline

\begin{tabular}[c]{@{}l@{}}Lin et al.\cite{lin2017survey}\end{tabular}
&\begin{tabular}[c]{@{}l@{}}2017\end{tabular}
&\begin{tabular}[c]{@{}l@{}}\checkmark{} \end{tabular}
&\begin{tabular}[c]{@{}l@{}}\checkmark{} \end{tabular}
&\begin{tabular}[c]{@{}l@{}}\checkmark{} \end{tabular}
&\begin{tabular}[c]{@{}l@{}}\checkmark{} \end{tabular}
&\begin{tabular}[c]{@{}l@{}}\xmark{} \end{tabular}
&\begin{tabular}[c]{@{}l@{}}\xmark{} \end{tabular}
&\begin{tabular}[c]{@{}l@{}}\xmark{} \end{tabular}
&\begin{tabular}[c]{@{}l@{}}\checkmark{} \end{tabular}
&\begin{tabular}[c]{@{}l@{}}\checkmark{} \end{tabular}
&\begin{tabular}[c]{@{}l@{}}\xmark{} \end{tabular}\\\hline


\begin{tabular}[c]{@{}l@{}}Miraz et al. \cite{miraz2018applications}\end{tabular}
&\begin{tabular}[c]{@{}l@{}}2018\end{tabular}
&\begin{tabular}[c]{@{}l@{}}\checkmark{} \end{tabular}
&\begin{tabular}[c]{@{}l@{}}\xmark{} \end{tabular}
&\begin{tabular}[c]{@{}l@{}}\xmark{} \end{tabular}
&\begin{tabular}[c]{@{}l@{}}\checkmark{} \end{tabular}
&\begin{tabular}[c]{@{}l@{}}\checkmark{} \end{tabular}
&\begin{tabular}[c]{@{}l@{}}\checkmark{} \end{tabular}
&\begin{tabular}[c]{@{}l@{}}\checkmark{} \end{tabular}
&\begin{tabular}[c]{@{}l@{}}\xmark{} \end{tabular}
&\begin{tabular}[c]{@{}l@{}}\xmark{} \end{tabular}
&\begin{tabular}[c]{@{}l@{}}\xmark{} \end{tabular}\\\hline
\begin{tabular}[c]{@{}l@{}}Yuan et al.\cite{yuan2018blockchain}\end{tabular}
&\begin{tabular}[c]{@{}l@{}}2018\end{tabular}
&\begin{tabular}[c]{@{}l@{}}\checkmark{} \end{tabular}
&\begin{tabular}[c]{@{}l@{}}\xmark{} \end{tabular}
&\begin{tabular}[c]{@{}l@{}}\xmark{} \end{tabular}
&\begin{tabular}[c]{@{}l@{}}\checkmark{} \end{tabular}
&\begin{tabular}[c]{@{}l@{}}\xmark{} \end{tabular}
&\begin{tabular}[c]{@{}l@{}}\xmark{} \end{tabular}
&\begin{tabular}[c]{@{}l@{}}\xmark{} \end{tabular}
&\begin{tabular}[c]{@{}l@{}}\xmark{} \end{tabular}
&\begin{tabular}[c]{@{}l@{}}\xmark{} \end{tabular}
&\begin{tabular}[c]{@{}l@{}}\xmark{} \end{tabular}\\\hline

\begin{tabular}[c]{@{}l@{}}Ali et al.\cite{ali2018applications}\end{tabular}
&\begin{tabular}[c]{@{}l@{}}2018\end{tabular}
&\begin{tabular}[c]{@{}l@{}}\checkmark{} \end{tabular}
&\begin{tabular}[c]{@{}l@{}}\xmark{} \end{tabular}
&\begin{tabular}[c]{@{}l@{}}\checkmark{} \end{tabular}
&\begin{tabular}[c]{@{}l@{}}\xmark{} \end{tabular}
&\begin{tabular}[c]{@{}l@{}}\checkmark{} \end{tabular}
&\begin{tabular}[c]{@{}l@{}}\checkmark{} \end{tabular}
&\begin{tabular}[c]{@{}l@{}}\checkmark{} \end{tabular}
&\begin{tabular}[c]{@{}l@{}}\checkmark{} \end{tabular}
&\begin{tabular}[c]{@{}l@{}}\checkmark{} \end{tabular}
&\begin{tabular}[c]{@{}l@{}}\xmark{} \end{tabular}\\\hline

\begin{tabular}[c]{@{}l@{}}Wust et al.\cite{wust2018you}\end{tabular}
&\begin{tabular}[c]{@{}l@{}}2018\end{tabular}
&\begin{tabular}[c]{@{}l@{}}\xmark{} \end{tabular}
&\begin{tabular}[c]{@{}l@{}}\xmark{} \end{tabular}
&\begin{tabular}[c]{@{}l@{}}\checkmark{} \end{tabular}
&\begin{tabular}[c]{@{}l@{}}\checkmark{} \end{tabular}
&\begin{tabular}[c]{@{}l@{}}\xmark{} \end{tabular}
&\begin{tabular}[c]{@{}l@{}}\checkmark{} \end{tabular}
&\begin{tabular}[c]{@{}l@{}}\checkmark{} \end{tabular}
&\begin{tabular}[c]{@{}l@{}}\checkmark{} \end{tabular}
&\begin{tabular}[c]{@{}l@{}}\xmark{} \end{tabular}
&\begin{tabular}[c]{@{}l@{}}\xmark{} \end{tabular}\\\hline


\begin{tabular}[c]{@{}l@{}}Salah et al. \cite{salahAccess} \end{tabular}
&\begin{tabular}[c]{@{}l@{}}2019\end{tabular}
&\begin{tabular}[c]{@{}l@{}}\xmark{}\end{tabular}
&\begin{tabular}[c]{@{}l@{}}\checkmark{}\end{tabular}
&\begin{tabular}[c]{@{}l@{}}\checkmark{}\end{tabular}
&\begin{tabular}[c]{@{}l@{}}\checkmark{}\end{tabular}
&\begin{tabular}[c]{@{}l@{}}\checkmark{}\end{tabular}
&\begin{tabular}[c]{@{}l@{}}\xmark{}\end{tabular}
&\begin{tabular}[c]{@{}l@{}}\checkmark{}\end{tabular}
&\begin{tabular}[c]{@{}l@{}}\checkmark{}\end{tabular}
&\begin{tabular}[c]{@{}l@{}}\checkmark{}\end{tabular}
&\begin{tabular}[c]{@{}l@{}}\xmark{}\end{tabular}\\\hline


\begin{tabular}[c]{@{}l@{}}Xie et al. \cite{xie2019survey} \end{tabular}
&\begin{tabular}[c]{@{}l@{}}2019\end{tabular}
&\begin{tabular}[c]{@{}l@{}} \checkmark{} \end{tabular}
&\begin{tabular}[c]{@{}l@{}} \checkmark{} \end{tabular}
&\begin{tabular}[c]{@{}l@{}} \checkmark{} \end{tabular}
&\begin{tabular}[c]{@{}l@{}} \checkmark{} \end{tabular}
&\begin{tabular}[c]{@{}l@{}} \checkmark{} \end{tabular}
&\begin{tabular}[c]{@{}l@{}} \checkmark{} \end{tabular}
&\begin{tabular}[c]{@{}l@{}} \xmark{} \end{tabular}
&\begin{tabular}[c]{@{}l@{}} \xmark{} \end{tabular}
&\begin{tabular}[c]{@{}l@{}} \xmark{} \end{tabular}
&\begin{tabular}[c]{@{}l@{}} \checkmark{} \end{tabular}\\\hline


\begin{tabular}[c]{@{}l@{}}Wang et al. \cite{wang2019survey} \end{tabular}
&\begin{tabular}[c]{@{}l@{}}2019\end{tabular}
&\begin{tabular}[c]{@{}l@{}} \checkmark{} \end{tabular}
&\begin{tabular}[c]{@{}l@{}} \xmark{} \end{tabular}
&\begin{tabular}[c]{@{}l@{}} \xmark{} \end{tabular}
&\begin{tabular}[c]{@{}l@{}} \checkmark{} \end{tabular}
&\begin{tabular}[c]{@{}l@{}} \xmark{} \end{tabular}
&\begin{tabular}[c]{@{}l@{}} \xmark{} \end{tabular}
&\begin{tabular}[c]{@{}l@{}} \xmark{} \end{tabular}
&\begin{tabular}[c]{@{}l@{}} \xmark{} \end{tabular}
&\begin{tabular}[c]{@{}l@{}} \checkmark{} \end{tabular}
&\begin{tabular}[c]{@{}l@{}} \xmark{} \end{tabular}\\\hline


\begin{tabular}[c]{@{}l@{}}Yang et al. \cite{yang2019survey} \end{tabular}
&\begin{tabular}[c]{@{}l@{}}2019\end{tabular}
&\begin{tabular}[c]{@{}l@{}} \checkmark{} \end{tabular}
&\begin{tabular}[c]{@{}l@{}} \checkmark{} \end{tabular}
&\begin{tabular}[c]{@{}l@{}} \checkmark{} \end{tabular}
&\begin{tabular}[c]{@{}l@{}} \checkmark{} \end{tabular}
&\begin{tabular}[c]{@{}l@{}} \checkmark{} \end{tabular}
&\begin{tabular}[c]{@{}l@{}} \xmark{} \end{tabular}
&\begin{tabular}[c]{@{}l@{}} \xmark{} \end{tabular}
&\begin{tabular}[c]{@{}l@{}} \xmark{} \end{tabular}
&\begin{tabular}[c]{@{}l@{}} \checkmark{} \end{tabular}
&\begin{tabular}[c]{@{}l@{}} \xmark{} \end{tabular}\\\hline


\begin{tabular}[c]{@{}l@{}}Yang et al. \cite{yang2019integrated} \end{tabular}
&\begin{tabular}[c]{@{}l@{}}2019\end{tabular}
&\begin{tabular}[c]{@{}l@{}} \checkmark{} \end{tabular}
&\begin{tabular}[c]{@{}l@{}} \checkmark{} \end{tabular}
&\begin{tabular}[c]{@{}l@{}} \xmark{} \end{tabular}
&\begin{tabular}[c]{@{}l@{}} \checkmark{} \end{tabular}
&\begin{tabular}[c]{@{}l@{}} \xmark{} \end{tabular}
&\begin{tabular}[c]{@{}l@{}} \checkmark{} \end{tabular}
&\begin{tabular}[c]{@{}l@{}} \xmark{} \end{tabular}
&\begin{tabular}[c]{@{}l@{}} \checkmark{} \end{tabular}
&\begin{tabular}[c]{@{}l@{}} \xmark{} \end{tabular}
&\begin{tabular}[c]{@{}l@{}} \xmark{} \end{tabular}\\\hline


\begin{tabular}[c]{@{}l@{}}Belotti et al. \cite{belotti2019vademecum} \end{tabular}
&\begin{tabular}[c]{@{}l@{}}2019\end{tabular}
&\begin{tabular}[c]{@{}l@{}} \xmark{} \end{tabular}
&\begin{tabular}[c]{@{}l@{}} \checkmark{} \end{tabular}
&\begin{tabular}[c]{@{}l@{}} \checkmark{} \end{tabular}
&\begin{tabular}[c]{@{}l@{}} \checkmark{} \end{tabular}
&\begin{tabular}[c]{@{}l@{}} \xmark{} \end{tabular}
&\begin{tabular}[c]{@{}l@{}} \checkmark{} \end{tabular}
&\begin{tabular}[c]{@{}l@{}} \xmark{} \end{tabular}
&\begin{tabular}[c]{@{}l@{}} \xmark{} \end{tabular}
&\begin{tabular}[c]{@{}l@{}} \checkmark{} \end{tabular}
&\begin{tabular}[c]{@{}l@{}} \xmark{} \end{tabular}\\\hline


\begin{tabular}[c]{@{}l@{}}Dai et al. \cite{dai2019blockchain} \end{tabular}
&\begin{tabular}[c]{@{}l@{}}2019\end{tabular}
&\begin{tabular}[c]{@{}l@{}} \checkmark{} \end{tabular}
&\begin{tabular}[c]{@{}l@{}} \checkmark{} \end{tabular}
&\begin{tabular}[c]{@{}l@{}} \xmark{} \end{tabular}
&\begin{tabular}[c]{@{}l@{}} \checkmark{} \end{tabular}
&\begin{tabular}[c]{@{}l@{}} \checkmark{} \end{tabular}
&\begin{tabular}[c]{@{}l@{}} \checkmark{} \end{tabular}
&\begin{tabular}[c]{@{}l@{}} \xmark{} \end{tabular}
&\begin{tabular}[c]{@{}l@{}} \checkmark{} \end{tabular}
&\begin{tabular}[c]{@{}l@{}} \checkmark{} \end{tabular}
&\begin{tabular}[c]{@{}l@{}} \xmark{} \end{tabular}\\\hline


\begin{tabular}[c]{@{}l@{}}Wu et al. \cite{wu2019comprehensive} \end{tabular}
&\begin{tabular}[c]{@{}l@{}}2019\end{tabular}
&\begin{tabular}[c]{@{}l@{}} \checkmark{} \end{tabular}
&\begin{tabular}[c]{@{}l@{}} \checkmark{} \end{tabular}
&\begin{tabular}[c]{@{}l@{}} \xmark{} \end{tabular}
&\begin{tabular}[c]{@{}l@{}} \checkmark{} \end{tabular}
&\begin{tabular}[c]{@{}l@{}} \xmark{} \end{tabular}
&\begin{tabular}[c]{@{}l@{}} \checkmark{} \end{tabular}
&\begin{tabular}[c]{@{}l@{}} \xmark{} \end{tabular}
&\begin{tabular}[c]{@{}l@{}} \checkmark{} \end{tabular}
&\begin{tabular}[c]{@{}l@{}} \xmark{} \end{tabular}
&\begin{tabular}[c]{@{}l@{}} \xmark{} \end{tabular}\\\hline


\begin{tabular}[c]{@{}l@{}}Viriyasitavat et al. \cite{viriyasitavat2019blockchain} \end{tabular}
&\begin{tabular}[c]{@{}l@{}}2019\end{tabular}
&\begin{tabular}[c]{@{}l@{}} \xmark{} \end{tabular}
&\begin{tabular}[c]{@{}l@{}} \checkmark{} \end{tabular}
&\begin{tabular}[c]{@{}l@{}} \xmark{} \end{tabular}
&\begin{tabular}[c]{@{}l@{}} \checkmark{} \end{tabular}
&\begin{tabular}[c]{@{}l@{}} \xmark{} \end{tabular}
&\begin{tabular}[c]{@{}l@{}} \checkmark{} \end{tabular}
&\begin{tabular}[c]{@{}l@{}} \xmark{} \end{tabular}
&\begin{tabular}[c]{@{}l@{}} \xmark{} \end{tabular}
&\begin{tabular}[c]{@{}l@{}} \xmark{} \end{tabular}
&\begin{tabular}[c]{@{}l@{}} \xmark{} \end{tabular}\\\hline


\begin{tabular}[c]{@{}l@{}}Mollah et al. \cite{mollah2020blockchain} \end{tabular}
&\begin{tabular}[c]{@{}l@{}}2020\end{tabular}
&\begin{tabular}[c]{@{}l@{}} \checkmark{} \end{tabular}
&\begin{tabular}[c]{@{}l@{}} \checkmark{} \end{tabular}
&\begin{tabular}[c]{@{}l@{}} \xmark{} \end{tabular}
&\begin{tabular}[c]{@{}l@{}} \checkmark{} \end{tabular}
&\begin{tabular}[c]{@{}l@{}} \checkmark{} \end{tabular}
&\begin{tabular}[c]{@{}l@{}} \xmark{} \end{tabular}
&\begin{tabular}[c]{@{}l@{}} \xmark{} \end{tabular}
&\begin{tabular}[c]{@{}l@{}} \checkmark{} \end{tabular}
&\begin{tabular}[c]{@{}l@{}} \checkmark{} \end{tabular}
&\begin{tabular}[c]{@{}l@{}} \checkmark{} \end{tabular}\\\hline


\begin{tabular}[c]{@{}l@{}}Liu. \cite{liu2020blockchain} \end{tabular}
&\begin{tabular}[c]{@{}l@{}}2020\end{tabular}
&\begin{tabular}[c]{@{}l@{}} \checkmark{} \end{tabular}
&\begin{tabular}[c]{@{}l@{}} \checkmark{} \end{tabular}
&\begin{tabular}[c]{@{}l@{}} \checkmark{} \end{tabular}
&\begin{tabular}[c]{@{}l@{}} \xmark{} \end{tabular}
&\begin{tabular}[c]{@{}l@{}} \checkmark{} \end{tabular}
&\begin{tabular}[c]{@{}l@{}} \xmark{} \end{tabular}
&\begin{tabular}[c]{@{}l@{}} \xmark{} \end{tabular}
&\begin{tabular}[c]{@{}l@{}} \checkmark{} \end{tabular}
&\begin{tabular}[c]{@{}l@{}} \xmark{} \end{tabular}
&\begin{tabular}[c]{@{}l@{}} \xmark{} \end{tabular}\\\hline


\begin{tabular}[c]{@{}l@{}}Neudecker et al. \cite{neudecker2018network} \end{tabular}
&\begin{tabular}[c]{@{}l@{}}2019\end{tabular}
&\begin{tabular}[c]{@{}l@{}} \checkmark{} \end{tabular}
&\begin{tabular}[c]{@{}l@{}} \checkmark{} \end{tabular}
&\begin{tabular}[c]{@{}l@{}} \xmark{} \end{tabular}
&\begin{tabular}[c]{@{}l@{}} \xmark{} \end{tabular}
&\begin{tabular}[c]{@{}l@{}} \checkmark{} \end{tabular}
&\begin{tabular}[c]{@{}l@{}} \xmark{} \end{tabular}
&\begin{tabular}[c]{@{}l@{}} \xmark{} \end{tabular}
&\begin{tabular}[c]{@{}l@{}} \xmark{} \end{tabular}
&\begin{tabular}[c]{@{}l@{}} \xmark{} \end{tabular}
&\begin{tabular}[c]{@{}l@{}} \xmark{} \end{tabular}\\\hline


\begin{tabular}[c]{@{}l@{}}Lao et al. \cite{lao2020survey} \end{tabular}
&\begin{tabular}[c]{@{}l@{}}2020\end{tabular}
&\begin{tabular}[c]{@{}l@{}} \checkmark{} \end{tabular}
&\begin{tabular}[c]{@{}l@{}} \checkmark{} \end{tabular}
&\begin{tabular}[c]{@{}l@{}} \xmark{} \end{tabular}
&\begin{tabular}[c]{@{}l@{}} \checkmark{} \end{tabular}
&\begin{tabular}[c]{@{}l@{}} \xmark{} \end{tabular}
&\begin{tabular}[c]{@{}l@{}} \checkmark{} \end{tabular}
&\begin{tabular}[c]{@{}l@{}} \xmark{} \end{tabular}
&\begin{tabular}[c]{@{}l@{}} \checkmark{} \end{tabular}
&\begin{tabular}[c]{@{}l@{}} \checkmark{} \end{tabular}
&\begin{tabular}[c]{@{}l@{}} \xmark{}  \end{tabular}\\\hline


\begin{tabular}[c]{@{}l@{}}Kolb et al. \cite{kolb2020core} \end{tabular}
&\begin{tabular}[c]{@{}l@{}}2020\end{tabular}
&\begin{tabular}[c]{@{}l@{}} \checkmark{} \end{tabular}
&\begin{tabular}[c]{@{}l@{}} \checkmark{} \end{tabular}
&\begin{tabular}[c]{@{}l@{}} \checkmark{} \end{tabular}
&\begin{tabular}[c]{@{}l@{}} \checkmark{} \end{tabular}
&\begin{tabular}[c]{@{}l@{}} \xmark{} \end{tabular}
&\begin{tabular}[c]{@{}l@{}} \xmark{} \end{tabular}
&\begin{tabular}[c]{@{}l@{}} \xmark{} \end{tabular}
&\begin{tabular}[c]{@{}l@{}} \checkmark{} \end{tabular}
&\begin{tabular}[c]{@{}l@{}} \checkmark{} \end{tabular}
&\begin{tabular}[c]{@{}l@{}} \xmark{} \end{tabular}\\\hline


\begin{tabular}[c]{@{}l@{}}Monrat et al. \cite{monrat2019survey} \end{tabular}
&\begin{tabular}[c]{@{}l@{}}2019\end{tabular}
&\begin{tabular}[c]{@{}l@{}} \checkmark{} \end{tabular}
&\begin{tabular}[c]{@{}l@{}} \checkmark{} \end{tabular}
&\begin{tabular}[c]{@{}l@{}} \xmark{} \end{tabular}
&\begin{tabular}[c]{@{}l@{}} \checkmark{} \end{tabular}
&\begin{tabular}[c]{@{}l@{}} \checkmark{} \end{tabular}
&\begin{tabular}[c]{@{}l@{}} \xmark{} \end{tabular}
&\begin{tabular}[c]{@{}l@{}} \xmark{} \end{tabular}
&\begin{tabular}[c]{@{}l@{}} \checkmark{} \end{tabular}
&\begin{tabular}[c]{@{}l@{}} \checkmark{} \end{tabular}
&\begin{tabular}[c]{@{}l@{}} \checkmark{} \end{tabular}\\\hline


\begin{tabular}[c]{@{}l@{}}Zhang et al. \cite{zhang2019security} \end{tabular}
&\begin{tabular}[c]{@{}l@{}}2019\end{tabular}
&\begin{tabular}[c]{@{}l@{}} \checkmark{} \end{tabular}
&\begin{tabular}[c]{@{}l@{}} \checkmark{} \end{tabular}
&\begin{tabular}[c]{@{}l@{}} \checkmark{} \end{tabular}
&\begin{tabular}[c]{@{}l@{}} \xmark{} \end{tabular}
&\begin{tabular}[c]{@{}l@{}} \xmark{} \end{tabular}
&\begin{tabular}[c]{@{}l@{}} \xmark{} \end{tabular}
&\begin{tabular}[c]{@{}l@{}} \checkmark{} \end{tabular}
&\begin{tabular}[c]{@{}l@{}} \checkmark{} \end{tabular}
&\begin{tabular}[c]{@{}l@{}} \checkmark{} \end{tabular}
&\begin{tabular}[c]{@{}l@{}} \xmark{} \end{tabular}\\\hline


\begin{tabular}[c]{@{}l@{}}Xiao et al. \cite{xiao2020survey} \end{tabular}
&\begin{tabular}[c]{@{}l@{}}2020\end{tabular}
&\begin{tabular}[c]{@{}l@{}} \xmark{} \end{tabular}
&\begin{tabular}[c]{@{}l@{}} \checkmark{} \end{tabular}
&\begin{tabular}[c]{@{}l@{}} \xmark{} \end{tabular}
&\begin{tabular}[c]{@{}l@{}} \xmark{} \end{tabular}
&\begin{tabular}[c]{@{}l@{}} \xmark{} \end{tabular}
&\begin{tabular}[c]{@{}l@{}} \xmark{} \end{tabular}
&\begin{tabular}[c]{@{}l@{}} \xmark{} \end{tabular}
&\begin{tabular}[c]{@{}l@{}} \checkmark{} \end{tabular}
&\begin{tabular}[c]{@{}l@{}} \checkmark{} \end{tabular}
&\begin{tabular}[c]{@{}l@{}} \xmark{} \end{tabular}\\\hline


\begin{tabular}[c]{@{}l@{}}Bodkhe et al. \cite{bodkhe2020survey} \end{tabular}
&\begin{tabular}[c]{@{}l@{}}2020\end{tabular}
&\begin{tabular}[c]{@{}l@{}} \checkmark{} \end{tabular}
&\begin{tabular}[c]{@{}l@{}} \checkmark{} \end{tabular}
&\begin{tabular}[c]{@{}l@{}} \xmark{} \end{tabular}
&\begin{tabular}[c]{@{}l@{}} \checkmark{} \end{tabular}
&\begin{tabular}[c]{@{}l@{}} \checkmark{} \end{tabular}
&\begin{tabular}[c]{@{}l@{}} \checkmark{} \end{tabular}
&\begin{tabular}[c]{@{}l@{}} \xmark{} \end{tabular}
&\begin{tabular}[c]{@{}l@{}} \xmark{} \end{tabular}
&\begin{tabular}[c]{@{}l@{}} \checkmark{} \end{tabular}
&\begin{tabular}[c]{@{}l@{}} \xmark{} \end{tabular}\\\hline


\begin{tabular}[c]{@{}l@{}}Al-Jaroodi et al. \cite{al2019blockchain} \end{tabular}
&\begin{tabular}[c]{@{}l@{}}2019\end{tabular}
&\begin{tabular}[c]{@{}l@{}} \xmark{} \end{tabular}
&\begin{tabular}[c]{@{}l@{}} \checkmark{} \end{tabular}
&\begin{tabular}[c]{@{}l@{}} \checkmark{} \end{tabular}
&\begin{tabular}[c]{@{}l@{}} \checkmark{} \end{tabular}
&\begin{tabular}[c]{@{}l@{}} \checkmark{} \end{tabular}
&\begin{tabular}[c]{@{}l@{}} \xmark{} \end{tabular}
&\begin{tabular}[c]{@{}l@{}} \xmark{} \end{tabular}
&\begin{tabular}[c]{@{}l@{}} \xmark{} \end{tabular}
&\begin{tabular}[c]{@{}l@{}} \xmark{} \end{tabular}
&\begin{tabular}[c]{@{}l@{}} \checkmark{} \end{tabular}\\\hline


\begin{tabular}[c]{@{}l@{}}Our Survey\end{tabular}
&\begin{tabular}[c]{@{}l@{}}2020\end{tabular}
&\begin{tabular}[c]{@{}l@{}}\checkmark{} \end{tabular}
&\begin{tabular}[c]{@{}l@{}}\checkmark{} \end{tabular}
&\begin{tabular}[c]{@{}l@{}}\checkmark{} \end{tabular}
&\begin{tabular}[c]{@{}l@{}}\checkmark{} \end{tabular}
&\begin{tabular}[c]{@{}l@{}}\checkmark{} \end{tabular}
&\begin{tabular}[c]{@{}l@{}}\checkmark{} \end{tabular}
&\begin{tabular}[c]{@{}l@{}}\checkmark{} \end{tabular}
&\begin{tabular}[c]{@{}l@{}}\checkmark{} \end{tabular}
&\begin{tabular}[c]{@{}l@{}}\checkmark{} \end{tabular}
&\begin{tabular}[c]{@{}l@{}}\checkmark{} \newline(distinguishing feature) \end{tabular}\\\hline


\end{tabular}
\caption{Comparative analysis of our survey with the existing survey literature pool}
\label{table:comparison}
\end{table*}


\section{Background}
\label{sec: background}
In this section, we provide the necessary background to understand what blockchain is and how it works. Our discussion in this section follows an evolutionary approach which means we start with Bitcoin \cite{nakamoto2008bitcoin} (the first incarnation of a blockchain-based financial application) and discuss how the technology evolved giving rise to other concepts and systems along the way.
\subsection{Blockchain and distributed ledger technology (DLT)}
\label{subsec: bc_and_dlt}
The original premise of blockchain is to establish trust in a peer-to-peer (P2P) network circumventing the need for any sort of third managing parties. As an example, Bitcoin introduced a P2P monetary value transfer system where no bank or any other financial institution is required to make a value-transfer transaction with anyone else on Bitcoin's blockchain network. Such a trust is in the form of verifiable mathematical evidence (more details on it follow in Section \ref{subsec: consensus}). The provision of this trust mechanism allows peers of a P2P network to transact with each other without necessarily trusting one another. Sometimes this is referred to as the \emph{trustless} property of blockchain. This trustlessness further implies that a party interested in transacting with another entity on blockchain does not necessarily have to know the real identity of it. This enables users of a public blockchain system (see Section \ref{subsec: pub_vs_priv_BC} for more details on public and private blockchains), such as Bitcoin, to remain anonymous. Further, a record of transactions among the peers are stored in a chain of a series of a data structure called \emph{blocks}, hence the name blockchain. Each peer of a blockchain  network maintains a copy of this record. Additionally, a consensus, taking into consideration the majority of the network peers, is also established on the state of the blockchain that all the peers of the network store. That is why, at times, blockchain is also referred to as the \emph{distributed ledger technology (DLT)}. Each instance of such a DLT, stored at each peer of the network, gets updated at the same time with no provision for retroactive mutations in the records.
\subsection{A clever use of hashing}
\label{subsec: hashing}
 \begin{figure*}[!ht]
\centering
\captionsetup{justification=centering}
\centerline{\includegraphics[width=.73\textwidth]{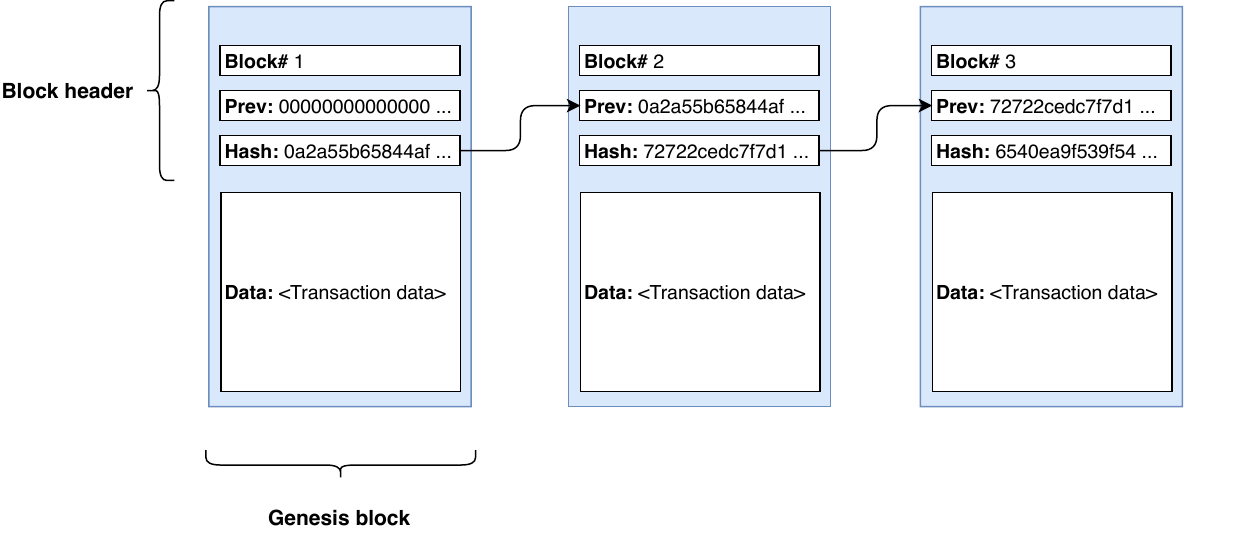}}
\caption{Hashing chains the blocks together and renders them immutable}
\label{fig: blockchain}
\end{figure*}
We now take a closer look at how hashing is used to \emph{chain} the blocks containing transaction records together and how such records are rendered immutable. A hash is defined as a unidirectional cryptographic function. A hash function usually takes an arbitrary input of an arbitrary length and outputs a seemingly random but fixed-length string of characters. Each such output is unique to the input given to this function and can be considered as the \emph{footprint} for the input. If the input is even so slightly changed then the output of the hash function almost always  changes completely and seemingly in a random fashion (there are, however, rare occasions where a \emph{collision} occurs when two distinct inputs to a hash function map to the same output) \cite{rogaway2004cryptographic}. This way hash of a piece of data can be used to verify the integrity of it. As an example, Secure Hash Algorithm 256 (SHA256) is a member of the family of SHA2 hash functions which is currently being deployed by many blockchain-based systems such as Bitcoin \cite{sha:online}. 

Figure \ref{fig: blockchain} shows a simple representation of an \emph{append-only} blockchain data structure making use of hashing. In this figure, the \emph{hash} field of each \emph{block} contains the hash value of all the contents of a given block (i.e., block number, \emph{previous hash}, shown as \emph{Prev} in Figure \ref{fig: blockchain}, and data). In this illustration, the most important field is the \emph{Prev} field. This field, in each block, contains the hash value of the block that comes before it. This chains the blocks together. Now, if the contents of a block are changed then this change is reflected, in addition to the hash of the block under consideration, in the portion of the blockchain that comes after the block being mutated. This way, hashing and the distribution of blockchain copies among the peers of a P2P network makes the records stored in a blockchain tamper evident. It can be noted in Figure \ref{fig: blockchain} that the first block in a blockchain is sometimes referred to as the \emph{genesis block} indicated by its Prev field initialized to contain all zeros.
\subsection{A coin: Transaction chain}
\label{subsec: tx_chain}
\begin{figure*}[!ht]
\centering
\captionsetup{justification=centering}
\centerline{\includegraphics[width=.62\textwidth]{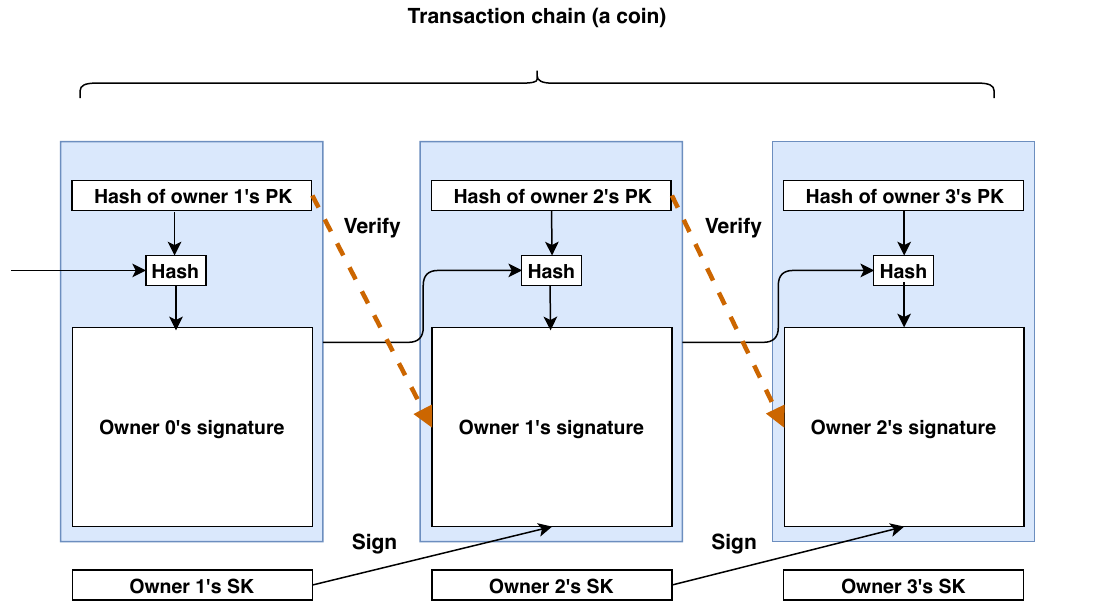}}
\caption{Transaction chain or a coin. Figure adapted from \cite{nakamoto2008bitcoin}}
\label{fig: tx_chain}
\end{figure*}

A transaction chain is shown in Figure \ref{fig: tx_chain}. It should be observed here that there is a difference between a transaction chain and a blockchain. Each block in a blockchain can contain multiple transaction chains. Each transaction chain in turn shows the value transferred from one peer of the network to another. Each such transaction chain is also sometimes referred to as a digital \emph{coin} or more generally as a \emph{token} . As an example Ethereum (discussed in Section \ref{subsec: smart_contracts}) allows one to define a custom token\footnote{\url{https://www.ethereum.org/token}}.

A transaction chain makes use of digital signatures, in addition to hashing like the way it is described above, to track the provenance of digital funds.

\subsection{Distributed consensus}
\label{subsec: consensus}
Distributed consensus is a mechanism through which peers of a distributed system collectively reach an agreement on the state of a collectively maintained record. In order to uphold the premise of decentralization, different blockchain-based systems deploy a particular flavour of distributed consensus. In this section, we first discuss the most popular and widely adopted consensus protocol called Proof-of-Work (PoW) mainly popularised by Bitcoin. We then build upon this discussion to describe and compare subsequent consensus mechanisms that have been deployed as different blockchain-based systems evolved and proliferated.
\vspace{2mm}
\subsubsection{Proof-of-Work (PoW)}
\label{subsubsec: pow}
PoW-based consensus mechanism was mainly popularized by Bitcoin \cite{nakamoto2008bitcoin}. PoW's main goal is to prevent double spending of a digital asset by providing a verifiable trust guarantee to a payee. Such a guarantee is provided in the form of publishing an integer called a \emph{nonce}. Finding a nonce is a computationally intensive process and is often referred to as \emph{mining}. The peer of a blockchain network that finds a nonce is called a \emph{miner}. Specifically, a nonce is an integer which, when hashed together with the contents of a block, outputs a hash matching a predefined pattern. Depending upon the underlying system, such a pattern is usually defined to start with a predefined number of zeros. The larger the number of leading zeros the harder (in computational terms) it is to find a nonce that produces a hash which matches such a pre-defined pattern. Sometimes this is referred to as the difficulty of mining. In principle, any peer node of a blockchain network can perform mining (i.e., collection of a set of transactions in a block to find the relevant nonce for it). PoW is a \emph{lottery-based} consensus mechanism, which implies that in a given large network, the peer who finds a nonce at a given time is decided randomly. Once a miner finds a nonce (or mines a block), the network awards such a node with a set number of \emph{cryptocurrency tokens} (such as bitcoins). This is how cryptocurrency is minted in cryptocurrency networks and is put into circulation in such networks.

\begin{figure*}[!ht]
\centering
\captionsetup{justification=centering}
\centerline{\includegraphics[width=.7\textwidth]{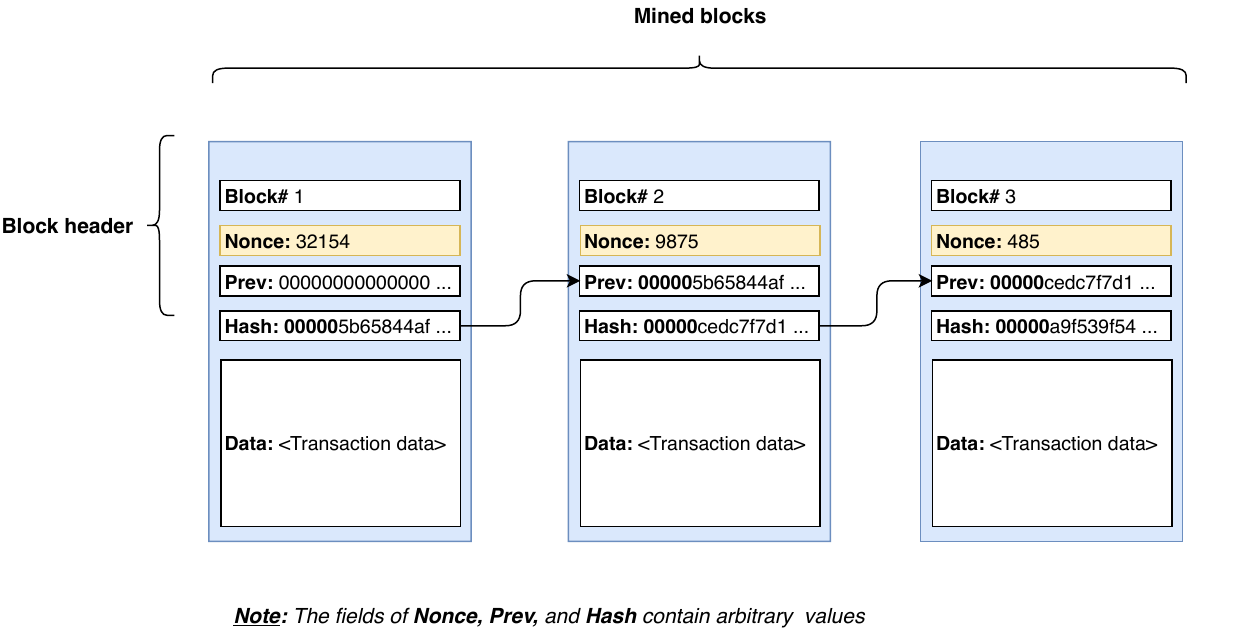}}
\caption{Mined blocks in a blockchain. Hash in each block now starts with \emph{five} zeros.}
\label{fig: mined_chain}
\end{figure*}

Furthermore, the mining process is based upon randomness, which renders adversarial tampering with the stored data in blockchain difficult as long as the majority of a network (in terms of computational resources) is honest. However, if an adversary (or a group of adversaries) gains more computational power than the honest portion of the network then it can potentially alter the records stored in a blockchain. Such an attack is sometimes referred to as a \textit{$51\%$ attack}. Figure \ref{fig: mined_chain} shows a chain of blocks with an extra field labeled as nonce. It should be noted in this figure that the hash of all the blocks (apart from the genesis block) starts from a set number of zeros. 
\vspace{2mm}
\subsubsection{Proof-of-Stake (PoS)}
\label{subsubsec: pos}
%
Blockchain-based systems, particularly Ethereum\footnote{\url{https://github.com/ethereum/wiki/wiki/Proof-of-Stake-FAQs}}, are considering an eventual shift to PoS- from PoW-based consensus. This is because of high computation, and in turn high energy costs associated with finding a nonce through mining.

In the PoS-based mechanism, the nodes with the largest stake (in monetary terms) in the underlying network have a greater say when it comes to proposing a new block to be appended to a blockchain. The monetary worth owned by such nodes is put at stake in order for them to behave honestly. An example of a PoS-based blockchain platform is Algorand \cite{chen2016algorand}, a permissionless blockchain platform (see Section \ref{subsec: pub_vs_priv_BC} for a discussion on public and private blockchains) that reduces the chances of forking (the undesirable process where two chains originate from a same block that reflects a conflict). Unlike PoW-based implementations, Algorand requires an insignificant amount of computation and generates a transaction history, which avoids forking with high probability. However, since PoS is still in its development phase, it does come with its fair share of issues. Most notable is the mismatch between the actual interest of nodes with the same stake in the underlying network\footnote{\url{https://tinyurl.com/poa-network}}.

\vspace{2mm}
\subsubsection{Proof-of-Authority (PoA)}
\label{subsubsec: poa}

Proof of authority (PoA) is another blockchain consensus approach mainly used to enable a comparatively fast transaction rate mostly in private blockchain settings \cite{de2018pbft, ekparinya2019attack}. PoA is derived from Byzantine Fault Tolerance based (BFT) consensus algorithms (see the next Section \ref{subsec: pbft} for details). Moreover, this PoA variant is mostly being used by the test networks mainly for experimentation (such as Rinkeby and Ropsten Ethereum networks). The idea of PoA is quite similar to PoS; in PoA  it is the identity (or reputation) of nodes that is put at stake instead of the monetary value owned by the nodes. This implies that PoA is mostly used to establish permissioned blockchains (see Section \ref{subsec: pub_vs_priv_BC}) where the identities of the peer nodes are known and they are given specific permissions to mine new blocks.
%
%
\vspace{2mm}
\subsubsection{Practical Byzantine Fault Tolerance}
\label{subsec: pbft}
Practical Byzantine Fault Tolerance-based consensus algorithm was first introduced for asynchronous systems (such as the Internet) to combat Byzantine faults \cite{castro2002practical}, such as arbitrary node behaviour that could imply software bugs, malfunctioning of a node, or an adversarial attack.
%
\label{subsubsec: bfs}
Byzantine faults are particularly of interest in the context of blockchain's peer-to-peer network. Byzantine faults imply an arbitrary behavior by peers of such networks due to adversarial malicious activities and software bugs that remain undetected particularly given the size and complexity of the software's (such as a set of smart contracts) source code\footnote{\url{https://tinyurl.com/the-dao-hack-explained}}.
\subsection{Smart contracts}
\label{subsec: smart_contracts}
One important aspect of blockchains is its use in enabling smart contracts \cite{luu2016making}. Smart contracts can simply be viewed as algorithmic enforcement of an agreement among, often, mutually non-trusting entities. More technically, a smart contract is a program that executes on blockchain in a distributed manner and possesses unique identification. It contains functions and state variables. These functions receive input parameters of the contract and get invoked when relevant transactions are made. The values of state variables are dependent on the logic contained in the functions \cite{ bahga2016blockchain}. These functions are normally written in high-level languages (such as Solidity or Python) \cite{ soliditydocs}. Compilers convert these programs into bytecode that is then deployed on a blockchain network. The functions contained within the bytecode of smart contracts are invoked when a node makes the relevant transaction aimed at the particular smart contract \cite{ bahga2016blockchain}. Smart contracts help automate the logic of an arbitrary value transfer system in an immutable manner where conditional transactions are recorded, executed, and distributed across the blockchain network. These contracts have the potential to reduce the legal (up to a certain extent) and enforcement costs while largely ruling out the need for central trusted or regulating authority \cite{bargar2016economics}. Smart contracts can create an environment of trust among the members of several contrasting and diverse communities \cite{swan2015blockchain}.

\emph{Ethereum}\footnote{https://ethereum.org/} was the first blockchain project that introduced and popularized the concept of smart contracts \cite{wood2014ethereum, kosba2016hawk}. It is an open-source, blockchain-based platform that enables one to develop and execute decentralized applications. One of Ethereum's goals is to ease the process of developing the decentralized applications called dApps \cite{ ethereum5, kakavand2016blockchain}. Ethereum can be considered as the next step, after Bitcoin, in the evolution of blockchain-based systems. Before Ethereum, most of the blockchain-based systems, mainly cryptocurrency-based projects, revolved around expanding on Bitcoin's core protocol and focusing on one specific application. Ethereum, however, generalizes and allows multiple such projects to coexist on a broader underlying blockchain-based compute resource.

Operations on Ethereum are performed by utilizing the Ethereum Virtual Machine (EVM). EVM is the implementation of the Ethereum protocol responsible for handling state transitions and carrying out computation tasks \cite{buterin2014next}. EVM provides the runtime environment for the execution of smart contracts \cite{bahga2016blockchain}. The EVM generated binary comprises smart contracts' opcode that gets deployed on the underlying blockchain.
%

%
\subsection{Public and private blockchains}
\label{subsec: pub_vs_priv_BC}
The underlying blockchains of Bitcoin, Ethereum and, in general, of most cryptocurrencies are open and public. This implies that anyone can join the blockchain network and transact with any other peer of the network. Moreover, such networks also encourage peers to stay anonymous. As an example in Bitcoin's network, peers are assigned addresses based on the hash of their public keys instead of based on their actual identities.

On the other hand, there are permissioned and private variants of blockchains as well. This concept was particularly popularized by Linux Foundation's Hyperledger Fabric (HLF) platform \footnote{\url{https://hyperledger-fabric.readthedocs.io/en/release-1.3/blockchain.html}}. This platform is proposed for business use cases where, in addition to data immutability and P2P consensus, transaction confidentiality is also required. Permissioned and private blockchain platforms such as HLF usually deploy a cryptographic membership service on top of their blockchain's immutable record keeping. Each peer in such a network can be uniquely identified based on its real-world identity. Proof-of-Authority (as discussed earlier) functions on the same principle of permissioned and private blockchains.

\subsection{Internet of value}
\label{subsec: int_of_value}
%
The value addition in businesses by blockchain technology is expected to grow to \$176 billion by 2025, according to Gartner\cite{www:www.gartner.com} Inc. Based on this technology, innovative payment channels are being introduced. One such example is Ripplenet \cite{www:ripple.com} that facilitates quick and lower-cost payments globally through its network of more than 300 financial institutions located in different geographical parts of the world.

\subsection{Digital assets}
\label{subsec: dig_assests}
%
A digital asset can be considered as the digital representation of a tradeable valuable that can be owned and used in a digital-value transfer system such as blockchain-based cryptocurrency networks. The use of digital assets is rising and evolving wave in the blockchain space. The potency to represent assets within a digitized system and carry out transactions via an open source blockchain technology is inspiring the creation of a whole new marketplace. The aim is to reduce the cost, risk, constrainsts, and fraud associated with the traditional trading systems. Digital asset tokens and the associated set of smart contracts can exemplify an arbitrary agreement among parties interested in a trade related to a digital asset. Such tokens further enhance efficacy in an end-to-end trading, services, and settlements towards a single coherent offering, and thus enable liquidity for previously illiquid markets .

An online blockchain-based game (developed on Ethereum network) of breeding digital cats called \emph{Cryptokitties} can be considered to understand the concept of a blockchain-based unique and tradeable digital asset. Cryptokitties is one of the earliest efforts to adopt blockchain technology for leisure and recreational activities. Most remarkably in December 2017, the popularity of the game congested the network of Ethereum, resulting in an all-time high volume of transactions .
CryptoKitties is an example of a non-fungible token (NFT) on the Ethereum-enabled blockchain network. The underlying logic that renders a Cryptokitty a unique tradeable asset is based on a smart contract stadnard called \emph{ERC721}. Cryptokitties can be regarded as unique and tradable ERC721 tokens where the value of these tokens can depreciate or increase according to the market. Hence, these Cryptokitties are secure against replication and cannot be transferred without the owner permission, i.e., even by the game creators.

In general, NFTs can be regarded as the tokenization (so that they can be rendered tradable on top of a blockchain-platform) of digital assets. Furthermore, ERC-721 provides a standard interface for NFT, where tokens represent a subset of Ethereum tokens. Since the initial publication of ERC-721 interface in 2017 as Ethereum Improvement Proposal (EIP), ERC-721-based tokens have allowed tokenization of ownership of any arbitrary data. It is important here to note that the key differentiator in NFTs is that every token is associated with a unique identifier, rendering each token unique to its respective owner. Lastly, unlike fungible ERC20 standard tokens that are interchangeable, such that users can create any amount of tokens using a single contract, ERC-721 standard requires each token to posses a different value within the same contract.


\textcolor{black}{The Ripple coin (XRP)\footnote{\url{https://ripple.com/xrp/}} is a further innovative option of tokens on the Ripple network used to establish transactional exchanges among parties that issue a new digital asset on the XRP ledger. Specifically, XRP can be transferred directly without a centralized party, rendering it a suitable solution in bridging different assets efficiently and speedily. Moreover, rather than leveraging the mining concept of blockchain, Ripple XRP adopts a unique and novel consensus mechanism via a network of servers in order to verify and validate transactions. This is achieved through a poll where servers on the network determine the authenticity and validity of all transactions based upon consensus.}

\textcolor{black}{
Moreover, in the realm of physical assets, blockchain technology can further enable digitisation of land registry system. Specifically, digitising registry systems via blockchain can enhance their reliability and transparency and reduce challenges of records' integrity. Deploying the distributed and shared database of blockchain can act as an incorruptible and unalterable repository of information for land registry records. An example of a use case related to land registry is \say{Blockchain Powered Land Registry in Ghana with BenBen}\footnote{\url{https://www.bigchaindb.com/usecases/government/benben/}}, which is a land registry system leveraging blockchain technology in order to help preserve property rights for citizens. BenBen\footnote{\url{http://www.benben.com.gh/}} has developed a top-of-stack land registry along with a verification platform for financial institutions, such that all transactions are captured and verified against the stored data.
This platform allows for a synchronized update of current registries and enable smart transactions and distribute private keys for users. As a result, a trusted and automated property transactions are enabled between all participating parties.
}


\subsection{Registration and digital identity}
\label{subsec: reg_id}
%

\textcolor{black}{
The concept of digital identity dates back to the beginning of the computer science era, which relates to issuer, user, and verifier as subjects of the digital identity system. However, issuance, storage, and presentation operations must further align with rigorous security requirements to fulfil blockchain operability specifications \cite{kuperberg2019blockchain}. These requirements include compatibility, unforgeability, integrity, scalability, performance/low latency, revocation, unlinkability, and selective disclosure. Schemes of privacy-enabled digital identity have been presented in the past, e.g., U-Prove and Idemix. However, these schemes are still not widely deployed and lack scalability and compatibility (i.e., assuming efficacious implementations requires a meta-system congregating multiple verifiers/issuers as well as credential schemas management. Furthermore, these traditional schemes require a global (centralized) third party, which must be trusted, for issuers data and parameters distribution and exchange.}

\textcolor{black}{
In order to address the aforementioned challenges in digital identity systems, Evernym, Inc. developed a practical digital identity scheme (of a global scale) called \say{Sovrin} \cite{khovratovich2017sovrin}. This scheme resolves operability and scalability issues based upon the use of permissioned blockchains and anonymous credentials concepts. The scheme further amalgamates revocation with anonymous credentials \cite{camenisch2002dynamic}, \cite{camenisch2009accumulator} for unforgeability, privacy, unlinkability and a distributed ledger, adopting practices from BFT \cite{lamport1982byzantine} and Ethereum \cite{wood2014ethereum} protocols.
\begin{itemize}
    \item Anonymous credentials for privacy: Idemix specification \cite{IDmixer:online}, \cite{camenisch2002design} is used as the anonymous credential module baseline. Unlike U-Prove, this module grants unlinkability by default and is built based upon the Charm framework \cite{Charm:online}, which offers a Python API for large integers, pairings, and signature mechanisms.
    \item Revocation feature and methods: Bilinear maps accumulators are used for revocation selection based upon \cite{camenisch2009accumulator}. However, the limitation here is that users need to be conscious about revoked credentials since the proof must be lively updated whenever issuer-specific data for the update is publicly communicated (i.e. prevent privacy leakage as non-revocation process/proof can reveal user's ID)
    \item Revocation with attribute-based sharding: A partitioning of credential IDs is adopted to thwart privacy leakage. The ID is partitioned into limited size shards $I_1$, $I_2$, ..., $I_n$, and the tail set for each shard becomes feasibly downloadable. The user therefore notify the verifier of their shard number so the latter can use the corresponding accumulator data. Additionally, a revocation-liveness parameter is also implemented in this module. Building upon this, the verifier determines the liveness of non-revocation proofs to be accepted (note that in order to restrain attacks against the revocation procedure, users are recommended to deny any specification requiring an accumulator younger than a day old.
\end{itemize}
}

\textcolor{black}{
Furthermore, various interoperability issues arise in blockchain networks include, but not limited to, energy consumption and regulation policies. Such issues are mainly due to the lack of standardized protocols for deploying blockchain-enabled mechanisms among different companies \cite{monrat2019survey}. While the number of companies interested in integrating blockchain technology has been dramatically evolving, standardization protocols to allow an efficient collaboration (among different blockchains) still do not exist which implies a lack of interoperability. Such an issue provides flexibility for blockchain developers to code with a variety of programming languages and platforms; nevertheless renders blockchain networks isolated and lack in-between interactions. A remarkable example here is the GitHub, which offers more than 6500 active blockchain-enabled projects (i.e., coded with different platforms and programming languages), protocols, and consensus algorithms. Hence a standard protocol is needed to permit collaborations within these developed applications and integration with existing blockchain systems \cite{monrat2019survey, jin2018towards}.}




\newcommand{\foo}{\makebox[0pt]{\textbullet}\hskip-0.5pt\vrule width 1pt\hspace{\labelsep}}
\begin{table}
\renewcommand\arraystretch{1.4}
\captionsetup{singlelinecheck=false,  labelfont=sc, labelsep=quad}
\caption{Timeline: Evolution of Blockchain}\vskip -1.5ex
\begin{tabular}{@{\,}r <{\hskip 2pt} !{\foo} >{\raggedright\arraybackslash}p{5cm}}
\toprule
\addlinespace[1.5ex]
2018 &  Blockchains potential got revamped by more investments in wide range of use cases\cite{usfsi20111:online}\\
2017 & Seven European banks, announced their program to develop a blockchain-based trade finance platform in collaboration with IBM \cite{IBMNewsr71:online}\\
2016 & Ethereum DAO code was compromised and hacked \cite{tikhomirov2017ethereum}, Emergence of permissioned blockchain solutions \cite{zheng2016blockchain}\\
2015 & Blockchain trial was initiated by NASDAQ \cite{zhu2016analysis}, Hyperledger project was started \cite{jacobovitz2016blockchain}\\
2014 & With crowdfunding the Ethereum Project was started \cite{zavolokina2016fintech}, Ethereum genesis block was created \cite{ozyilmaz2017integrating}, \cite{Launchin35:online}\\
2013 & Ethereum, a blockchain-based distributed computing platform was proposed \cite{Historyo98:online}\\
2012 & Coinbase, started as brokerage for Bitcoin \cite{Coinbase59:online}\\
2011 & Silk Road launched with Bitcoin as payment method \cite{hendrickson2016political}, BitPay first Blockchain-based wallet \cite{bitpayAp82:online}, Emergence of other cryptocurrencies like Swiftcoin \cite{Cryptocu96:online, BNAKLaun81:online, bruno2017system}, Litecoin \cite{zhao2015cryptocurrency}\\
2010 & First Bitcoin cryptocurrency exchange Mt. Gox started working \cite{ron2013quantitative}, \cite{yermack2015bitcoin}\\
2009 & First Bitcoin block was created \cite{wang2015exploring}, \cite{Bitcoins11:online}\\
2008 & Bitcoin's whitepaper was published by Satoshi \cite{nakamoto2008bitcoin}\\
\end{tabular}
\end{table}

\section{Blockchain-based Network Applications} 
\label{app}
\begin{figure*}[!t]
\centering
\captionsetup{justification=centering}
\centerline{\includegraphics[width=.7\textwidth]{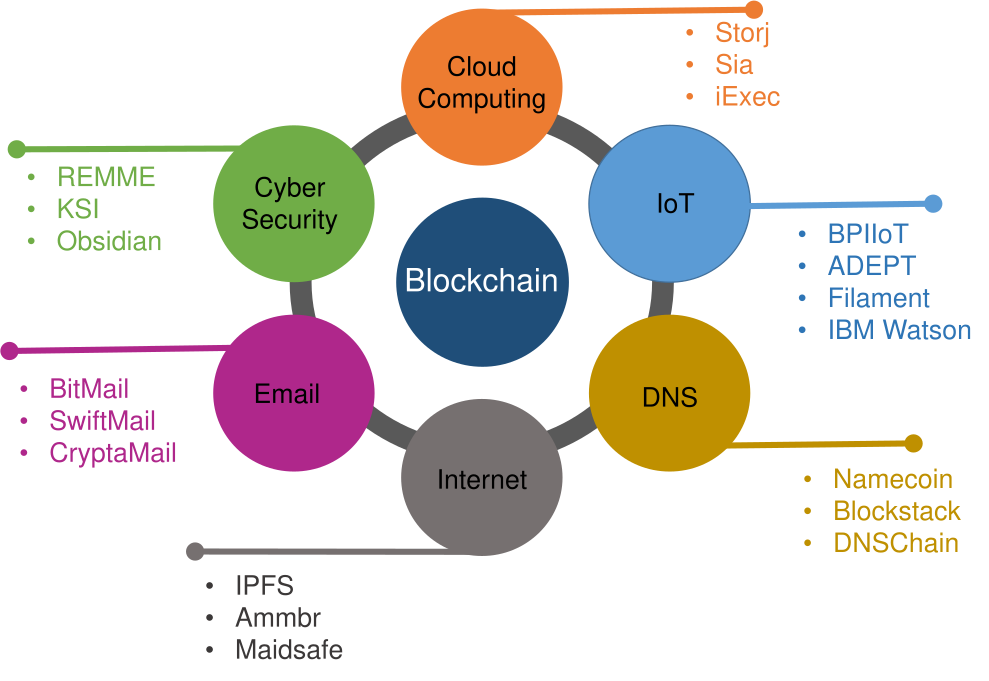}}
\caption{Examples of blockchain-based network applications and their solutions}
\label{fig:appp}
\end{figure*}

Other than cryptocurrencies, blockchain finds its applications in various other fields, particularly those that require more transparency and trust in their record-keeping. Some blockchain-based network applications with their platforms are shown in Fig. \ref{fig:appp}.

\subsection{Software-Defined Networks} \label{subsec: dist_internet}

\textcolor{black}{Software-Defined Networking (SDN) is an evolving networking technology that detaches data plane from control traffic. In such a technology, networking resources are managed by a centralized controller acting as the networking operating system (NOS) \cite{xie2019survey}. However, scalability is a major constraint in the single SDN-enabled networking environments, and thus the adoption of blockchain technology with SDN can help with facilitation of multi-domain SDNs interconnection and communication paving. For example, Sharma et al. \cite{sharma2017software} a fog-based solution is proposed leveraging multiple/distributed SDN controllers capabilities. Namely, in order to improve reliability and scalability, the blockchain technology is deployed here to distributively interconnect multiple SDN controllers. The presented solution is a decentralized cloud architecture based upon integration of SDN-enabled edge computing and blackchain technology, where the fog nodes are placed at the network edge. The architecture is distributed as three layers, cloud, device, and fog.}

\textcolor{black}{Blockchain in this solution is mainly used to record the QoS, service pool, and payments, while the proof-of-service plays the role of a consensus mechanism to control the service usage. However, this solution has not been implemented yet and security of fog nodes enabling communication across IoT entities remains an open research problem. Further studies such as, Sharma et al. \cite{sharma2017distblocknet} developed a blockchain-enabled distributed and secure SDN framework, where all controllers operate as blockchain entities to control the flow tables in SDN switching devices over the SDN data plane. Lastly, QiU et al. \cite{qiu2018blockchain} proposed distributed software-defined industrial IoT (SDIIoT) using the permissioned blockchain to improve security, reliability, and traceability across all distributed devices. This solution further resolves the limitation in permissioned Blockchain throughput and manages access operations to computational resources.}

\textcolor{black}{Furthermore, emergence of SDN and Network Function Virtualization (NFV) can provide virtualized edge platforms for future Internet development (IoT in particular). Virtual nodes in such virtualized platforms are dynamically managed and can render IoT-based shared edge feasible along with virtualized assets \cite{samaniego2016hosting}. However, the configuration assets in SDN are handled and maintained by a centralized control module, which therefore enables sophisticated centralized attack surfaces \cite{ali2018applications}. A remarkable solution was presented in \cite{sharma2017distblocknet} to resolve such a challenge through a decentralization of SDN control layer via blockchain technology. However, the security of the virtualized IoT assets using blockchain is yet to remain a major concern yielding an interesting future research direction \cite{ferrag2018blockchain}.}

\subsection{The Decentralized Internet}
\label{subsec: dist_internet}
The Internet has enabled the evolution of a number of applications such as mobile health, education, e-commerce, online social systems, and digital financial services. However many parts of the world are still deprived of the Internet's boons due to the existence of a digital divide \cite{gnangnon2017does, nekrasov2017limits, weidmann2016digital, park2017digital}. Moreover, the existing Internet infrastructure is predominantly centralized creating monopolies in the provision of services to its users \cite{klein2002icann, purkayastha2014us}. Distributed denial of service (DDoS) attacks on DNS servers\footnote{https://www.wired.com/2016/10/internet-outage-ddos-dns-dyn}, certificate authority compromises (as mentioned in Section \ref{Cert_Auth}), cybersecurity-related incidents \cite{jasper2017cyber, patrick2017need, karchefsky2017toward} and similar other service disruptions are rife mainly because of the largely centralized nature of the current Internet and the services that it provides \cite{Blocksta20:online}. Whereas, the decentralized approach to the online service provisioning gives more control to the users (or the edges of the Internet) and ensures fair participation and sharing of the resources. It is believed that decentralization of the communication infrastructure may bridge the gap of the digital divide and make the Internet services reachable to the remaining unconnected portion of the planet \cite{AmmbrWhi61:online}. 

In this section, we try to re-imagine different components of the Internet through the perspective of Blockchain's premise of decentralization and distributed trust.
\vspace{2mm}
\subsubsection{Decentralized naming systems}
\label{subsubsec: naming_system}
 Domain name system (DNS) is an example of online namespace system. Its primary goal is to resolve each unique hostname to an IP address(es) and vice versa. Presently, the largely centralized nature of DNS raises the odds for single-point failures and makes such systems prone to malpractice and malicious activities by the main stakeholders and governments. In the past, the seizure of hundreds of domain names by governments or the regulatory institutions have turned scientists, activists, and enthusiasts to think about possible alternatives to this largely centralized namespace system \cite{WikiLeak85:online, FourRoun21:online, bendrath2011end, denardis2012hidden, kalodner2015empirical}. 
 
Most applications place a demand for a namespace system that can ensure security during the provision of such identifiers. Blockchain can enable a namespace system by making use of global, tamper-resistant, and append-only ledgers and thereby guarantee the integrity, availability, uniqueness, and security of name-value pairs. While some challenges remain to be solved, the blockchain technology can successfully provide the essential basis for the construction and governance of secure and distributed naming services \cite{angieri2019distributed}. Such blockchain-based networks further encourage the inclusion of honest network peers since for a sufficiently large blockchain network, it becomes very difficult and costly for the adversarial elements to tinker with the blockchain records \cite{ali2016blockstack}.

In 2011, an experimental open-source startup called Namecoin came into being providing distributed DNS services based on blockchain technology with the aim of improved security mechanism, decentralization, confidentiality, and agility \cite{Namecoin94:online, Namecoin51:online}. Namecoin is designed to work on top of a blockchain and as an alternative to the existing conventional DNS root servers for the storage of registered domain names \cite{ali2016blockstack}. Being a blockchain-based system (with secretly held private keys corresponding to the registered domain names) it is immune to censorship or seizure of the registered domain name accounts. Similarly, any change in domain names, recorded on a blockchain, requires proof-of-work by the longest chain of honest network peers (see Section \ref{subsubsec: pow} for details), which in turn is in control of the highest computing pool \cite{kalodner2015empirical, back2002hashcash}. 

Another blockchain-based namespace system called Blockstack, inspired by the Namecoin network, improves upon various performance limitations of Namecoin (for a detailed analysis of Namecoin, please see \cite{kalodner2015empirical}) most importantly security and scalability \cite{ali2016blockstack}. The aspect of security was particularly improved by Blockstack by migrating from Namecoin's blockchain to Bitcoin's larger blockchain. The reason being the bigger size of Bitcoin's network, which makes it harder (as compared to Namecoin's relatively smaller network) for a 51\% attack \cite{kroll2013economics} (see Section \ref{subsubsec: pow}). One of the distinguishing features of Blockstack system is the introduction of a \textit{virtualchain} \cite{kirkmanusing}. Virtualchain is a logical overlay layer that sits on top of a production blockchain such as Bitcoin. Virtualchain eases the process of modifying the underlying blockchain without requiring actual consensus-breaking changes to it. Blockstack system facilitates users to register unique human-readable usernames and employs the distributed PKI system to bind user identities with arbitrary data values. This new registration system thus functions without the requirement of any centrally trusted third party \cite{ali2016blockstack, Blocksta20:online}. Blockstack enables users to own and control their data and access to this data at all times.


%


\subsubsection{Routing in the decentralized Internet}
\label{subsubsec: routing}
The interoperability of many still distinct (and largely isolated and self contained) blockchain networks will pose a problem in future if they are to come together to enable a wide-spread adoption of blockchain-powered decentralized web. There is a need for a routing mechanism that can take into account different characteristics of different blockchain networks and route a transaction from one network to a potentially different one and back. The main problem in inter-blockchain network routing is of verification of blockchain records among different blockchain networks and the provision of communication between any two peers belonging to any two distinct blockchain networks. In a single network this problem gets trivial with all the peers agreeing to follow the same consensus protocol (for example PoW). The motivation to enable interoperability among different blockchain networks can be taken from the concept of a \emph{lightweight client} of a blockchain network. Such clients are able to verify the existence of a record of a transaction in a blockchain network without downloading the entire bulk of blockchain data. The lightweight clients do so by making use of a technique called Simple Payment Verficiation (SPV)\footnote{\url{http://docs.electrum.org/en/latest/spv.html}} \cite{nakamoto2008bitcoin} which allows a client to verify the existence of a transaction record only by downloading the comparatively lightweight, block headers, in the form of a Merkle branch, in comparison to the entire blockchain data. Following a similar principle, \emph{Blocknet}\footnote{\url{https://blocknet.co/}} proposes a solution for inter-blockchain routing infrastructure \cite{blocknet:online}. Blocknet achieves interoperability by making use of two main components namely XBridge and XRouter. XBridge is responsible for implementing the \emph{exchange} functionality which implies enabling of atomic swaps of tokens between two blockchains. XRouter on the other hand implements \emph{communication} functionality and in unison with XBridge and making use of SPV a transaction can then be performed between two peers belonging to different blockchain networks.

Another project that proposes a solution to enable cross-ledger payments is called \emph{Interledger}\footnote{\url{https://interledger.org/}} \cite{interledger:online}. Interledger presents the concept of \emph{connectors} that act as decentralized exchanges between two distinct blockchain ledgers and route transactions (or \emph{packets of money} as per Interledger's vernacular). Interledger takes its inspiration from IP routing and instead of IP addresses it makes use of an ILP (Interledger packet) address. ILP packets differ from the \emph{best-effort} IP routing in the way that ILP packets can not be lost or stolen since in the case of ILP, funds with real monetary value are transferred instead of data. This is achieved by making use of Hashed Timelock Agreements (HTLA)\footnote{\url{https://interledger.org/rfcs/0022-hashed-timelock-agreements/}} in combination with SPV to settle cross ledger payment claims. HTLAs work across the ledgers and enable conditional transfers. Conditional transfers involve a \emph{preparation step} whereby a transfer is first prepared which implies that a sender's funds are put on hold by a ledger's contract until a condition is met which manifests itself in the form of a digest of a cryptographic hash function. Its incumbent on a recipient to present this digest in the form of a \emph{preimage} within a certain time window. If the time expires the funds are automatically released to the sender. This way, by making use of HTLAs the funds can not be lost in transit.

In conclusion, we see the problem of blockchain interoperability as akin to the Border Gateway Protocol's (BGP) routing problem where different Autonomous Systems (ASes) interoperate with each other with a mutually agreed upon control plane information. In our opinion these two problems seem to fit well together. Both domains (i.e., BGP routing and blockchain interoperability) can motivate solutions in each other. As an example, in our opinion, it would be beneficial if BGP attributes such as AS prefixes with corresponding control plane information (such as peering agreements) are stored in an immutable manner in a blocckhain-based database for routing checks. There will, however, be scalability and latency concerns as a blockchain's transaction rate must keep up with the dynamic nature of the changing network topologies in different ASes. Still, storage of network topological graphs with peering agreements will create an opportunity for a more trusted, transparent, and auditable routing decisions with a lesser chance for censorship and collusion.
\subsection{Decentralized Email}
\label{subsec: dist_email}
Today, electronic mail (email) is a common form of communication among many that usually consists of a mail client and an associated server.  There are various protocols such as SMTP, ESMTP, POP, and IMAP for formatting, processing, delivering, and displaying email messages by ensuring interoperability among different mail clients and servers. The security of an email system relies on a continuous process of planning and management. Email messages pass through the non-trusted external networks that are often beyond the control of an email provider's security system. These email messages, without appropriate security safeguards, can potentially be read, modified, and copied at any point along their path \cite{stine2010mail}.   Melissa, Sasser worm and other embedded hyperlinks and viruses have damaged millions of computers and their data \cite{ferguson2005fostering}. Email solutions (such as Yahoo) have suffered from data breaches in the past and have resultantly urged their users to change their password keys \cite{emailyahoo}. In order to improve on these centralized email systems to better safeguard the users' private and sensitive information, a radical change in the underlying technology seems imperative.

One of the solutions to address the vulnerabilities of the email system described above can be in the form of a blockchain-powered decentralized and distributed email system. Email addresses, in a similar way to DNS address assignment as discussed in the last section, can be assigned to the users over blockchain technology.  In this system, there is no centralized controlling server in order to gain access to personal data and records. Most importantly,
email communication using blockchain technology is not under the influence of government authorities that could exploit the centralized email providers such as ISPs and technology giants such as Google, Amazon, and Facebook, etc. John McAfee Swiftmail\footnote{http://johnmcafeeswiftmail.com/} is a blockchain-based email solution with $256$-bit end-to-end encryption for the protection of data. CryptaMail\footnote{http://www.cryptamail.com/} is another blockchain-based email service that claims $100\%$ security based on the decentralized system without third party involvement. Gmelius blockchain architecture is a hybrid system that offers a scalable and cost-effective framework that anchors email associated data into the Ethereum \cite{bersieremail}.

\textcolor{black}{Lastly, given the above research efforts on blockchain technology support for decentralized email systems, yet the Quality of Service (QoS) remains another significant concern of interest among future research trends. The email service is delay-sensitive and does not tolerate failures, whereas blockchain transactions can experience delays (or can even be ignored in a time period).}




%

\subsection{Blockchain for the Internet-of-Things (IoT)}
\label{subsec: bc_for_iot}
The Internet of Things (IoT) broadly speaking is a network of everyday objects in which the IoT devices capture or generate enormous amounts of data and send it over the network \cite{xia2012internet}. This interconnection of a large number of IoT devices is known to cause many privacy and security issues \cite{fernandez2018review, ahmed2017comprehensive, oravec2017emerging,sicari2015security}, including, but not limited to, authentication, privacy preserving, and data tampering/false data injection. The IoT-based social, such as health-related, applications often end up monitoring and collecting sensitive personal information. When such information is exposed to third parties, such as health-care providers, the prospects of inadvertent or malicious privacy compromises become highly probable \cite{ukil2014iot}. Compliance with the privacy and security rules and policies for a particular application is a significant challenge in IoT-based systems \cite{pasquier2018data}. In such systems, blockchain-based solutions can help in addressing the issues related to security and privacy. Besides the by-design existence of some implementation constraints of energy, delay, and computation overhead in IoT devices, businesses have started initiatives to use blockchain into their various domains such as in production and supply chain management \cite{kshetri2017can, huckle2016internet}. For example, the IBM Watson IoT platform\footnote{\url{ibm.co/2rJWCPC}} empowers the users to put their data on blockchain ledgers, which can later be used in shared transactions among different members of an IoT-related business consortium. This way members of such consortium can take part in verifying transactions against IoT data, dispute resolution, and accountability mechanism in a trusted, transparent, and mutually agreed upon manner. The data collected from devices in an IoT network is formatted into such API formats that are understandable to blockchain smart contracts. The IBM Watson IoT platform enables a business solution to manage, analyze, and customize IoT data, according to a pre-agreed policy, to be shared among permissioned clients, members, and smart contracts \cite{kshetri2017can}.

The importance of IoT can be gauged by observing the manufacturing industry, which is increasingly adopting IoT-based solutions for machine diagnostics, manufacturing automation, and health management of industrial machines \cite{bahga2016blockchain}. Cloud-powered manufacturing systems along with IoT technology help in the provisioning of manufacturing resources to the clients as per the existing demand. This usually requires the involvement of a centrally trusted third party. A blockchain-based platform called Blockchain Platform for Industrial Internet of Things (BPIIoT) is a trustless P2P network where the exchange of services may take place without the need for a central trusted third party \cite{bahga2016blockchain}. BPIIoT provides a platform for the development of dApps pertaining to P2P manufacturing applications. BPIIoT improves on a similar project called Slock.it\footnote{\url{https://slock.it/landing.html}}, according to the authors of \cite{bahga2016blockchain}, being generic in terms of dApp development. BPIIoT's platform consists of a single-board computer that provides a bridge to both cloud and blockchain services. BPIIoT enables customer-to-machine and machine-to-machine transactions without the involvement of third parties. For more details on the applications of blockchain for the Internet of things (IoT), the interested readers are referred to a comprehensive survey on this topic \cite{ali2018applications}.

\begin{table*}[!ht]
\centering
\scriptsize
\begin{tabular}{ |m{2.4cm} | m{3.8cm} |m{10.9cm}|}
\hline
\textbf{Scope}
& \textbf{Example(s)}
& \textbf{Description}   
 \\ \hline
\begin{tabular}[c]{@{}l@{}}Cryptocurrency\end{tabular}
&\begin{tabular}[c]{@{}l@{}}Bitcoin, Bcash, Iota, OmiseGO, \\Litecoin, Ripple, Dash, Zcash, Monero\end{tabular}
&\begin{tabular}[c]{@{}l@{}}Decentralized peer-to-peer electronic cash system for online payments.\end{tabular}\\\hline

\begin{tabular}[c]{@{}l@{}}Smart Contract\end{tabular}
&\begin{tabular}[c]{@{}l@{}}Ethereum \cite{wood2014ethereum}, Ripple\cite{luu2016making}\end{tabular}
&\begin{tabular}[c]{@{}l@{}}Occurrence of certain events triggers transfers of different things, i.e., security deposit payment, \\saving wallets, decentralized gambling, wills etc. \end{tabular}\\\hline

\begin{tabular}[c]{@{}l@{}}Cloud Services\end{tabular}
&\begin{tabular}[c]{@{}l@{}}Abuse Prevention\cite{szefer2013bitdeposit}\end{tabular}
&\begin{tabular}[c]{@{}l@{}}Defence to stop attacks and service abuses in cloud computing applications. \end{tabular}\\\hline

\begin{tabular}[c]{@{}l@{}}Message Exchange\end{tabular}
&\begin{tabular}[c]{@{}l@{}}Bitmessage\cite{warren2012bitmessage}\end{tabular}
&\begin{tabular}[c]{@{}l@{}}Secure system to send and receive messages. \end{tabular}\\\hline

\begin{tabular}[c]{@{}l@{}}Identity and Privacy\end{tabular}
&\begin{tabular}[c]{@{}l@{}}ChainAnchor\cite{shrier2016blockchain}\end{tabular}
&\begin{tabular}[c]{@{}l@{}}Trusted, privacy-preserving, identity management system. \end{tabular}\\\hline

\begin{tabular}[c]{@{}l@{}}Voting System\end{tabular}
&\begin{tabular}[c]{@{}l@{}}Electronic Vote\cite{noizat2015blockchain}\end{tabular}
&\begin{tabular}[c]{@{}l@{}}Electronic vote transaction system for a voter to spend the vote in favor of one or more candidate recipients. \end{tabular}\\\hline

\begin{tabular}[c]{@{}l@{}}Digital Content\end{tabular}
&\begin{tabular}[c]{@{}l@{}}Content Distribution\cite{kishigami2015blockchain}\end{tabular}
&\begin{tabular}[c]{@{}l@{}}Decentralized and peer-to-peer digital content management system with rights management mechanism. \end{tabular}\\\hline

\begin{tabular}[c]{@{}l@{}}Health\end{tabular}
&\begin{tabular}[c]{@{}l@{}}Patient Data\cite{peterson2016blockchain}\end{tabular}
&\begin{tabular}[c]{@{}l@{}}Patient data sharing system based on blockchain technology. \end{tabular}\\\hline

\begin{tabular}[c]{@{}l@{}}Transportation\end{tabular}
&\begin{tabular}[c]{@{}l@{}}Vehicle Communication\cite{dorri2017automotive}\end{tabular}
&\begin{tabular}[c]{@{}l@{}}Secure vehicle to vehicle communication system. \end{tabular}\\\hline

\begin{tabular}[c]{@{}l@{}}Agriculture\end{tabular}
&\begin{tabular}[c]{@{}l@{}}ICT E-Agriculture\cite{lin2017blockchain}\end{tabular}
&\begin{tabular}[c]{@{}l@{}}Distributed ledger system to safeguarded transparent data management. \end{tabular}\\\hline

\begin{tabular}[c]{@{}l@{}}Software\end{tabular}
&\begin{tabular}[c]{@{}l@{}}Software Connector\cite{xu2016blockchain}\end{tabular}
&\begin{tabular}[c]{@{}l@{}}Software components states sharing system without trusting a central integration point.\end{tabular}\\\hline

\begin{tabular}[c]{@{}l@{}}Micro Finance\end{tabular}
&\begin{tabular}[c]{@{}l@{}}Stellar\cite{mattila2016blockchain}\end{tabular}
&\begin{tabular}[c]{@{}l@{}}Creates services and financial products using blockchain architecture.\end{tabular}\\\hline

\begin{tabular}[c]{@{}l@{}}E-Commerce\end{tabular}
&\begin{tabular}[c]{@{}l@{}}OpenBazaar\cite{OpenBaza43:online}\end{tabular}
&\begin{tabular}[c]{@{}l@{}}Provides trading platform for users where they can make free transactions among themselves.\end{tabular}\\\hline

\begin{tabular}[c]{@{}l@{}}Mobile Banking \end{tabular}
&\begin{tabular}[c]{@{}l@{}}Atlas\cite{Atlas85:online}\end{tabular}
&\begin{tabular}[c]{@{}l@{}}Atlas provides platform for mobile banking and connects world communities through it.\end{tabular}\\\hline

\begin{tabular}[c]{@{}l@{}}Storage \end{tabular}
&\begin{tabular}[c]{@{}l@{}}Sia\cite{Sia60:online}\end{tabular}
&\begin{tabular}[c]{@{}l@{}}A cloud storage platforms, enables anyone to make money.\end{tabular}\\\hline

\begin{tabular}[c]{@{}l@{}}DNS \end{tabular}
&\begin{tabular}[c]{@{}l@{}}Namecoin\cite{wei2017review}\end{tabular}
&\begin{tabular}[c]{@{}l@{}}A blockchain-based domain name system.\end{tabular}\\\hline

\begin{tabular}[c]{@{}l@{}}Document Management \end{tabular}
&\begin{tabular}[c]{@{}l@{}}Blockcerts\cite{HomeBloc62:online}\end{tabular}
&\begin{tabular}[c]{@{}l@{}}Issue and verify certificates for academic, professional, workforce and civic records.\end{tabular}\\\hline

\begin{tabular}[c]{@{}l@{}}Storage \end{tabular}
&\begin{tabular}[c]{@{}l@{}}BigchainDB, MaidSafe,\\ Filecoin\cite{Bigchain57:online}\cite{MaidSafe99:online}\cite{Filecoin58:online} \end{tabular}
&\begin{tabular}[c]{@{}l@{}}Scalable storage which supports diverse applications, platforms, industries and use cases.\end{tabular}\\\hline

\begin{tabular}[c]{@{}l@{}}Business and Economy\end{tabular}
&\begin{tabular}[c]{@{}l@{}}IBM Blockchain Platform\cite{IBMBlock70:online} \end{tabular}
&\begin{tabular}[c]{@{}l@{}}Integrated platform designed for creation and acceleration of blockchain based businesses.\end{tabular}\\\hline

\begin{tabular}[c]{@{}l@{}}Internet of Things (IoT) \end{tabular}
&\begin{tabular}[c]{@{}l@{}}IBM Watson IoT\cite{IBMWatso80:online}\end{tabular}
&\begin{tabular}[c]{@{}l@{}}Accountability and security in blockchain-based internet of things.\end{tabular}\\\hline

\end{tabular}
\caption{Examples of blockchain-based applications}
\label{table: Apps}
\end{table*}

\begin{table*}[!ht]
\centering
\scriptsize
\begin{tabular}{ |m{2.4cm} | m{3.8cm} |m{10.9cm}|}
\hline
\textbf{Scope}
& \textbf{Startups}
& \textbf{Description}   
 \\ \hline
 
\begin{tabular}[c]{@{}l@{}}IoT and Economics\end{tabular}
&\begin{tabular}[c]{@{}l@{}}Chronicled\cite{Chronicl81:online}\end{tabular}
&\begin{tabular}[c]{@{}l@{}}Provides trusted data, ensures data provenence of IoT devices and helps in business process automation\end{tabular}\\\hline

\begin{tabular}[c]{@{}l@{}}Security and Intelligence\end{tabular}
&\begin{tabular}[c]{@{}l@{}}Elliptic\cite{Elliptic79:online}\end{tabular}
&\begin{tabular}[c]{@{}l@{}}Necessary intelligence information to security agencies and financial departments.\end{tabular}\\\hline

\begin{tabular}[c]{@{}l@{}}Data Security\end{tabular}
&\begin{tabular}[c]{@{}l@{}}LuxTrust\cite{LuxTrust43:online}\end{tabular}
&\begin{tabular}[c]{@{}l@{}}Provides security to customer's electronic data and digital identity.\end{tabular}\\\hline

\begin{tabular}[c]{@{}l@{}}Regulatory Compliance\end{tabular}
&\begin{tabular}[c]{@{}l@{}}GuardTime\cite{DataCent57:online}\end{tabular}
&\begin{tabular}[c]{@{}l@{}}Data protection regulatory compliance software.\end{tabular}\\\hline

\begin{tabular}[c]{@{}l@{}}Financial\end{tabular}
&\begin{tabular}[c]{@{}l@{}}Augur\cite{Decentra22:online}\end{tabular}
&\begin{tabular}[c]{@{}l@{}}A market forecasting tool to increase profitability.\end{tabular}\\\hline

\begin{tabular}[c]{@{}l@{}}Transportation\end{tabular}
&\begin{tabular}[c]{@{}l@{}}Lazooz\cite{LaZooz5:online}\end{tabular}
&\begin{tabular}[c]{@{}l@{}} Real-time ridesharing services.\end{tabular}\\\hline

\begin{tabular}[c]{@{}l@{}}Property Records\end{tabular}
&\begin{tabular}[c]{@{}l@{}}Ubiquity\cite{UBITQUIT21:online}\end{tabular}
&\begin{tabular}[c]{@{}l@{}}Provide service for secure ownership record of property.\end{tabular}\\\hline

\begin{tabular}[c]{@{}l@{}}Process Compliance\end{tabular}
&\begin{tabular}[c]{@{}l@{}}Startumn\cite{Stratumn57:online}\end{tabular}
&\begin{tabular}[c]{@{}l@{}}Ensures process integrity and improves regulatory compliance.\end{tabular}\\\hline

\begin{tabular}[c]{@{}l@{}}Music\end{tabular}
&\begin{tabular}[c]{@{}l@{}}Mycelia\cite{Myceliaf27:online}\end{tabular}
&\begin{tabular}[c]{@{}l@{}}Music industry online services.\end{tabular}\\\hline

\begin{tabular}[c]{@{}l@{}}Asset Management\end{tabular}
&\begin{tabular}[c]{@{}l@{}}Gem\cite{Introduc12:online}\end{tabular}
&\begin{tabular}[c]{@{}l@{}}Secure identification of assets.\end{tabular}\\\hline

\begin{tabular}[c]{@{}l@{}}Data Security\end{tabular}
&\begin{tabular}[c]{@{}l@{}}Tieriom\cite{TierionB18:online}\end{tabular}
&\begin{tabular}[c]{@{}l@{}}Data protection service.\end{tabular}\\\hline

\begin{tabular}[c]{@{}l@{}}Tracking and Ownership\end{tabular}
&\begin{tabular}[c]{@{}l@{}}Provenance\cite{Provenan44:online}\end{tabular}
&\begin{tabular}[c]{@{}l@{}}Maintain digital history of things.\end{tabular}\\\hline

\begin{tabular}[c]{@{}l@{}}Music\end{tabular}
&\begin{tabular}[c]{@{}l@{}}Ujo Music\cite{UJO40:online}\end{tabular}
&\begin{tabular}[c]{@{}l@{}}An online music store.\end{tabular}\\\hline

\begin{tabular}[c]{@{}l@{}}Smart Contracts\end{tabular}
&\begin{tabular}[c]{@{}l@{}}SkuChain\cite{Skuchain28:online}\end{tabular}
&\begin{tabular}[c]{@{}l@{}}Offers services like: Smart contracts, provenance of things, Inventory Management.\end{tabular}\\\hline

\begin{tabular}[c]{@{}l@{}}Storage\end{tabular}
&\begin{tabular}[c]{@{}l@{}}Storj\cite{StorjDec1:online}\end{tabular}
&\begin{tabular}[c]{@{}l@{}}A distributed storage platform.\end{tabular}\\\hline

\begin{tabular}[c]{@{}l@{}}E-commerce \end{tabular}
&\begin{tabular}[c]{@{}l@{}}Gyft\cite{GyftBloc41:online} \end{tabular}
&\begin{tabular}[c]{@{}l@{}}An online gift transfer platform.\end{tabular}\\\hline

\begin{tabular}[c]{@{}l@{}}Firearms\end{tabular}
&\begin{tabular}[c]{@{}l@{}}BlockSafe\cite{Blocksaf95:online} \end{tabular}
&\begin{tabular}[c]{@{}l@{}}A secure and privacy enabled firearm solution.\end{tabular}\\\hline

\begin{tabular}[c]{@{}l@{}}Health and Environment\end{tabular}
&\begin{tabular}[c]{@{}l@{}}BitGive\cite{BitgiveF58:online}\end{tabular}
&\begin{tabular}[c]{@{}l@{}}By using blockchain technology it works for the improvement of public health and environment worldwide.\end{tabular}\\\hline

\end{tabular}
\caption{Examples of blockchain-based startups}
\label{table: Appss}
\end{table*}

Another IoT project, managed by IBM in collaboration with Samsung, is the blockchain-powered and Ethereum-based Autonomous Decentralized Peer-to-Peer Telemetry (ADEPT) system. Ethereum is a blockchain-based generalized technology that can be considered as the compute framework for trustful messaging. Contracts authored under this framework endorse the rules designed for interaction between network nodes and thus are considered more secure. It also provides developers with a platform for building applications integrated with the Ethereum message passing framework \cite{wood2014ethereum}. ADEPT realizes a decentralized IoT solution by following the three principles: i) P2P messaging, ii) distributed file sharing, and iii) autonomous coordination among the devices of IoT network. ADEPT makes use of Telehash (an encrypted mesh networking protocol)\footnote{\url{http://telehash.org}}, BitTorrent, and Ethereum respectively to realize the three principles just described. Ethereum's blockchain enables device owners of ADEPT's IoT network to automate rules of engagement, the registration and authentication processes, and interactions among themselves in a decentralized and trusted manner. This can be achieved in one of two ways namely: i) proximity-based: taking into consideration physical, temporal or social distance and ii) consensus-based: taking into consideration selection, validation, or blacklisting criterion \cite{Empoweri53:online, Adepttec30:online, karstconnecting}. 

Among other works is Filament, a blockchain-based technology stack that enables IoT devices to discover, register, manage, and communicate in a decentralized manner \cite{Filament18:online, worner2016bitcoin}. In \cite{bocek2017blockchains}, a system named modum.io\footnote{\url{https://modum.io}} has been presented, which utilizes blockchain-based IoT devices to ensure the immutability of the transactions related to physical products and facilitates in the regularization of the supply-chain management process in the various fields \cite{gonczol2020blockchain}. 
\textcolor{black}{Given the growth of blockchain aligns with transactions storage by users, robust miners must handle consensus protocols in the blockchain. Hence as discussed earlier, various energy efficacious consensus algorithms were presented to store only recent transactions (e.g., mini-blockchain \cite{francca2015homomorphic}, proof-of-stake \cite{bitshares:online}, and proof-of-space and delegated proof-of-space \cite{dziembowski2015proofs, fan2018roll}). The challenge however in IoT devices is the resource and power constraints that render them typically unable to fulfill the essential power consumption and computation in handling consensus and blockchain storage. Hence elaborating power efficient consensus mechanisms is a grand research challenge over IoT-enabled blockchain.}

\textcolor{black}{Remarkable work has recently been presented to address constrained resources based upon enabling blockchain for IoT environments. Most notably, Xu et al. \cite{xu2017intelligent} proposed a smart resource management for cloud datacenters (where billions of IoT devices transfer data to the cloud using virtualization technologies via Internet connection) by leveraging blockchain technology. Namely, the proposed mechanism minimizes energy consumption cost that is achieved through enabling users to sign transactions with their private keys, whereas neighbor users are capable to validate or reject broadcast transactions. Sharma et al. \cite{sharma2017software} further presented a cloud architecture based upon emerging blockchain technology with fog computing and software-defined networks (SDN). Specifically, blockchain capabilities are deployed here to ensure availability and scalability of networking-enabled services, while SDN controllers of fog hosts grant efficient management PIs to network operators. Further studies such as Xia et al. \cite{xia2017medshare} presented a data sharing system leveraging blockchain technology named MeDShare. This proposed solution operates based upon three key layers; user, data query, and data structuring and provenance layers.}

\textcolor{black}{Besides these efforts, Jiang et al. \cite{jiang2020searchain} presented Searchain, a keyword search system that intends to improve efficiency in data storage and privacy over heterogeneous IoT-enabled storage resources. Specifically, Searchain grants a private keyword search in decentralized storage systems based upon two key modules, blockchain of ordered blocks and P2P architecture-based transaction hosts. Tapas et al. \cite{tapas2018blockchain} further addressed security challenges in IoT-enabled blockchain, namely authorization and delegation. The proposed solution is designed and integrated as smart contracts handler in the Ethereum system and furtherly offers authorization and access control management over IoT devices. Moreover, Alphand et al. \cite{alphand2018iotchain} presented a further security architecture to enforce authorization and access control to IoT devices through blockchain technology. The proposed solution, named as IoTChain, delivers an efficacious multicasting of IoT resources based upon a conjunctive integration of the ACE authorization framework \cite{seitz2017authentication} and the OSCAR architecture \cite{vuvcinic2015oscar}.}

\subsection{Blockchain-based Content Distribution}
\label{subsec: bc_cdn}
Content distribution networks (CDNs) are an effective approach to improve Internet service quality by replicating the content at different strategic geographic locations in the form of data centers. Users can request and access data from the closest replica server instead of always fetching it from the data-originating server. Generally, large companies such as Netflix and Google's YouTube service, have their own dedicated CDNs, while smaller organizations can rent CDN space from other companies like Akamai. BitTorrent is a P2P content distribution protocol that enables the propagation of data using networks of computers for downloading and uploading simultaneously without a central server \cite{pouwelse2005bittorrent}. BitTorrent's network consists of a large number of peers, which complicates the task of traffic management. The other major issue with the current CDNs is that the content creators receive an inadequate share of the revenue, especially in digital content distribution sector \cite{CDNIssus}. Similarly, the media sector is also significantly suffering because the content can be easily copied and distributed. 

Blockchain technology can be the solution with the necessary ingredients to significantly resolve the challenges related to content distribution. It can stabilize the rights management related issues for studios and artists by providing a better way of content control. This can enable a more agile method for content delivery with a more trusted, autonomous, and intelligent network. In a blockchain-based CDN, the participants can independently verify a record and its origin without the need for a centralized authority for verification. Blockchain can store all the record related to the content (e.g., its origin), and share over the network in an immutable form along with the provision of enabling a monetization system to empower the content creators.

DECENT\footnote{https://decent.ch/}, as an example, is a blockchain-based CDN that provides secure content distribution and maintains the reputations of the content creator with a  mechanism for the payment between authors and client nodes also in place. Content (e.g., ebooks, videos, and audio) is released cryptographically over the global DECENT network and other nodes can then purchase them with DECENT tokens. SingularDTV\footnote{\url{https://singulardtv.com/}} is a media industry initiative in which an Ethereum-based entertainment studio is developed that can enable rights management as well as P2P distribution to empower artists and creators. 

\subsection{Distributed Cloud Storage} \label{subsec: dist_cloud_storage}
\begin{table*}[!ht]
\centering
\scriptsize
\begin{tabular}{ |m{5cm} |m{10.6cm}|}
\hline
\textbf{Platform}
& \textbf{Description}   
 \\ \hline

\begin{tabular}[c]{@{}l@{}}Swarm \cite{Swarm:online} \end{tabular}
&\begin{tabular}[c]{@{}l@{}}An open Infrastructure for Digital Securities \end{tabular}\\\hline

\begin{tabular}[c]{@{}l@{}}InterPlanetary File System (IPFS) \cite{IPFS:online} \end{tabular}
&\begin{tabular}[c]{@{}l@{}}A protocol and peer-to-peer network for storing and sharing data in a distributed file system \end{tabular}\\\hline

\begin{tabular}[c]{@{}l@{}}Sia \cite{Sia60:online} \end{tabular}
&\begin{tabular}[c]{@{}l@{}} A platform for securing storage transactions with smart contracts \end{tabular}\\\hline

\begin{tabular}[c]{@{}l@{}}MaidSafe \cite{MaidSafe99:online} \end{tabular}
&\begin{tabular}[c]{@{}l@{}}A decentralized platform for application development via a proof-of-resources protocol \end{tabular}\\\hline

\begin{tabular}[c]{@{}l@{}}Storj \cite{StorjDec95:online} \end{tabular}
&\begin{tabular}[c]{@{}l@{}}A decentralized file storage solution over P2P network using blockchain hash table \end{tabular}\\\hline

\begin{tabular}[c]{@{}l@{}}Filecoin \cite{Filecoin58:online} \end{tabular}
&\begin{tabular}[c]{@{}l@{}}A digital payment system and blockchain-based cooperative digital storage \end{tabular}\\\hline


\begin{tabular}[c]{@{}l@{}}BlockScores/NextCloud \cite{BlockScores:online} \cite{NextCloud:online} \end{tabular}
&\begin{tabular}[c]{@{}l@{}}An application for blockchain and smart contract interacting via secure leaderboards \end{tabular}\\\hline


\end{tabular}
\caption{\textcolor{black}{Examples of blockchain platforms for distributed cloud storage}}
\label{table: Appss}
\end{table*}

Today, consumers and enterprises face the storage and management problems caused by an ever-increasing volume of data on non-volatile data storage systems. Despite the popularity of cloud storage solutions (such as Dropbox and Google Drive), the control, security, and privacy of data remain major concerns \cite{crosby2016blockchain}. It is largely due to the current model being adopted by the cloud storage systems that often puts them under a centralized institutional authority. In this model, data is transferred over TCP/IP from a client to the host servers in the legacy client-server model \cite{wilkinson2014metadisk}. The information thieves, censorship agencies and spies can potentially tamper with or copy the stored confidential files from hosting servers through technological means, legal tactics and political strategies \cite{kaufman2009data,kandukuri2009cloud,subashini2011survey,wang2009enabling,wang2010privacy}. 

Such problems, mostly caused by central and identifiable points in the current cloud storage systems\footnote{\url{https://newsroom.fb.com/news/2018/09/security-update/}}, can potentially be solved using decentralization and (transparent and trusted execution in the form of) automation based on a trust agreement between a client and a host service provider. There exist some storage solutions such as MaidSafe\footnote{\url{https://maidsafe.net}} and Tornet\footnote{\url{https://github.com/bytemaster/tornet}} that outline possible alternatives for a decentralized cloud, but security, scalability,
and cost efficiency of these solutions still remain in question. Therefore, a cloud storage system with trusted and verifiable security guarantees, high redundancy, and scalability, is required that should be economically viable while being practical at the same time. Blockchain-based cloud storage solutions inherit characteristics such as decentralization, anonymity, and trusted execution of transactions among the members of a trust agreement and can pave the way for a verifiable and trusted cloud computing era \cite{gai2020blockchain, wang2019efficient}. 

Storj\footnote{\url{https://storj.io}} is a blockchain-based P2P distributed data storage platform that enables users to tailor their data sharing and storage as per individual agreements with other network peers and the third party service providers. Entities can earn cryptocurrency-based micro-payments by sharing the unused disk space and Internet bandwidth of their computing devices. In the context of distributed cloud, Dong et al. \cite{dong2017betrayal} proposed a  game-theoretic, smart-contract-based verifiable cloud-computing framework. This enables the clients to analyze collusion between two different clouds by making them perform the same computing task. In this framework, the users use smart contracts to simulate distrust, tension, and betrayal between the clouds to detect, and in turn, avoid cheating and collusion. Similarly, Sia\footnote{\url{https://sia.tech}} is another blockchain-based cloud storage platform. Sia platform automates trusted service level agreements (SLAs) between a user and storage provider using smart contracts. It is an open source platform that splits users' data into encrypted fragments and distributes them across a P2P network that increases network resilience and reduces downtime. Unlike the traditional storage solutions, the data in this scheme becomes more secure in the sense that one can only access this data if in possession of associated cryptographic keys. Another important work is Filecoin \cite{filecoinWhitepaper}. Filecoin realizes the concept of distributed storage network in terms of an algorithmic marketplace for storage. Filecoin is built as an incentive layer on top of another distributed file system called Inter-Planetary File System (IPFS). The miners in Filecoin host the storage space with the mining capability determined by the storage capacity a miner possesses. Filecoin enables verifiable markets, which dictates how and where data is written to and read from. Each read/write transaction is powered by the underlying cryptocurrency called Filecoin.


\subsection{Applications in Online Social Networks}
\label{dist_OSNs}
The engagement of people with online social networks (OSNs) has increased greatly in recent years \cite{Numbero75:online}. Users often put trust in these OSNs and share their personal details with their online social community. Privacy and security concerns however still remain an issue with many OSNs. Any breach of trust has the potential to detriment a user's virtual and, often in turn, real-world identities \cite{fire2014online}. As an example, in one of the biggest data breaches\footnote{https://www.nytimes.com/2018/03/19/technology/facebook-cambridge-analytica-explained.html}, a data firm named Cambridge Analytica got the access to personal information of more than 50 million Facebook (an online social network) users during 2016 US presidential campaign. The firm provided software tools to analyze/predict American voters' behavior/personalities and influenced their choices of the ballot\footnote{https://www.theguardian.com/news/2018/mar/17/cambridge-analytica-facebook-influence-us-election}.

Decentralization, transparency, and P2P consensus gives blockchain the potential to address most of these aforementioned security and privacy concerns prevalent in OSNs  \cite{mitblock86:online}. As an example, a blockchain-based social media platform named ``Steem''\footnote{https://steem.io} gives online community an opportunity to have a say on the nature of the  content that gets popular on a social network. Steem enables users to earn rewards on the basis of votes received by the community against their contributions \cite{SteemWhi91:online}. This encourages an honest participation of community peers in maintaining the quality of the overall network. Such OSN systems can further be made self-healing by a blockchain-based ``reputation system'', such as the one proposed by Dennis et al. \cite{dennis2015rep}. This system keeps records of users' reputation based on their transaction history. In our opinion, such techniques, while not being free of some ethical concerns, greatly reduce the snooping and policing by the centralized authorities such as governments\footnote{\url{https://www.theguardian.com/world/2013/jun/06/us-tech-giants-nsa-data}}. 

\subsection{Cybersecurity}

A study on cybercrime \cite{2017Cost74:online} conducted on some organizations, says that information loss remained the major cost component and increased from 35\% in 2015 to 43\% in 2017.  Blockchains in particular can be a costly target for cyberattacks \cite{IECBlock59, myers2017block}. As an example, DDoS attacks on a blockchain system can take the form of flooding the network with small transactions. Still such transactions must be paid for (in the units of gas) in order for them to be confirmed by the network \cite{IECBlock59}. The operations that require very (disproportionately) low gas costs are vulnerable to exploitation by attacker\footnote{\url{https://www.coindesk.com/so-ethereums-blockchain-is-still-under-attack/}}. However, when it comes to the execution of smart contracts then there is a large attack surface area \cite{liu2019survey}. This is because, often, a set of smart contracts is deployed to automate an application with all of its members working in unison. If one member of such a set malfunctions it can then trigger a domino effect rendering the whole set malfunction \cite{IECBlock59}. As an example an ambitious ethereum-based project implemented called decentralized autonomous organization (DAO) got hacked resulting in the theft of about $60$M Ether\footnote{\url{https://tinyurl.com/DAOattack}}. Such attacks can further be avoided by providing further trust guarantees for the code and logic of the smart contract itself. For instance, Tezos proposes the concept of a \emph{self amending ledger} and to make the deployment of a smart contract more trusted it provides formal proofs of the code of a smart contract in order to secure the trust of all the parties interested in the execution of this smart contact \cite{tezosWhitePaper}.

Based on blockchain technology, REMME\footnote{https://www.remme.io} is a password authentication system for safeguarding the confidential credential information from cyberattacks and at the same time disregarding the need to remember passwords \cite{WhitePap7:online}. 

Estonian cryptographer Ahto Buldas co-founded an information security company named Guardtime\footnote{https://guardtime.com} in 2007. This company has been working to secure sensitive records using blockchain technology. The company has designed a Keyless Signature Infrastructure (KSI) \cite{KSIdatas0:online} against the commonly used Public Key Infrastructure (PKI). In this new infrastructure, centralized Certificate Authority (CA) uses asymmetric encryption and manages public keys. Thus helping in reducing the risk of informational asset loss from cybersecurity-related incidents.

Obsidian\footnote{https://obsidianplatform.com} is also blockchain technology based platform for secure message exchange without any provisioning of centralized management mechanism. In this system, the meta-data about the undergoing communications is spread out in distributed ledgers and cannot be collected at centralized locations. Hence, in the context of cybersecurity, it decreases the chance of surveillance or tracking and in this way addresses privacy issues \cite{Obsidian56:online}. 

\subsection{Public Key Infrastructure (PKI): Certificate Authority (CA)}
\label{Cert_Auth}
Public Key Infrastructure (PKI) establishes a link between identities like domain names to a cryptographic public key with help of certificates \cite{polyzosblockchain, conti2018survey}. Among traditional approaches to PKIs, the most common choice is the use of Certificate Authority (CA) that serves as a trusted third party and manages the distribution digital certificates over the network. This creates a single point of failure in such PKIs in practice \cite{fromknecht2014decentralized}. There have been many incidents when these centralized CA's have been compromised---e.g., the DigiNotar attack: 531 fraudulent certificates issued \cite{prins2011diginotar} \cite{fisher2012final}; Trustwave's issuance of digital ``skeleton key'' for surveillance \cite{Trustwav26:online}; Debian's predictable random number generator in the OpenSSL package \cite{DebianSe99:online}; Stuxnet malware: compromise on code-signing certificates \cite{w32stuxn78:online} \cite{seltzer2013securing}; Duqu malware: stealing of digital certificates along with the private keys \cite{bencsath2011duqu, bencsath2012cousins, bencsath2012duqu, faisal2012stuxnet}; and the console hacking of Playstation 3 with compromised private keys \cite{schmid2015ecdsa}. 

Developing a blockchain-based PKI is a feasible alternative to the existing PKIs, which can provide the required security properties \cite{axon2015privacy}. In a blockchain-based implementation of the PKI system, the user identities are bound to public-keys using distributed public ledgers \cite{ali2016blockstack}.
A blockchain-based decentralized PKI system called ``CertCoin'' for secure identity management and retention has been in use. This system trusts the majority of peer network users instead of any central trusted party. It has two different mechanisms for verification of the known public key and the lookup for a new public key, which are supported by decentralized efficient data structures \cite{fromknecht2014decentralized}. In \cite{qin2017cecoin}, another blockchain-based distributed PKI scheme has been proposed that resolves the single point failure issue. This scheme ensures validity and ownership consistency of public-key certificates by miner's proof-of-work. It uses Merkle Patricia tree (see for details \cite{Patricia69:online}) for efficient accessibility of certificates without relying on any central trusted third party. Similarly other blockchain-based PKIs have been discussed in \cite{axonpb,al2017scpki,tewarix509cloud,matsumoto2016ikp}.
\subsection{Other Applications}

Using the blockchain technology, a company named Factom has started a land registration project with the Government of Honduras to ensure integrity and correctness of the information. Using the same technology, they have engaged in projects related to smart cities, document verification, and the finance industry \cite{underwood2016blockchain}. 

In another application, a blockchain-based startup Everledger is working on bringing transparency to the supply chain of diamonds, which was previously perceived as complex, risky and prone to carrying false and incomplete information. Everledger has been designed to reduce fraudulent modifications in the records to help financial institutions, businesses, and insurance companies with actual details of information \cite{yeoh2017regulatory}.

A bitcoin-based startup Abra for transferring money to anyone with minimal charges of transaction. No intermediate party gets involved in this transaction \cite{AbraBuyS84:online}. Blockchain is being considered as a novel software connector, which can provide a decentralized alternative to existing centralized systems resulting in quality attributes. For example, Xu et al. \cite{xu2016blockchain} found that blockchain can improve information transparency and traceability as a software connector.

Openchain\footnote{https://www.openchain.org} is a distributed ledger based system, which helps in the management of digital assets while ensuring their robustness, security, and scalability. AKASHA\footnote{https://akasha.world} provides people with a platform to publish and share their content online. Participants of this system get rewarded for their content based on the votes against their entries.

OpenBazaar\footnote{https://www.openbazaar.org} is a blockchain-based platform, which facilitates people to make transactions freely among themselves. Users of this system cannot censor the transactions or freeze the payments. Users also enjoy the flexibility of sharing information as much as they want. However, the buyers and sellers can engage intermediate moderators to resolve any dispute that may arise between the involved parties \cite{mattila2016blockchain}.

\section{Challenges and the Road Ahead}
\label{sec: challenges}
The blockchain is expected to drive economic changes on a global scale by revolutionizing industry and commerce by redefining how digital trust mechanisms through distributed consensus mechanisms and transparent tamper-evident recordkeeping. The disruption of blockchain is evident, and people are beginning to adopt this distributed ledger technology. There are, however, various hurdles that are slowing down the rate of blockchain's adoption. Some of these challenges are discussed below and with pointers to how these challenges might find a solution in the future. 

\subsection{Governance, Operational \& Regulatory Issues}
\label{subsec: governance}
Blockchain has great potential to enable efficient and secure real-time transactions across a large number of industries by providing financial services visibility along a supply chain and streamlining government authorities and consumers. Blockchain technology is still far from being adopted en masse due to some unsolved challenges of standards and regulation. Although it’s hard to regulate the development of the blockchain technology itself, blockchain-based activities (such as financial services, smart contract, etc.) should be regulated \cite{cermeno2016blockchain}.  To support its emergence and commercial implementation, the development of standards and regulations are required to establish market confidence and trust. These regulations can also be used for law enforcement to monitor fraudulent activities e.g., money laundering. 

In May 2016, a complex set of smart contracts named Decentralized Autonomous Organization (DAO) was built on top of Ethereum blockchain. It was a crowd-funding platform for defining organization rules\footnote{https://www.coindesk.com/understanding-dao-hack-journalists}. After this smart contract's creation, there was a period of funding during which users could earn its restrictive ownership by purchasing Ether (i.e., the underlying cryptocurrecy). After the completion of that funding period, the DAO started its operation in which the restrictive owners (also called members) casted their votes for the usage of collected funds. Initially, this operation was very successful and raised over \$150M from 11,000 members within a one month duration \cite{bharadwaj2016blockchain}. In June 2016, almost \$70M were drained after a hack making use of a recursive call exploit. The hackers used this exploit to get Ether back from DAO repeatedly before its actual balance update\footnote{https://www.cryptocompare.com/coins/guides/the-dao-the-hack-the-soft-fork-and-the-hard-fork}. Another such incident happened in May 2017, when the WannaCry ransomware cyberattack targeted computers, encrypted their data and demanded the ransom money in cryptocurrency. In total, an amount higher than \pounds 108,000 was paid in Bitcoin cryptocurrency by the victims. The impact of this cyberattack was reportedly seen in 150 countries worldwide\footnote{https://www.theguardian.com/technology/2017/may/12/global-cyber-attack-ransomware-nsa-uk-nhs}.

If blockchain is to get widely adopted, centralized regulatory agencies, such as governmental agencies and multinational corporations, may be unable to control and shape the activities based on blockchain technology \cite{wright2015decentralized}. Because blockchain has no specific location and each node may subject to a different geographic jurisdiction and therefore different applicable laws and legal requirements. There is no central administration for each distributed ledger, therefore, territorial regulations constitute a problem \cite{sebastian2016blockchain}. As a result, there is an increased need to focus on the regulation of this cross-border nature of technology. 

In the \textit{Roadmap for Blockchain Standards Report} \cite{Roadmapf79}, it has been emphasized that there is a need to establish international standards regarding blockchain terminology, interoperability (between blockchain systems), user privacy, security, user identity, governance and risk related issues so that people's confidence in blockchain-based businesses may be developed. The report has further highlighted the need for collaboration among committees and experts in order to further strengthen the regulated use of the blockchain technology.

In \cite{de2017understanding}, it has been described that there are many interpretations of the blockchain technology in literature and formal blockchain terminologies are yet to be defined, i.e., permissioned blockchains vs. private distributed ledgers are few of those used interchangeably. In this \cite{deshpande2017understanding} literature, the importance of standards in paving the way for interoperability between multiple blockchain platforms and applications, have been discussed. The author is of the view that developing such standards for ensuring interoperability can help in minimizing the risk of fragmented blockchain systems.


At first, the organizations who have been governing the Internet, considered blockchain technologies as beyond their scope but this opinion changed later \cite{tapscott2016blockchain}. The World Wide Web Consortium (W3C) has been discussing online payments by utilizing the blockchain's potential\footnote{https://goo.gl/NjVLri}. The Internet Governance Forum (IGF) has been arranging sessions on blockchain technology to devise a distributed governance framework\footnote{https://goo.gl/9pPeiQ}. The Her Majesty's Revenue and Customs (HMRC) issued a policy paper describing the tax treatment for the income earned from Bitcoin (blockchain-based cryptocurrency) and other cryptocurrencies-related activities \cite{Revenuea86}. The Financial Crimes Enforcement Network (FinCEN) has recommended that decentralized currencies should follow the money laundering regulations \cite{network2013application} \cite{guadamuz2015blockchains}. 

The European Securities Market Authority (ESMA) has issued a paper\cite{2016773d76} in which the benefits and risks of the blockchain technology in securities markets have been discussed. The UK Treasury has issued a report \cite{walport2016distributed}, which has emphasized the need for Government to make efforts for the necessary regulatory framework in parallel to new blockchain-based developments. Moreover, other US regulatory authorities and agencies like Securities and Exchange Commission (SEC), Commodity Futures Trading Commission (CFTC), Internal Revenue Service (IRS) and Federal Trade Commission (FTC) have been working to make regulations pertaining to blockchain-based businesses and applications \cite{kakavand2016blockchain}.

\vspace{2mm}
\subsubsection{Blockchain and GDPR}
\label{subsubsec: bc_gdpr}
The European General Data Protection Regulation (GDPR) was adopted in 2016 by the European Parliament and the European Council \cite{GDPRdoc}. Since then, two years were given to the businesses to prepare themselves to comply with the regulation. In this section, we discuss where does the compliance with GDPR put the blockchain technology? Will the original premise of decentralization and immutability be able to sustain under the GDPR ramifications particularly when we consider the \emph{right to be forgotten} clause of GDPR? In what follows we first provide a brief overview of the GDPR, the duties it puts on businesses, the rights it gives to the users', and finally what are its ramifications on the blockchain technology in general?

After its legislation, the GDPR came into effect on May 25, 2018, and is applicable to any kind of information that can be associated with an either identified or identifiable living person\footnote{\url{https://gdpr-info.eu/art-4-gdpr/}}. Some of the examples of identifiable information include names, unique code number, IP address, single or multiple identifying characteristics. Further, GDPR applies throughout the lifecycle of personal data i.e., from data collection, to data processing through to the ultimate disposal of this data. GDPR-compliant businesses are bound to collect only data for the clearly stated purposes and process it with the users' consent. After the use of personal data for the said purpose, according to GDPR, the businesses are incumbent to delete the personal information from their local storage. However, this excludes data pertaining to a deceased person and processing of such data is at the disposal of local policies in place at a particular geographic region \cite{GDPRdoc}.

GDPR gives users certain rights when they interact with businesses which provide a service based on their personal data collection and processing. These rights include: 

\begin{enumerate}
\item \textbf{Awareness:} This entails that the users' must be informed about how their personal data will be used; 
\item \textbf{Access:} The users must be able to access copies of their data collected by a business or a service provider free of charge; 

\item \textbf{Correction:} If a user finds some inaccuracies in her data held by a company then she must be able to flag it as disputed; 

\item \textbf{Deletion:} A user must be able to make a company delete all the information pertaining to her whenever she chooses. (This right is sometimes referred to as the \emph{right to be forgotten}); 

\item \textbf{Restriction:} If a user is in a process of assessing the accurateness of her data use she must be able  to restrict the access to her data during the process; 

\item \textbf{Objection:} A user must be able to object to the uses of her data if she disagrees with some of the automated decisions involving her (such as marketing ads or shopping recommendations).
\end{enumerate}

It can be observed that many of these rights seem to fit quite well with the blockchain's premise of decentralization, tamper-evident record keeping, transparency, and auditability. There are however a few nuances which we discuss next.

There are two important terms that GDPR defines namely \emph{data controller} and \emph{data processor} which require special attention when dealing with blockchain-based projects. Both of these entities take part in users' personal data processing with their specific consent. There is, however, a nuance in the way these two entities function. Controller is an entity which sets the purposes and means for data processing. Controllers can take the shape of a natural or legal person, authority, or an agency. Data processors, on the other hand, is similarly a natural or legal person, authority, or an agency that processes personal data on a controller's behalf strictly following the rules specified by the corresponding controller. There should also be an agreement between a controller and a processor clearly defining their roles and functions \cite{bacon2017blockchain}. Given the users' rights, as mentioned above, one of the underlying principles of GDPR is auditability which provides the provision to hold the process and the entities  involved in personal data processing accountable for their responsibilities, functions, and actions\footnote{\url{https://thenextweb.com/syndication/2018/07/26/gdpr-blockchain-cryptocurrency/}}. In the decentralized environment of blockchain the important issue is related to specifying who gets to be a data controller and who a processor \cite{bacon2017blockchain, cnil:online}. 

In terms of blockchain, we consider a number of scenarios (self open, self private, open, private, consensus protocols) to answer the questions related to deciding the roles of controllers and processors. First, an entity (a business for instance) can choose to make use of the open and permissionless blockchain. In this scenario, such an entity can potentially write the core blockchain protocol and make it open source. Further, such entities can also deploy a set of smart contracts defining data processing rules and interactions among nodes of the network. We conjecture that this way such an entity can assume the role of data controller. Further, anyone can download the client software and become a node in the overall blockchain's P2P network. It has been a common practice that the open and public blockchains make use of PoW-based consensus mechanism. This implies that any node in the network can process transactions and validate them by including them in a mined block and ultimately appending that block to the overall blockchain. As we discussed earlier that PoW-based mining is a lottery-based process which means that it is a random event that a node in a network finds a nonce hence mining and ultimately appending this block to the blockchain. In this scenario, it is not a trivial task to decide who is the processor. Since potentially all the nodes in such a PoW-based network process data at the same time. We conjecture, either the whole network should be considered as a processor or the responsibility of being a processor should be weighted as per the processing power of either individual nodes or pool of such nodes (which are sometimes referred to as mining pools).

The second scenario is of private and permissioned blockchain. In this scenario, a number of entities can come together to form a consortium and then automate the dynamics of such a consortium using a permissioned version of blockchain. In this setup the entities can make use of a PoA-based consensus mechanism or Hyperledger's channel-based permissioned blockchain \footnote{\url{https://hyperledger-fabric.readthedocs.io/en/release-1.3/channels.html}}. Further, such entities can rent storage and computational resources from a third party cloud provider and hence rendering them as data processors. On the other hand, the consortium as a whole can assume the role of a data controller. Again, we conjecture, if the consortium makes use of a consensus mechanism such as PoS then as far as accountability is concerned then each node can be held accountable according to the stake value that such node holds in the overall network.

\subsubsection{Right to be forgotten}
\label{subsubsec: RTBF}
Although many of the principles outlined in GDPR such as data auditability fit quite well with blockchain's premise, the main bone of contention, however, in the way of making blockchain-based decentralized solutions in compliance with GDPR is the so called \emph{right to be forgotten}. This right dictates that a user must be able to instruct a business at any time to remove personal data pertaining to her. Further, as discussed above, the businesses are incumbent to delete personal data after a set duration of time. This right seems at odds with the blockchain's data integrity guarantees. The situation gets worse as blockchain-based solutions are usually distributed with multiple copies of the records stored at different nodes of the network.

As paradoxical as deletion and integrity seem at first glance there are, however, proposals to reconcile these two seemingly opposing principles\footnote{\url{https://thenextweb.com/syndication/2018/07/26/gdpr-blockchain-cryptocurrency/}}. One of the solutions could be to encrypt each data entry in blockchain with a key pair and only store the ciphertext on a blockchain. This way deletion can be achieved by simply deleting the corresponding private key while still preserving the ciphertext on blockchain. In some geographical jurisdictions (such as in Britain\footnote{\url{https://bit.ly/1VBf6Y8}}) the interpretation of GDPR does recognize such methods of digital deletion. However, such techniques do not provide a future-proof guarantee since with the advent of new and faster technologies and techniques such as quantum computing, such encryption methods pose a risk for future data breaches\footnote{\url{https://www.bundesblock.de/wp-content/uploads/2018/05/GDPR_Position_Paper_v1.0.pdf}}.

Another proposal is to only store hashes of data on blockchain while storing the actual data in an off-chain storage. This way the deletion can be achieved by deleting the off-chain stored record while keeping the hash of it intact on blockchain. An argument against this technique is that the hash of a blob of data can still qualify as personal data since if an entity possesses this blob then she can easily reconstruct the hash and decipher what was stored on blockchain in the first place. To get around this problem, we can use \emph{hash peppering} whereby a random and secretly kept nonce is appended to the blob of data before taking its hash and storing it on blockchain. This, however, implies keeping the nonces well protected and secret and does imply some level of trust on third parties that are responsible for peppered hashing of data.

Another way can be to make use of a technique similar to the way \emph{channels} are implemented in Hyperledger Fabric\footnote{\url{https://hyperledger-fabric.readthedocs.io/en/release-1.3/channels.html}}. Channels can be understood as confidential and permissioned islands of smaller blockchain instances on top of a larger blockchain infrastructure. A blockchain instance pertaining to a channel can be audited in the same way a public and open blockchain opens itself to auditing. However, the actual contents of transactional records are encrypted and one can not decipher the nature of the business being automated in a channel's instance of blockchain. This way by deleting the cryptographic information related to such a channel the whole instance of the corresponding blockchain can be rendered redundant.
\subsection{Scalability Issues}
\label{sec: scalabilityIssues}
%
Scalability is one of the major concerns in the way of wide spread adoption of blockchain-based technological solutions. We discuss this concern with following three different perspectives.
\vspace{1mm}
\subsubsection{Transaction throughput}
\label{subsubsec: throughput}
Although the Bitcoin is a popular blockchain-based global cryptocurrency, scaling it to handle the large transaction volumes worldwide raises some concerns. Among other things, the transaction processing rate of Bitcoin is affected by (1) the available network bandwidth, and (2) the network delay affects. Miners with high bandwidth and with less network delay can broadcast their blocks among peer nodes with ease and speed, while on the other hand low bandwidth miners with limited computational resources possess less probability of getting their fair share in a successful execution of proof-of-work \cite{sompolinsky2013accelerating}.

Bitcoin has seen an increasing interest, which has raised questions about its scalability. Scalability was one of the reasons that led to the creation of Bitcoin Cash\footnote{\url{https://www.investopedia.com/tech/bitcoin-vs-bitcoin-cash-whats-difference/}}; a forked version of Bitcoin but with a larger block size to allow more transactions per block. 

The blockchain-based systems are usually self-managed and accept transaction blocks after approximate intervals of time. The throughputs of these transactions are mainly based on block interval and maximum block size \cite{croman2016scaling}. It has been predicted that if the blocks size were to continue to grow at the same rate then it might attain a value close to its maximum capacity level by 2017 and this could be a significant scalability concern \cite{BlogTrad16:online}. 

Increasing the block size does imply a higher transaction throughput, however, this will also mean that the larger blocks would require more time to reach to the peer nodes of the network resulting in higher latency when it comes to proposing new blocks or reaching consensus on the state of a blockchain. On the other hand, the latency would decrease with decreased block interval but at the cost of potential disagreement in the system \cite{eyal2016bitcoin}. Similarly, other consensus protocols such as PoS-based consensus (as meniotned in Section \ref{subsubsec: pos}) are in development phase which are aimed to addressed the scalability and energy concerns.
\vspace{1mm}
\subsubsection{Storage}
\label{subsubsec: storage}
In addition to the block size scalability concern, the storage capacity of peer nodes is another issue. The transaction rate has a direct relation with the storage capacity of the participating nodes. With more nodes joining the network, the transaction rate would likely be higher and will require more storage space on the peer nodes, which might be seen as a limitation from the perspective of the consumers \cite{wattenhoferfast}.

It has been identified that blockchain technology is not limited to cryptocurrencies, but there are various blockchain-based prototype applications that are being used in domains such as IoT, Botnet, P2P broadcast protocols, smart property, and others. This shows the potential of blockchain technology for various other industries. Currently, the size of blockchain-based applications, in terms of their user base, is relatively small. Bitcoin is the largest solution based on the blockchain, but the transaction rate in bitcoin's network in comparison to the traditional digital payment solutions is considerably lower. However, in future, blockchain-based solutions could be used by millions or trillions of individuals and the number of transactions would increase drastically. Because of the distributed storage characteristic inherent in blockchains, it will put increasing pressure on storage nodes, which could result in increased synchronization delay, power consumption, and server costs. We believe that more research is required in order to address these these scalability issues. 

\subsubsection{The Lightning Network and Sharding}
\label{subsubsec: lightNetSharding}
The scalability issue can, up to some extent, be addressed by distributing the transaction execution process into multiple steps \cite{yu2020survey}. To ensure scalability, the execution of transactions can be performed outside the blockchain, whereas the validation should take place within the blockchain network. This would decrease the transaction confirmation time.  For example, the Lightning Network is able to perform 45000 transactions per second by executing the transactions outside the blockchain \cite{scal}. 

Another possible solution could be a decentralized database that can be used by both public and private blockchain and deploying sharding (which implies horizontal partitioning of records given large databases) and then merging the shardes at regular intervals\footnote{\url{https://medium.com/edchain/what-is-sharding-in-blockchain-8afd9ed4cff0}} \cite{databasesss, luu2016secure}. A decentralized database would be able to process millions of writes per second with the storage capacity of petabytes and latency in sub-seconds. This will also allow more nodes to be added to the platform, which would increase the performance and make the capacity scalabile.

\subsection{Security and Privacy Concerns}
\label{subsec: securityConcerns}
Besides security being in the system by design of the blockchain-based transactions, privacy remains a concern in applications and platforms \cite{davids2016research}. The blockchain technology has been considered as privacy-preserver and rated well in this context \cite{duffield2014darkcoin}\cite{de2016interplay}\cite{wang2016maturity}. However, third-party web trackers have been observed deanonymizing users of cryptocurrencies. These trackers fetch user's identity and purchase information from shopping websites to be used for advertisement and analysis purpose. Normally, these trackers have sufficient information required to uniquely identify the blockchain-based transaction along with user's identity \cite{goldfeder2017cookie}. 

It has been widely believed that blockchain is safe as its transactions are executed with generated addresses instead of real identities \cite{zheng2016blockchain}. Besides this, in \cite{meiklejohn2013fistful} \cite{kosba2016hawk}, it has been shown that the blockchain transactions do not ensure privacy since the transaction balances and values against public key(s) remain available for all.

\begin{figure*}[!ht]
\centering
\captionsetup{justification=centering}
\centerline{\includegraphics[width=.7\textwidth]{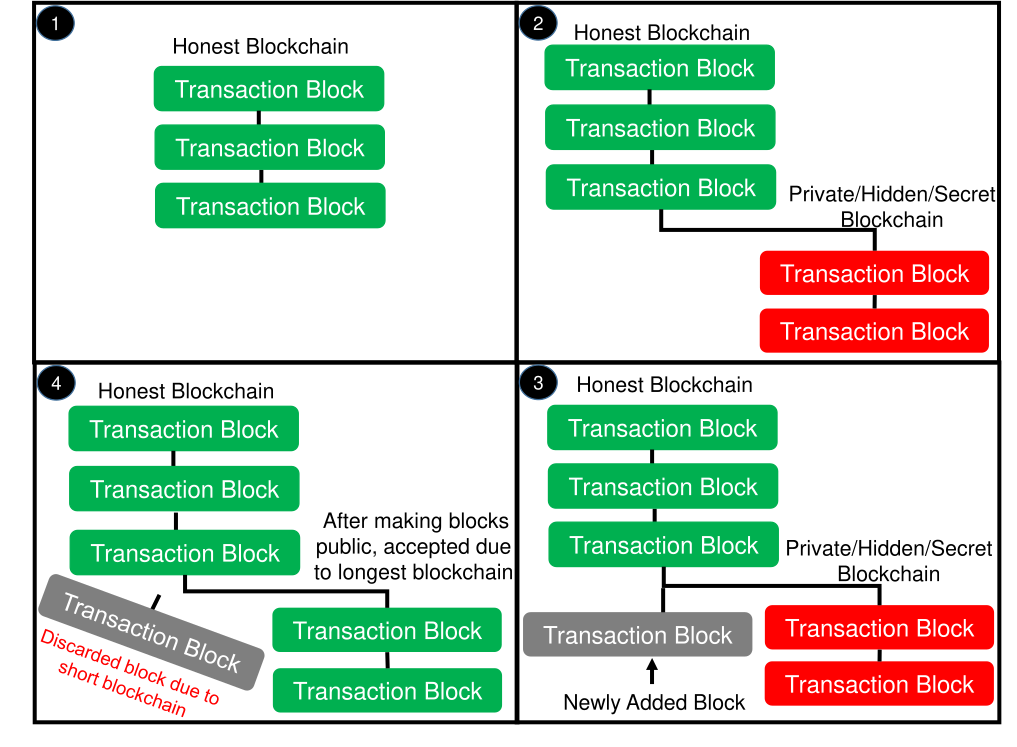}}
\caption{Workflow of selfish mining attack}
\label{fig:selfish}
\end{figure*}


In addition to the privacy-related issues, there are some security concerns related to blockchain technology. There are certain scenarios that may affect the expected behavior of the blockchain system. Consider the case where a miner-A successfully generates two blocks but does not disclose it to the peer honest network nodes, instead withholds these. We may call these as secret/hidden or private blocks. The miner-A releases these secret blocks when some honest nodes complete mining of a new block (say grey block). After the release of secret blocks, the miner-A successfully adds his two secret blocks in the blockchain network (since the miner-A holds the the longest chain of honest network nodes), whereas the newly added grey block does not remain a part of honest blockchain because the grey block does not hold the longest chain of honest network nodes \cite{pass2017fruitchains} \cite{eyal2014majority}. This type of attack is called selfish mining attack (see Figure \ref{fig:selfish} ) and this results in the undermining of the fair share of the block mining rewards 

51\% attack \cite{bradbury2013problem, bastiaan2015preventing, yli2016current} is another type of attack on blockchain systems. In this attack, a miner having more than half (i.e, 51\%) of network node's computational resources dominates the blockchain system in terms of transaction generation, approval, and verification and thus paves the way for fraudulent transactions generation \cite{xu2016blockchains}.

\subsection{Sustainability Issues}
\label{subsec: sustainabilityIssues}
Blockchain has attained an extraordinary amount of interest and attention and a large number of industries are adopting this virtual digital ledger.  However, it is still unclear that any particular solution of blockchain can attain a certain level of adoption for their sustainability. As a new technology, blockchain still facing operational, technical and its adoption-related issues. Similarly, there are also some aspects of blockchain technology that may need further modification or development to attain its anticipated potential. For example, although blockchain does provide a reliable cryptocurrency mechanism, it also adds latency to the network since the verification of the transaction requires consensus, which requires a certain amount of computation and a certain amount of time.

The sustainability of blockchain is still uncertain for international development projects, especially in developing countries. These projects require a very large infrastructure and involve various stakeholders, cross-border organizations, governments, and public or private parties. In these scenarios, the practicality of blockchain is unclear and it is the time to explore how blockchain will facilitate and sustain in such projects. Therefore, sustainability scientists and blockchain developers must discuss problems and solutions. More research is needed to find energy efficient approaches for Bitcoin mining. Behavioral and psychological research is required to attain people's trust in technology for cryptography. Most importantly, lawyers and programmers must collaborate to formulate smart contracts and dictionaries will be necessary that connect computer codes and legal languages.




\subsection{Anonymity}
In a blockchain system, the users utilize generated addresses, which are mostly in the form of public keys, for their unique identification over the blockchain network. The blockchain users can generate their multiple addresses in order to avoid the revelation of their real identities. These addresses are generated in the form of cryptographic keys. The said keys are then used to send and receive blockchain based transactions \cite{koshy2014analysis}. 
 
Moreover, there is no central storage system for preserving the user's private identification details in the blockchain network. By this way, the privacy in blockchain system is maintained up-to certain extent, however, the user's privacy protection is not guaranteed since the transaction amount details and the blockchain-based cryptographic keys (i.e., used for user identification) along with their respective balances, are publicly visible \cite{zheng2016blockchain}. 

The blockchain-based applications still do not completely guarantee the preservation of transactional anonymity.
The transactional transparency is impacted due to the lack of strong anonymity support for the end users \cite{de2016interplay}. In \cite{meiklejohn2013fistful}, the author showed that the movements of blockchain-based transactions are traceable and thus do not possess enough anonymity \cite{herrera2015research}. Few other anonymity tracing techniques are discussed in \cite{reid2013analysis} \cite{ron2013quantitative}.


\subsection{Use of Artificial Intelligence and Machine Learning}

Recent advancements in blockchain technology are making new ways for the involvement of AI and machine learning (ML) that can help to solve many challenges of blockchain with several important future applications \cite{rahouti2018bitcoin}. Blockchains is a technology that is being used to verify, execute and record the transaction. AI can help in understanding, recognizing, assessment decision making in the blockchain. Whereas ML techniques could help to find ways to improve decision making and smart contracts. For instance, AI can help to build an intelligent oracle without the control of the third party. This would learn and train itself to make the smart contract smarter \cite{zheng2016blockchain}. The integration of AI and ML with blockchain will potentially create a new paradigm by accelerating the analysis enormous amount of data. Examples include automation of tokens creation, recommender systems, security enhancement, etc.

\subsubsection{Use of Big Data Analytics}
\label{subsubsec: BD}

Recently, many companies are focusing to adopt the blockchain technology in their frameworks. This is creating new types of data for analysis by the powerful tools of big data. There is a huge number of blocks---increasing rapidly and constantly throughout the globe. Each block is full of information (i.e., details of every financial transaction) that can be used for analysis to explore thousands of patterns and trends. The blockchain is a technology that provides integrity, but not analysis. By using big data, it will be possible to detect nefarious users with whom business would be dangerous. Big data can also provide real-time fraud detection based on the users' records and history. The risky transactions or malicious users can be detected quickly by using big data analytics. This will result in cost reduction for real-time transaction \cite{bigdata}. Furthermore, user trading patterns can also be used to predict trading behaviors and potential partners for trade with the help of big data analytics \cite{zheng2016blockchain}. A good resource to conduct big data analysis on (real-time updated) data related to Ethereum and Bitcoin's blockchain is by using Google's BigQuery\footnote{\url{https://cloud.google.com/blog/products/gcp/bitcoin-in-bigquery-blockchain-analytics-on-public-data}},\footnote{\url{https://cloud.google.com/blog/products/data-analytics/ethereum-bigquery-public-dataset-smart-contract-analytics}}. For more details on the applications of blockchain for enabling AI, the interested readers are referred to a comprehensive survey on this topic \cite{salahAccess}.

\subsection{Usability and Key Management}

One of the primary challenges that any new technology faces is the usability. This issue is more acute in blockchain because of new architecture and high stakes. The transaction flow should be visible to users to analyze the whole transaction flows. This will improve the usability and help the individuals to understand and analyze the whole blockchain network \cite{yli2016current}. There are some systems such as Bitconeview \cite{di2015bitconeview} and  Bitiodine \cite{spagnuolo2014bitiodine} that proved to be very effective for the detection and analysis of blockchain-related patterns. These systems also help to improve security and privacy-related concerns. 

It has also been reported in the challenges and limitations of blockchain that the bitcoin API is difficult to use for the developments \cite{swan2015blockchain}. Bitcoin users have to deal with public key cryptography that differs from the password-based authentication system. The usability of bitcoin key management also presents fundamental challenges for end users \cite{eskandariusability}. This requires more research in the future to provide more ease to the end users and the developers. 



\section{Conclusion}

\label{con}
 In this paper, we provide a study on blockchain-based network applications, discuss their applicability, sustainability and scalability challenges. We also discuss some of the most prevalent and important legal ramifications of working with blockchain-based solutions. Additionally, this paper suggests some future directions that will be helpful to support sustainable blockchain-based solutions. At the time of writing, we believe that, blockchain is still in its infancy implying there will be sometime spent before it gets ubiquitous and widely adopted. However, the aim of this study is to provide a guiding reference manual in a generic form to both the researches and practitioners of the filed so that a more informed decision can be made either for conducting similar research or designing a blockchain-based solution. 


\begin{thebibliography}{100}
\bibitem{ali2017trust}
M.~Ali, ``Trust-to-trust design of a new internet,'' Ph.D. dissertation,
  Princeton University, 2017.

\bibitem{hamida2017blockchain}
E.~B. Hamida, K.~L. Brousmiche, H.~Levard, and E.~Thea, ``Blockchain for
  enterprise: Overview, opportunities and challenges,'' \emph{ICWMC 2017},
  p.~91, 2017.

\bibitem{swan2015blockchain}
M.~Swan, \emph{Blockchain: Blueprint for a new economy}.\hskip 1em plus 0.5em
  minus 0.4em\relax O'Reilly Media, Inc., 2015.

\bibitem{mori2016financial}
T.~Mori, ``Financial technology: blockchain and securities settlement,''
  \emph{Journal of Securities Operations \& Custody}, vol.~8, no.~3, pp.
  208--227, 2016.

\bibitem{WEFReali3:online}
D.~Tapscott and A.~Tapscott, ``Realizing the potential of blockchain: A multi
  stakeholder approach to the stewardship of blockchain and cryptocurrencies,''
  \url{http://www3.weforum.org/docs/WEF_Realizing_Potential_Blockchain.pdf},
  (Accessed on 20-May-2020).

\bibitem{peters2016understanding}
G.~W. Peters and E.~Panayi, ``Understanding modern banking ledgers through
  blockchain technologies: Future of transaction processing and smart contracts
  on the internet of money,'' in \emph{Banking Beyond Banks and Money}.\hskip
  1em plus 0.5em minus 0.4em\relax Springer, 2016, pp. 239--278.

\bibitem{facebook_breach:online}
N.~Badshah, ``Facebook to contact 87 million users affected by data breach,''
  \url{https://www.theguardian.com/technology/2018/apr/08/facebook-to-contact-the-87-million-users-affected-by-data-breach},
  April 2018, (Accessed on 20-May-2020).

\bibitem{ali2018applications}
M.~S. Ali, M.~Vecchio, M.~Pincheira, K.~Dolui, F.~Antonelli, and M.~H. Rehmani,
  ``Applications of blockchains in the internet of things: A comprehensive
  survey,'' \emph{IEEE Communications Surveys \& Tutorials}, 2018.

\bibitem{ferrag2018blockchain}
M.~A. Ferrag, M.~Derdour, M.~Mukherjee, A.~Derhab, L.~Maglaras, and H.~Janicke,
  ``Blockchain technologies for the internet of things: Research issues and
  challenges,'' \emph{IEEE Internet of Things Journal}, 2018.

\bibitem{zheng2016blockchain}
Z.~Zheng, S.~Xie, H.-N. Dai, and H.~Wang, ``Blockchain challenges and
  opportunities: A survey,'' \emph{Work Paper}, 2016.

\bibitem{guo2016blockchain}
Y.~Guo and C.~Liang, ``Blockchain application and outlook in the banking
  industry,'' \emph{Financial Innovation}, vol.~2, no.~1, p.~24, 2016.

\bibitem{yli2016current}
J.~Yli-Huumo, D.~Ko, S.~Choi, S.~Park, and K.~Smolander, ``Where is current
  research on blockchain technology?---a systematic review,'' \emph{PloS one},
  vol.~11, no.~10, p. e0163477, 2016.

\bibitem{pilkington201611}
M.~Pilkington, ``11 blockchain technology: principles and applications,''
  \emph{Research handbook on digital transformations}, p. 225, 2016.

\bibitem{nofer2017blockchain}
M.~Nofer, P.~Gomber, O.~Hinz, and D.~Schiereck, ``Blockchain,'' \emph{Business
  \& Information Systems Engineering}, vol.~59, no.~3, pp. 183--187, 2017.

\bibitem{zheng2017overview}
Z.~Zheng, S.~Xie, H.~Dai, X.~Chen, and H.~Wang, ``An overview of blockchain
  technology: Architecture, consensus, and future trends,'' in \emph{Big Data
  (BigData Congress), 2017 IEEE International Congress on}.\hskip 1em plus
  0.5em minus 0.4em\relax IEEE, 2017, pp. 557--564.

\bibitem{lin2017survey}
I.-C. Lin and T.-C. Liao, ``A survey of blockchain security issues and
  challenges.'' \emph{IJ Network Security}, vol.~19, no.~5, pp. 653--659, 2017.

\bibitem{miraz2018applications}
D.~Miraz and M.~Ali, ``Applications of blockchain technology beyond
  cryptocurrency,'' \emph{Annals of Emerging Technologies in Computing
  (AETiC)}, vol.~2, pp. 1--6, 01 2018.

\bibitem{yuan2018blockchain}
Y.~Yuan and F.-Y. Wang, ``Blockchain and cryptocurrencies: Model, techniques,
  and applications,'' \emph{IEEE Transactions on Systems, Man, and Cybernetics:
  Systems}, vol.~48, no.~9, pp. 1421--1428, 2018.

\bibitem{wust2018you}
K.~W{\"u}st and A.~Gervais, ``Do you need a blockchain?'' in \emph{2018 Crypto
  Valley Conference on Blockchain Technology (CVCBT)}.\hskip 1em plus 0.5em
  minus 0.4em\relax IEEE, 2018, pp. 45--54.

\bibitem{salahAccess}
K.~Salah, M.~H. Rehman, N.~Nizamuddin, and A.~Al-Fuqaha, ``Blockchain for {AI}:
  Review and open research challenges,'' \emph{IEEE Access}, pp. 1--1, 2019.

\bibitem{xie2019survey}
J.~Xie, H.~Tang, T.~Huang, F.~R. Yu, R.~Xie, J.~Liu, and Y.~Liu, ``A survey of
  blockchain technology applied to smart cities: Research issues and
  challenges,'' \emph{IEEE Communications Surveys \& Tutorials}, vol.~21,
  no.~3, pp. 2794--2830, 2019.

\bibitem{wang2019survey}
W.~Wang, D.~T. Hoang, P.~Hu, Z.~Xiong, D.~Niyato, P.~Wang, Y.~Wen, and D.~I.
  Kim, ``A survey on consensus mechanisms and mining strategy management in
  blockchain networks,'' \emph{IEEE Access}, vol.~7, pp. 22\,328--22\,370,
  2019.

\bibitem{yang2019survey}
W.~Yang, E.~Aghasian, S.~Garg, D.~Herbert, L.~Disiuta, and B.~Kang, ``A survey
  on blockchain-based internet service architecture: Requirements, challenges,
  trends, and future,'' \emph{IEEE Access}, vol.~7, pp. 75\,845--75\,872, 2019.

\bibitem{yang2019integrated}
R.~Yang, F.~R. Yu, P.~Si, Z.~Yang, and Y.~Zhang, ``Integrated blockchain and
  edge computing systems: A survey, some research issues and challenges,''
  \emph{IEEE Communications Surveys \& Tutorials}, vol.~21, no.~2, pp.
  1508--1532, 2019.

\bibitem{belotti2019vademecum}
M.~Belotti, N.~Bo{\v{z}}i{\'c}, G.~Pujolle, and S.~Secci, ``A vademecum on
  blockchain technologies: When, which, and how,'' \emph{IEEE Communications
  Surveys \& Tutorials}, vol.~21, no.~4, pp. 3796--3838, 2019.

\bibitem{dai2019blockchain}
H.-N. Dai, Z.~Zheng, and Y.~Zhang, ``Blockchain for internet of things: A
  survey,'' \emph{IEEE Internet of Things Journal}, vol.~6, no.~5, pp.
  8076--8094, 2019.

\bibitem{wu2019comprehensive}
M.~Wu, K.~Wang, X.~Cai, S.~Guo, M.~Guo, and C.~Rong, ``A comprehensive survey
  of blockchain: From theory to iot applications and beyond,'' \emph{IEEE
  Internet of Things Journal}, vol.~6, no.~5, pp. 8114--8154, 2019.

\bibitem{viriyasitavat2019blockchain}
W.~Viriyasitavat, L.~Da~Xu, Z.~Bi, and D.~Hoonsopon, ``Blockchain technology
  for applications in internet of things—mapping from system design
  perspective,'' \emph{IEEE Internet of Things Journal}, vol.~6, no.~5, pp.
  8155--8168, 2019.

\bibitem{mollah2020blockchain}
M.~B. Mollah, J.~Zhao, D.~Niyato, K.-Y. Lam, X.~Zhang, A.~M. Ghias, L.~H. Koh,
  and L.~Yang, ``Blockchain for future smart grid: A comprehensive survey,''
  \emph{IEEE Internet of Things Journal}, 2020.

\bibitem{liu2020blockchain}
Y.~Liu, F.~R. Yu, X.~Li, H.~Ji, and V.~C. Leung, ``Blockchain and machine
  learning for communications and networking systems,'' \emph{IEEE
  Communications Surveys \& Tutorials}, 2020.

\bibitem{neudecker2018network}
T.~Neudecker and H.~Hartenstein, ``Network layer aspects of permissionless
  blockchains,'' \emph{IEEE Communications Surveys \& Tutorials}, vol.~21,
  no.~1, pp. 838--857, 2019.

\bibitem{lao2020survey}
L.~Lao, Z.~Li, S.~Hou, B.~Xiao, S.~Guo, and Y.~Yang, ``A survey of iot
  applications in blockchain systems: Architecture, consensus, and traffic
  modeling,'' \emph{ACM Computing Surveys (CSUR)}, vol.~53, no.~1, pp. 1--32,
  2020.

\bibitem{kolb2020core}
J.~Kolb, M.~AbdelBaky, R.~H. Katz, and D.~E. Culler, ``Core concepts,
  challenges, and future directions in blockchain: A centralized tutorial,''
  \emph{ACM Computing Surveys (CSUR)}, vol.~53, no.~1, pp. 1--39, 2020.

\bibitem{monrat2019survey}
A.~A. Monrat, O.~Schel{\'e}n, and K.~Andersson, ``A survey of blockchain from
  the perspectives of applications, challenges, and opportunities,'' \emph{IEEE
  Access}, vol.~7, pp. 117\,134--117\,151, 2019.

\bibitem{zhang2019security}
R.~Zhang, R.~Xue, and L.~Liu, ``Security and privacy on blockchain,'' \emph{ACM
  Computing Surveys (CSUR)}, vol.~52, no.~3, pp. 1--34, 2019.

\bibitem{xiao2020survey}
Y.~Xiao, N.~Zhang, W.~Lou, and Y.~T. Hou, ``A survey of distributed consensus
  protocols for blockchain networks,'' \emph{IEEE Communications Surveys \&
  Tutorials}, 2020.

\bibitem{bodkhe2020survey}
U.~Bodkhe, D.~Mehta, S.~Tanwar, P.~Bhattacharya, P.~K. Singh, and W.-C. Hong,
  ``A survey on decentralized consensus mechanisms for cyber physical
  systems,'' \emph{IEEE Access}, vol.~8, pp. 54\,371--54\,401, 2020.

\bibitem{al2019blockchain}
J.~Al-Jaroodi and N.~Mohamed, ``Blockchain in industries: A survey,''
  \emph{IEEE Access}, vol.~7, pp. 36\,500--36\,515, 2019.

\bibitem{nakamoto2008bitcoin}
S.~Nakamoto, ``Bitcoin: A peer-to-peer electronic cash system,'' 2008.

\bibitem{rogaway2004cryptographic}
P.~Rogaway and T.~Shrimpton, ``Cryptographic hash-function basics: Definitions,
  implications, and separations for preimage resistance, second-preimage
  resistance, and collision resistance,'' in \emph{International workshop on
  fast software encryption}.\hskip 1em plus 0.5em minus 0.4em\relax Springer,
  2004, pp. 371--388.

\bibitem{sha:online}
``Descriptions of sha-256, sha-384, and sha-512,''
  \url{https://web.archive.org/web/20130526224224/http://csrc.nist.gov/groups/STM/cavp/documents/shs/sha256-384-512.pdf},
  (Accessed on 20-May-2020).

\bibitem{chen2016algorand}
J.~Chen and S.~Micali, ``Algorand,'' \emph{arXiv preprint arXiv:1607.01341},
  2016.

\bibitem{de2018pbft}
S.~De~Angelis, L.~Aniello, R.~Baldoni, F.~Lombardi, A.~Margheri, and
  V.~Sassone, ``Pbft vs proof-of-authority: Applying the cap theorem to
  permissioned blockchain,'' 2018.

\bibitem{ekparinya2019attack}
P.~Ekparinya, V.~Gramoli, and G.~Jourjon, ``The attack of the clones against
  proof-of-authority,'' \emph{arXiv preprint arXiv:1902.10244}, 2019.

\bibitem{castro2002practical}
M.~Castro and B.~Liskov, ``Practical byzantine fault tolerance and proactive
  recovery,'' \emph{ACM Transactions on Computer Systems (TOCS)}, vol.~20,
  no.~4, pp. 398--461, 2002.

\bibitem{luu2016making}
L.~Luu, D.-H. Chu, H.~Olickel, P.~Saxena, and A.~Hobor, ``Making smart
  contracts smarter,'' in \emph{Proceedings of the 2016 ACM SIGSAC Conference
  on Computer and Communications Security}.\hskip 1em plus 0.5em minus
  0.4em\relax ACM, 2016, pp. 254--269.

\bibitem{bahga2016blockchain}
A.~Bahga and V.~K. Madisetti, ``{Blockchain platform for industrial Internet of
  Things},'' \emph{J. Softw. Eng. Appl}, vol.~9, no.~10, p. 533, 2016.

\bibitem{soliditydocs}
``Solidity — solidity 0.6.9 documentation,''
  \url{https://solidity.readthedocs.io/en/develop/}, (Accessed on 20-May-2020).

\bibitem{bargar2016economics}
D.~Bargar, ``{The Economics of the Blockchain: A study of its engineering and
  transaction services marketplace},'' Ph.D. dissertation, Clemson University,
  2016.

\bibitem{wood2014ethereum}
G.~Wood, ``Ethereum: A secure decentralised generalised transaction ledger,''
  \emph{Ethereum project yellow paper}, vol. 151, pp. 1--32, 2014.

\bibitem{kosba2016hawk}
A.~Kosba, A.~Miller, E.~Shi, Z.~Wen, and C.~Papamanthou, ``Hawk: The blockchain
  model of cryptography and privacy-preserving smart contracts,'' in \emph{IEEE
  Symposium on Security and Privacy (SP), 2016}.\hskip 1em plus 0.5em minus
  0.4em\relax IEEE, 2016, pp. 839--858.

\bibitem{ethereum5}
``ethereum-homestead.pdf,''
  \url{https://buildmedia.readthedocs.org/media/pdf/ethereum-homestead/latest/ethereum-homestead.pdf},
  (Accessed on 20-May-2020).

\bibitem{kakavand2016blockchain}
H.~Kakavand and N.~Kost De~Sevres, ``The blockchain revolution: An analysis of
  regulation and technology related to distributed ledger technologies,'' 2016.

\bibitem{buterin2014next}
V.~Buterin \emph{et~al.}, ``A next-generation smart contract and decentralized
  application platform,'' \emph{White Paper}, 2014.

\bibitem{www:www.gartner.com}
``Three things cios need to know about the blockchain business value
  forecast,''
  \url{https://www.gartner.com/en/documents/3776763/three-things-cios-need-to-know-about-the-blockchain-busi},
  (Accessed on 20-May-2020).

\bibitem{www:ripple.com}
``Instantly move money to all corners of the world | ripple,''
  \url{https://ripple.com/}, (Accessed on 20-May-2020).

\bibitem{kuperberg2019blockchain}
M.~Kuperberg, ``Blockchain-based identity management: A survey from the
  enterprise and ecosystem perspective,'' \emph{IEEE Transactions on
  Engineering Management}, 2019.

\bibitem{khovratovich2017sovrin}
D.~Khovratovich and J.~Law, ``Sovrin: digital identities in the blockchain
  era,'' \emph{Github Commit by jasonalaw October}, vol.~17, 2017.

\bibitem{camenisch2002dynamic}
J.~Camenisch and A.~Lysyanskaya, ``Dynamic accumulators and application to
  efficient revocation of anonymous credentials,'' in \emph{Annual
  International Cryptology Conference}.\hskip 1em plus 0.5em minus 0.4em\relax
  Springer, 2002, pp. 61--76.

\bibitem{camenisch2009accumulator}
J.~Camenisch, M.~Kohlweiss, and C.~Soriente, ``An accumulator based on bilinear
  maps and efficient revocation for anonymous credentials,'' in
  \emph{International workshop on public key cryptography}.\hskip 1em plus
  0.5em minus 0.4em\relax Springer, 2009, pp. 481--500.

\bibitem{lamport1982byzantine}
L.~Lamport, R.~Shostak, and M.~Pease, ``The byzantine generals problem,''
  \emph{ACM Transactions on Programming Languages and Systems (TOPLAS)},
  vol.~4, no.~3, pp. 382--401, 1982.

\bibitem{IDmixer:online}
``Specification of the identity mixer cryptographic library version 2.3.0,”
  2009,''
  \url{https://domino.research.ibm.com/library/cyberdig.nsf/papers/EEB54FF3B91C1D648525759B004FBBB1/File/rz3730_revised.pdf},
  (Accessed on 15-{May}-2020).

\bibitem{camenisch2002design}
J.~Camenisch and E.~Van~Herreweghen, ``Design and implementation of the idemix
  anonymous credential system,'' in \emph{Proceedings of the 9th ACM conference
  on Computer and communications security}, 2002, pp. 21--30.

\bibitem{Charm:online}
``A tool for rapid cryptographic prototyping,'' \url{http://charm-crypto.com/},
  (Accessed on 15-{May}-2020).

\bibitem{jin2018towards}
H.~Jin, X.~Dai, and J.~Xiao, ``Towards a novel architecture for enabling
  interoperability amongst multiple blockchains,'' in \emph{2018 IEEE 38th
  International Conference on Distributed Computing Systems (ICDCS)}.\hskip 1em
  plus 0.5em minus 0.4em\relax IEEE, 2018, pp. 1203--1211.

\bibitem{usfsi20111:online}
``Breaking blockchain open deloitte’s 2018 global blockchain survey,''
  \url{https://www2.deloitte.com/content/dam/Deloitte/us/Documents/financial-services/us-fsi-2018-global-blockchain-survey-report.pdf},
  (Accessed on 20-May-2020).

\bibitem{IBMNewsr71:online}
``Ibm news room - 2017-06-26 seven major european banks select ibm to bring
  blockchain-based trade finance to small and medium enterprises - united
  states,'' \url{https://www-03.ibm.com/press/us/en/pressrelease/52706.wss},
  (20-May-2020).

\bibitem{tikhomirov2017ethereum}
S.~Tikhomirov, ``Ethereum: state of knowledge and research perspectives,'' in
  \emph{International Symposium on Foundations and Practice of Security}.\hskip
  1em plus 0.5em minus 0.4em\relax Springer, 2017, pp. 206--221.

\bibitem{zhu2016analysis}
H.~Zhu and Z.~Z. Zhou, ``Analysis and outlook of applications of blockchain
  technology to equity crowdfunding in china,'' \emph{Financial innovation},
  vol.~2, no.~1, p.~29, 2016.

\bibitem{jacobovitz2016blockchain}
O.~Jacobovitz, ``Blockchain for identity management,'' \emph{The Lynne and
  William Frankel Center for Computer Science Department of Computer Science.
  Ben-Gurion University, Beer Sheva Google Scholar}, 2016.

\bibitem{zavolokina2016fintech}
L.~Zavolokina, M.~Dolata, and G.~Schwabe, ``Fintech transformation: How
  it-enabled innovations shape the financial sector,'' in \emph{International
  Workshop on Enterprise Applications and Services in the Finance
  Industry}.\hskip 1em plus 0.5em minus 0.4em\relax Springer, 2016, pp. 75--88.

\bibitem{ozyilmaz2017integrating}
K.~R. {\"O}zy{\i}lmaz and A.~Yurdakul, ``Integrating low-power iot devices to a
  blockchain-based infrastructure: work-in-progress,'' in \emph{Proceedings of
  the Thirteenth ACM International Conference on Embedded Software 2017
  Companion}.\hskip 1em plus 0.5em minus 0.4em\relax ACM, 2017, p.~13.

\bibitem{Launchin35:online}
``Launching the ether sale | ethereum foundation blog,''
  \url{https://blog.ethereum.org/2014/07/22/launching-the-ether-sale/},
  (Accessed on 20-May-2020).

\bibitem{Historyo98:online}
``History of ethereum — ethereum homestead 0.1 documentation,''
  \url{http://ethdocs.org/en/latest/introduction/history-of-ethereum.html},
  (Accessed on 20-May-2020).

\bibitem{Coinbase59:online}
``Coinbase definition,''
  \url{https://www.investopedia.com/terms/c/coinbase.asp}, (Accessed on
  20-May-2020).

\bibitem{hendrickson2016political}
J.~R. Hendrickson, T.~L. Hogan, and W.~J. Luther, ``The political economy of
  bitcoin,'' \emph{Economic Inquiry}, vol.~54, no.~2, pp. 925--939, 2016.

\bibitem{bitpayAp82:online}
``Bitcoin payment gateway api v0.3,''
  \url{https://bitpay.com/downloads/bitpayApi-0.3.pdf}, (Accessed on
  20-May-2020).

\bibitem{Cryptocu96:online}
``Cryptocurrencies timeline: a history of digital money,''
  \url{https://www.telegraph.co.uk/technology/digital-money/the-history-of-cryptocurrency/},
  (Accessed on 20-May-2020).

\bibitem{BNAKLaun81:online}
``Bnak launches swiftcoin, electronic currency that is safer than cash |
  business wire,''
  \url{https://www.businesswire.com/news/home/20121119005937/en/BNAK-Launches-Swiftcoin-Electronic-Currency-Safer-Cash},
  (Accessed on 20-May-2020).

\bibitem{bruno2017system}
D.~B. Bruno, ``System and method for providing a cryptographic platform for
  exchanging debt securities denominated in virtual currencies,'' Jul.~27 2017,
  uS Patent App. 15/483,190.

\bibitem{zhao2015cryptocurrency}
Y.~Zhao, ``Cryptocurrency brings new battles into the currency market,''
  \emph{Future Internet (FI) and Innovative Internet Technologies and Mobile
  Communications (IITM)}, vol.~91, 2015.

\bibitem{ron2013quantitative}
D.~Ron and A.~Shamir, ``Quantitative analysis of the full bitcoin transaction
  graph,'' in \emph{International Conference on Financial Cryptography and Data
  Security}.\hskip 1em plus 0.5em minus 0.4em\relax Springer, 2013, pp. 6--24.

\bibitem{yermack2015bitcoin}
D.~Yermack, ``Is bitcoin a real currency? an economic appraisal,'' in
  \emph{Handbook of digital currency}.\hskip 1em plus 0.5em minus 0.4em\relax
  Elsevier, 2015, pp. 31--43.

\bibitem{wang2015exploring}
L.~Wang and Y.~Liu, ``Exploring miner evolution in bitcoin network,'' in
  \emph{International Conference on Passive and Active Network
  Measurement}.\hskip 1em plus 0.5em minus 0.4em\relax Springer, 2015, pp.
  290--302.

\bibitem{Bitcoins11:online}
``Bitcoin's quirky genesis block turns eight years old today | featured bitcoin
  news,''
  \url{https://news.bitcoin.com/bitcoins-quirky-genesis-block-turns-eight-years-old-today/},
  (Accessed on 20-May-2020).

\bibitem{sharma2017software}
P.~K. Sharma, M.-Y. Chen, and J.~H. Park, ``A software defined fog node based
  distributed blockchain cloud architecture for iot,'' \emph{Ieee Access},
  vol.~6, pp. 115--124, 2017.

\bibitem{sharma2017distblocknet}
P.~K. Sharma, S.~Singh, Y.-S. Jeong, and J.~H. Park, ``Distblocknet: A
  distributed blockchains-based secure sdn architecture for iot networks,''
  \emph{IEEE Communications Magazine}, vol.~55, no.~9, pp. 78--85, 2017.

\bibitem{qiu2018blockchain}
C.~Qiu, F.~R. Yu, H.~Yao, C.~Jiang, F.~Xu, and C.~Zhao, ``Blockchain-based
  software-defined industrial internet of things: A dueling deep q-learning
  approach,'' \emph{IEEE Internet of Things Journal}, vol.~6, no.~3, pp.
  4627--4639, 2018.

\bibitem{samaniego2016hosting}
M.~Samaniego and R.~Deters, ``Hosting virtual iot resources on edge-hosts with
  blockchain,'' in \emph{2016 IEEE International Conference on Computer and
  Information Technology (CIT)}.\hskip 1em plus 0.5em minus 0.4em\relax IEEE,
  2016, pp. 116--119.

\bibitem{gnangnon2017does}
S.~K. Gnangnon and H.~Iyer, ``Does bridging the {I}nternet access divide
  contribute to enhancing countries' integration into the global trade in
  services markets?'' \emph{Telecommunications Policy}, 2017.

\bibitem{nekrasov2017limits}
M.~Nekrasov, L.~Parks, and E.~Belding, ``Limits to {I}nternet freedoms: Being
  heard in an increasingly authoritarian world,'' in \emph{Proceedings of the
  2017 Workshop on Computing Within Limits}.\hskip 1em plus 0.5em minus
  0.4em\relax ACM, 2017, pp. 119--128.

\bibitem{weidmann2016digital}
N.~B. Weidmann, S.~Benitez-Baleato, P.~Hunziker, E.~Glatz, and
  X.~Dimitropoulos, ``Digital discrimination: Political bias in internet
  service provision across ethnic groups,'' \emph{Science}, vol. 353, no. 6304,
  pp. 1151--1155, 2016.

\bibitem{park2017digital}
S.~Park, ``{Digital inequalities in rural Australia: A double jeopardy of
  remoteness and social exclusion},'' \emph{Journal of Rural Studies}, vol.~54,
  pp. 399--407, 2017.

\bibitem{klein2002icann}
H.~Klein, ``Icann and internet governance: Leveraging technical coordination to
  realize global public policy,'' \emph{The Information Society}, vol.~18,
  no.~3, pp. 193--207, 2002.

\bibitem{purkayastha2014us}
P.~Purkayastha and R.~Bailey, ``{US control of the Internet: Problems facing
  the movement to international governance},'' \emph{Monthly Review}, vol.~66,
  no.~3, p. 103, 2014.

\bibitem{jasper2017cyber}
S.~Jasper and J.~Wirtz, ``Cyber security,'' in \emph{The Palgrave Handbook of
  Security, Risk and Intelligence}.\hskip 1em plus 0.5em minus 0.4em\relax
  Springer, 2017, pp. 157--176.

\bibitem{patrick2017need}
H.~Patrick and Z.~Fields, ``A need for cyber security creativity,''
  \emph{Collective Creativity for Responsible and Sustainable Business
  Practice}, pp. 42--61, 2017.

\bibitem{karchefsky2017toward}
S.~Karchefsky and H.~R. Rao, ``Toward a safer tomorrow: Cybersecurity and
  critical infrastructure,'' in \emph{The Palgrave Handbook of Managing
  Continuous Business Transformation}.\hskip 1em plus 0.5em minus 0.4em\relax
  Springer, 2017, pp. 335--352.

\bibitem{Blocksta20:online}
``Blockstack, building the decentralized computing network,''
  \url{https://blockstack.org/}, (Accessed on 20-May-2020 ).

\bibitem{AmmbrWhi61:online}
``Ammbr whitepaper v1.0,''
  \url{http://ammbr.com/docs/20171018/Ammbr_Whitepaper_v2.1_18Oct2017.pdf},
  (Accessed on 20-May-2020).

\bibitem{WikiLeak85:online}
``Wikileaks.org taken down by us dns provider | netcraft news,''
  \url{https://news.netcraft.com/archives/2010/12/03/wikileaks-org-taken-down-by-us-dns-provider.html},
  (Accessed on 20-May-2020).

\bibitem{FourRoun21:online}
``Four rounds of ice domain name seizures and related controversies and
  opposition – berkeley technology law journal,''
  \url{https://bit.ly/36qsA5H}, (Accessed on 20-May-2020).

\bibitem{bendrath2011end}
R.~Bendrath and M.~Mueller, ``The end of the net as we know it? deep packet
  inspection and internet governance,'' \emph{New Media \& Society}, vol.~13,
  no.~7, pp. 1142--1160, 2011.

\bibitem{denardis2012hidden}
L.~DeNardis, ``Hidden levers of internet control: An infrastructure-based
  theory of internet governance,'' \emph{Information, Communication \&
  Society}, vol.~15, no.~5, pp. 720--738, 2012.

\bibitem{kalodner2015empirical}
H.~A. Kalodner, M.~Carlsten, P.~Ellenbogen, J.~Bonneau, and A.~Narayanan, ``{An
  Empirical Study of Namecoin and Lessons for Decentralized Namespace
  Design},'' in \emph{WEIS}.\hskip 1em plus 0.5em minus 0.4em\relax Citeseer,
  2015.

\bibitem{angieri2019distributed}
S.~Angieri, A.~Garc{\'\i}a-Mart{\'\i}nez, B.~Liu, Z.~Yan, C.~Wang, and
  M.~Bagnulo, ``A distributed autonomous organization for internet address
  management,'' \emph{IEEE Transactions on Engineering Management}, 2019.

\bibitem{ali2016blockstack}
M.~Ali, J.~C. Nelson, R.~Shea, and M.~J. Freedman, ``Blockstack: A global
  naming and storage system secured by blockchains.'' in \emph{USENIX Annual
  Technical Conference}, 2016, pp. 181--194.

\bibitem{Namecoin94:online}
``Namecoin,'' \url{https://www.namecoin.org/}, (Accessed on 20-May-2020).

\bibitem{Namecoin51:online}
``Namecoin - wikipedia,'' \url{https://en.wikipedia.org/wiki/Namecoin},
  (Accessed on 20-May-2020).

\bibitem{back2002hashcash}
A.~Back \emph{et~al.}, ``Hashcash-a denial of service counter-measure,'' 2002.

\bibitem{kroll2013economics}
J.~A. Kroll, I.~C. Davey, and E.~W. Felten, ``{The economics of Bitcoin mining,
  or Bitcoin in the presence of adversaries},'' in \emph{Proceedings of WEIS},
  vol. 2013, 2013, p.~11.

\bibitem{kirkmanusing}
S.~S. Kirkman and R.~Newman, ``Using smart contracts and blockchains to support
  consumer trust across distributed clouds.''

\bibitem{blocknet:online}
``The blocknet: Design specification,''
  \url{https://www.blocknet.co/wp-content/uploads/whitepaper/Blocknet_Whitepaper.pdf},
  (Accessed on 20-May-2020).

\bibitem{interledger:online}
E.~S. Stefan~Thomas, ``A protocol for interledger payments,''
  \url{https://interledger.org/interledger.pdf}, (Accessed on 11/01/2019).

\bibitem{stine2010mail}
K.~Stine and M.~Scholl, ``E-mail security. an overview of threats and
  safeguards.'' \emph{Journal of AHIMA}, vol.~81, no.~4, pp. 28--30, 2010.

\bibitem{ferguson2005fostering}
A.~J. Ferguson, ``Fostering e-mail security awareness: The west point
  carronade,'' \emph{Educase Quarterly}, vol.~28, no.~1, pp. 54--57, 2005.

\bibitem{emailyahoo}
C.~Taylor, ``Blockchain \& email, access date: 06-{Oct}-2018,''
  \url{http://finteknews.com/blockchain-email/}, 2016.

\bibitem{bersieremail}
F.~Bersier and R.~Bischof, ``Email stamping: Gmelius blockchain architecture,''
  https://gmelius.com/email-stamping-blockchain.pdf, 2017, (Accessed on
  06-{Oct}-2018).

\bibitem{xia2012internet}
F.~Xia, L.~T. Yang, L.~Wang, and A.~Vinel, ``Internet of things,''
  \emph{International Journal of Communication Systems}, vol.~25, no.~9, p.
  1101, 2012.

\bibitem{fernandez2018review}
T.~M. Fern{\'a}ndez-Caram{\'e}s and P.~Fraga-Lamas, ``A review on the use of
  blockchain for the internet of things,'' \emph{IEEE Access}, vol.~6, pp.
  32\,979--33\,001, 2018.

\bibitem{ahmed2017comprehensive}
A.~W. Ahmed, M.~M. Ahmed, O.~A. Khan, and M.~A. Shah, ``A comprehensive
  analysis on the security threats and their countermeasures of {{IoT}},''
  \emph{International Journal of Advanced Computer Science and Applications},
  vol.~8, no.~7, pp. 489--501, 2017.

\bibitem{oravec2017emerging}
J.~A. Oravec, ``Emerging “cyber hygiene” practices for the internet of
  things ({IoT}): Professional issues in consulting clients and educating users
  on {IoT} privacy and security,'' in \emph{Professional Communication
  Conference (ProComm), 2017 IEEE International}.\hskip 1em plus 0.5em minus
  0.4em\relax IEEE, 2017, pp. 1--5.

\bibitem{sicari2015security}
S.~Sicari, A.~Rizzardi, L.~A. Grieco, and A.~Coen-Porisini, ``Security, privacy
  and trust in internet of things: The road ahead,'' \emph{Computer Networks},
  vol.~76, pp. 146--164, 2015.

\bibitem{ukil2014iot}
A.~Ukil, S.~Bandyopadhyay, and A.~Pal, ``{IoT}-privacy: To be private or not to
  be private,'' in \emph{Computer Communications Workshops (INFOCOM WKSHPS),
  2014 IEEE Conference on}.\hskip 1em plus 0.5em minus 0.4em\relax IEEE, 2014,
  pp. 123--124.

\bibitem{pasquier2018data}
T.~Pasquier, J.~Singh, J.~Powles, D.~Eyers, M.~Seltzer, and J.~Bacon, ``Data
  provenance to audit compliance with privacy policy in the internet of
  things,'' \emph{Personal and Ubiquitous Computing}, vol.~22, no.~2, pp.
  333--344, 2018.

\bibitem{kshetri2017can}
N.~Kshetri, ``Can blockchain strengthen the internet of things?'' \emph{IT
  Professional}, vol.~19, no.~4, pp. 68--72, 2017.

\bibitem{huckle2016internet}
S.~Huckle, R.~Bhattacharya, M.~White, and N.~Beloff, ``Internet of things,
  blockchain and shared economy applications,'' \emph{Procedia computer
  science}, vol.~98, pp. 461--466, 2016.

\bibitem{szefer2013bitdeposit}
J.~Szefer and R.~B. Lee, ``Bitdeposit: Deterring attacks and abuses of cloud
  computing services through economic measures,'' in \emph{Cluster, Cloud and
  Grid Computing (CCGrid), 2013 13th IEEE/ACM International Symposium
  on}.\hskip 1em plus 0.5em minus 0.4em\relax IEEE, 2013, pp. 630--635.

\bibitem{warren2012bitmessage}
J.~Warren, ``Bitmessage: A peer-to-peer message authentication and delivery
  system,'' \emph{white paper (27 November 2012),
  \url{https://bitmessage.org/bitmessage.pdf}}, 2012.

\bibitem{shrier2016blockchain}
D.~Shrier, D.~Sharma, and A.~Pentland, ``Blockchain \& financial services: The
  fifth horizon of networked innovation,'' 2016.

\bibitem{noizat2015blockchain}
P.~Noizat, ``Blockchain electronic vote,'' \emph{Handbook of Digital Currency:
  Bitcoin, Innovation, Financial Instruments, and Big Data}, p. 453, 2015.

\bibitem{kishigami2015blockchain}
J.~Kishigami, S.~Fujimura, H.~Watanabe, A.~Nakadaira, and A.~Akutsu, ``The
  blockchain-based digital content distribution system,'' in \emph{Big Data and
  Cloud Computing (BDCloud), 2015 IEEE Fifth International Conference
  on}.\hskip 1em plus 0.5em minus 0.4em\relax IEEE, 2015, pp. 187--190.

\bibitem{peterson2016blockchain}
K.~Peterson, R.~Deeduvanu, P.~Kanjamala, and K.~Boles, ``A blockchain-based
  approach to health information exchange networks,'' in \emph{Proc. NIST
  Workshop Blockchain Healthcare}, vol.~1, 2016, pp. 1--10.

\bibitem{dorri2017automotive}
A.~Dorri, M.~Steger, S.~S. Kanhere, and R.~Jurdak, ``Blockchain: A distributed
  solution to automotive security and privacy,'' \emph{IEEE Communications
  Magazine}, vol.~55, no.~12, pp. 119--125, 2017.

\bibitem{lin2017blockchain}
Y.-P. Lin, J.~R. Petway, J.~Anthony, H.~Mukhtar, S.-W. Liao, C.-F. Chou, and
  Y.-F. Ho, ``Blockchain: The evolutionary next step for ict e-agriculture,''
  \emph{Environments}, vol.~4, no.~3, p.~50, 2017.

\bibitem{xu2016blockchain}
X.~Xu, C.~Pautasso, L.~Zhu, V.~Gramoli, A.~Ponomarev, A.~B. Tran, and S.~Chen,
  ``The blockchain as a software connector,'' in \emph{13th Working IEEE/IFIP
  Conference on Software Architecture (WICSA), 2016}.\hskip 1em plus 0.5em
  minus 0.4em\relax IEEE, 2016, pp. 182--191.

\bibitem{mattila2016blockchain}
J.~Mattila \emph{et~al.}, ``The blockchain phenomenon--the disruptive potential
  of distributed consensus architectures,'' The Research Institute of the
  Finnish Economy, Tech. Rep., 2016.

\bibitem{OpenBaza43:online}
``Openbazaar: Online marketplace | peer-to-peer ecommerce,''
  \url{https://www.openbazaar.org/}, (Accessed on 06-{Oct}-2018).

\bibitem{Atlas85:online}
``Atlas,'' \url{https://atlas.money/}, (Accessed on 06-{Oct}-2018).

\bibitem{Sia60:online}
``Sia,'' \url{http://sia.tech/}, (Accessed on 06-{Oct}-2018).

\bibitem{wei2017review}
H.~Wei-hong, A.~Meng, S.~Lin, X.~Jia-gui, and L.~Yang, ``Review of
  blockchain-based dns alternatives,'' vol.~3, no.~3, pp. 71--77, 2017.

\bibitem{HomeBloc62:online}
``Home | blockchain education network (ben),''
  \url{https://blockchainedu.org/}, (Accessed on 06-{Oct}-2018).

\bibitem{Bigchain57:online}
``Bigchaindb---the scalable blockchain database.''
  \url{https://www.bigchaindb.com/}, (Accessed on 06-{Oct}-2018).

\bibitem{MaidSafe99:online}
``Maidsafe---the new decentralized internet,'' \url{https://maidsafe.net/},
  (Accessed on 06-{Oct}-2018).

\bibitem{Filecoin58:online}
``Filecoin,'' \url{https://filecoin.io/}, (Accessed on 06-{Oct}-2018).

\bibitem{IBMBlock70:online}
``Ibm blockchain,'' \url{https://www.ibm.com/blockchain/}, (Accessed on
  06-{Oct}-2018).

\bibitem{IBMWatso80:online}
``Ibm watson {{IoT}}---private blockchain,''
  \url{https://www.ibm.com/internet-of-things/platform/private-blockchain/},
  (Accessed on 06-{Oct}-20187).

\bibitem{Chronicl81:online}
``Chronicled,'' \url{https://www.chronicled.com/}, (Accessed on 06-{Oct}-2018).

\bibitem{Elliptic79:online}
``Elliptic,'' \url{https://www.elliptic.co/}, (Accessed on 06-{Oct}-2018).

\bibitem{LuxTrust43:online}
``Luxtrust,'' \url{https://www.luxtrust.lu/}, (Accessed on 06-{Oct}-2018).

\bibitem{DataCent57:online}
``Data-centric security | guardtime industrial blockchain,''
  \url{https://guardtime.com/}, (Accessed on 06-{Oct}-2018).

\bibitem{Decentra22:online}
``Decentralized prediction markets | augur project,'' \url{https://augur.net/},
  (Accessed on 06-{Oct}-2018).

\bibitem{LaZooz5:online}
``Lazooz,'' \url{http://lazooz.org/}, (Accessed on 06-{Oct}-2018).

\bibitem{UBITQUIT21:online}
``Ubitquity - the first blockchain-secured platform for real estate
  recordkeeping,'' \url{https://www.ubitquity.io/web/index.html}, (Accessed on
  06-{Oct}-2018).

\bibitem{Stratumn57:online}
``Stratumn | trust the process,'' \url{https://stratumn.com/}, (Accessed on
  06-{Oct}-2018).

\bibitem{Myceliaf27:online}
``Mycelia for music - for a fairtrade music industry,''
  \url{http://myceliaformusic.org/}, (Accessed on 06-{Oct}-2018).

\bibitem{Introduc12:online}
``Introducing gemos, your blockchain operating system.'' \url{https://gem.co/},
  (Accessed on 06-{Oct}-2018).

\bibitem{TierionB18:online}
``Tierion - blockchain proof engine | api,'' \url{https://tierion.com/},
  (Accessed on 06-{Oct}-2018).

\bibitem{Provenan44:online}
``Provenance | technology,'' \url{https://www.provenance.org/technology},
  (Accessed on 06-{Oct}-2018).

\bibitem{UJO40:online}
``Ujo,'' \url{https://ujomusic.com/}, (Accessed on 06-{Oct}-2018).

\bibitem{Skuchain28:online}
``Skuchain - turn information into capital | turn information into capital,''
  \url{http://www.skuchain.com/}, (Accessed on 06-{Oct}-2018).

\bibitem{StorjDec1:online}
``Storj - decentralized cloud storage,'' \url{https://storj.io/}, (Accessed on
  06-{Oct}-2018).

\bibitem{GyftBloc41:online}
``Gyft block - building gift cards 2.0 on blockchain technology,''
  \url{https://block.gyft.com/}, (Accessed on 06-{Oct}-2018).

\bibitem{Blocksaf95:online}
``Blocksafe™ - blockchain centric enhanced firearm network,''
  \url{http://www.blocksafefoundation.com/}, (Accessed on 06-{Oct}-2018).

\bibitem{BitgiveF58:online}
``Bitgive foundation,'' \url{https://www.bitgivefoundation.org/}, (Accessed on
  06-{Oct}-2018).

\bibitem{Empoweri53:online}
IBM, ``Empowering the edge,'' \url{https://tinyurl.com/IBM-edge-report},
  (Accessed on 06-{Oct}-2018).

\bibitem{Adepttec30:online}
``Adept tech paper v10.3,'' \url{https://tinyurl.com/adept-white-paper},
  (Accessed on 06-{Oct}-2018).

\bibitem{karstconnecting}
J.~J. Karst and G.~Brodar, ``Connecting multiple devices with blockchain in the
  internet of things.''

\bibitem{Filament18:online}
``Filament foundations.pages,'' \url{https://tinyurl.com/filament-report},
  (Accessed on 06-{Oct}-2018).

\bibitem{worner2016bitcoin}
D.~W{\"o}rner, T.~Von~Bomhard, Y.-P. Schreier, and D.~Bilgeri, ``The bitcoin
  ecosystem: Disruption beyond financial services?'' 2016.

\bibitem{bocek2017blockchains}
T.~Bocek, B.~B. Rodrigues, T.~Strasser, and B.~Stiller, ``Blockchains
  everywhere-a use-case of blockchains in the pharma supply-chain,'' in
  \emph{Integrated Network and Service Management (IM), 2017 IFIP/IEEE
  Symposium on}.\hskip 1em plus 0.5em minus 0.4em\relax IEEE, 2017, pp.
  772--777.

\bibitem{gonczol2020blockchain}
P.~Gonczol, P.~Katsikouli, L.~Herskind, and N.~Dragoni, ``Blockchain
  implementations and use cases for supply chains-a survey,'' \emph{Ieee
  Access}, vol.~8, pp. 11\,856--11\,871, 2020.

\bibitem{francca2015homomorphic}
B.~Fran{\c{c}}a, ``Homomorphic mini-blockchain scheme,'' 2015.

\bibitem{bitshares:online}
``Dpos description on bitshares,''
  \url{ttps://how.bitshares.works/en/master/technology/dpos.html}, (Accessed on
  27-{May}-2020).

\bibitem{dziembowski2015proofs}
S.~Dziembowski, S.~Faust, V.~Kolmogorov, and K.~Pietrzak, ``Proofs of space,''
  in \emph{Annual Cryptology Conference}.\hskip 1em plus 0.5em minus
  0.4em\relax Springer, 2015, pp. 585--605.

\bibitem{fan2018roll}
X.~Fan and Q.~Chai, ``Roll-dpos: a randomized delegated proof of stake scheme
  for scalable blockchain-based internet of things systems,'' in
  \emph{Proceedings of the 15th EAI International Conference on Mobile and
  Ubiquitous Systems: Computing, Networking and Services}, 2018, pp. 482--484.

\bibitem{xu2017intelligent}
C.~Xu, K.~Wang, and M.~Guo, ``Intelligent resource management in
  blockchain-based cloud datacenters,'' \emph{IEEE Cloud Computing}, vol.~4,
  no.~6, pp. 50--59, 2017.

\bibitem{xia2017medshare}
Q.~Xia, E.~B. Sifah, K.~O. Asamoah, J.~Gao, X.~Du, and M.~Guizani, ``Medshare:
  Trust-less medical data sharing among cloud service providers via
  blockchain,'' \emph{IEEE Access}, vol.~5, pp. 14\,757--14\,767, 2017.

\bibitem{jiang2020searchain}
P.~Jiang, F.~Guo, K.~Liang, J.~Lai, and Q.~Wen, ``Searchain: Blockchain-based
  private keyword search in decentralized storage,'' \emph{Future Generation
  Computer Systems}, vol. 107, pp. 781--792, 2020.

\bibitem{tapas2018blockchain}
N.~Tapas, G.~Merlino, and F.~Longo, ``Blockchain-based iot-cloud authorization
  and delegation,'' in \emph{2018 IEEE International Conference on Smart
  Computing (SMARTCOMP)}.\hskip 1em plus 0.5em minus 0.4em\relax IEEE, 2018,
  pp. 411--416.

\bibitem{alphand2018iotchain}
O.~Alphand, M.~Amoretti, T.~Claeys, S.~Dall'Asta, A.~Duda, G.~Ferrari,
  F.~Rousseau, B.~Tourancheau, L.~Veltri, and F.~Zanichelli, ``Iotchain: A
  blockchain security architecture for the internet of things,'' in \emph{2018
  IEEE Wireless Communications and Networking Conference (WCNC)}.\hskip 1em
  plus 0.5em minus 0.4em\relax IEEE, 2018, pp. 1--6.

\bibitem{seitz2017authentication}
L.~Seitz, G.~Selander, E.~Wahlstroem, S.~Erdtman, and H.~Tschofenig,
  ``Authentication and authorization for constrained environments (ace),''
  \emph{Internet Engineering Task Force, Internet-Draft
  draft-ietf-aceoauth-authz-07}, 2017.

\bibitem{vuvcinic2015oscar}
M.~Vu{\v{c}}ini{\'c}, B.~Tourancheau, F.~Rousseau, A.~Duda, L.~Damon, and
  R.~Guizzetti, ``Oscar: Object security architecture for the internet of
  things,'' \emph{Ad Hoc Networks}, vol.~32, pp. 3--16, 2015.

\bibitem{pouwelse2005bittorrent}
J.~Pouwelse, P.~Garbacki, D.~Epema, and H.~Sips, ``The bittorrent {P2P}
  file-sharing system: Measurements and analysis,'' in \emph{IPTPS},
  vol.~5.\hskip 1em plus 0.5em minus 0.4em\relax Springer, 2005, pp. 205--216.

\bibitem{CDNIssus}
R.~Aitken, ``Can decent's `crypto-fuelled' blockchain revolutionize content \&
  data distribution? accessed on 06-{Oct}-2018,'' \url{https://goo.gl/hCtEm1},
  2017.

\bibitem{Swarm:online}
``Filecoin,'' \url{https://www.swarm.fund/}, (Accessed on 15-{May}-2020).

\bibitem{IPFS:online}
``Filecoin,'' \url{https://ipfs.io/}, (Accessed on 15-{May}-2020).

\bibitem{StorjDec95:online}
``Storj - decentralized cloud storage,'' \url{https://storj.io/}, (Accessed on
  06-{Oct}-2018).

\bibitem{BlockScores:online}
``Filecoin,'' \url{http://blockscores.com/}, (Accessed on 15-{May}-2020).

\bibitem{NextCloud:online}
``Filecoin,'' \url{https://nextcloud.com/}, (Accessed on 15-{May}-2020).

\bibitem{crosby2016blockchain}
M.~Crosby, P.~Pattanayak, S.~Verma, and V.~Kalyanaraman, ``Blockchain
  technology: Beyond bitcoin,'' \emph{Applied Innovation}, vol.~2, pp. 6--10,
  2016.

\bibitem{wilkinson2014metadisk}
S.~Wilkinson, J.~Lowry, and T.~Boshevski, ``Metadisk a blockchain-based
  decentralized file storage application,'' Technical Report, Available:
  \url{http://metadisk.org/metadisk.pdf}, Tech. Rep., 2014.

\bibitem{kaufman2009data}
L.~M. Kaufman, ``Data security in the world of cloud computing,'' \emph{IEEE
  Security \& Privacy}, vol.~7, no.~4, 2009.

\bibitem{kandukuri2009cloud}
B.~R. Kandukuri, A.~Rakshit \emph{et~al.}, ``Cloud security issues,'' in
  \emph{Services Computing, 2009. SCC'09. IEEE International Conference
  on}.\hskip 1em plus 0.5em minus 0.4em\relax IEEE, 2009, pp. 517--520.

\bibitem{subashini2011survey}
S.~Subashini and V.~Kavitha, ``A survey on security issues in service delivery
  models of cloud computing,'' \emph{Journal of network and computer
  applications}, vol.~34, no.~1, pp. 1--11, 2011.

\bibitem{wang2009enabling}
Q.~Wang, C.~Wang, J.~Li, K.~Ren, and W.~Lou, ``Enabling public verifiability
  and data dynamics for storage security in cloud computing,'' \emph{Computer
  Security--ESORICS 2009}, pp. 355--370, 2009.

\bibitem{wang2010privacy}
C.~Wang, Q.~Wang, K.~Ren, and W.~Lou, ``Privacy-preserving public auditing for
  data storage security in cloud computing,'' in \emph{Infocom 2010}.\hskip 1em
  plus 0.5em minus 0.4em\relax IEEE, 2010, pp. 1--9.

\bibitem{gai2020blockchain}
K.~Gai, J.~Guo, L.~Zhu, and S.~Yu, ``Blockchain meets cloud computing: A
  survey,'' \emph{IEEE Communications Surveys \& Tutorials}, 2020.

\bibitem{wang2019efficient}
Q.~Wang, H.~Wang, and B.~Zheng, ``An efficient distributed storage strategy for
  blockchain,'' in \emph{Proceedings of the ACM Turing Celebration
  Conference-China}, 2019, pp. 1--5.

\bibitem{dong2017betrayal}
C.~Dong, Y.~Wang, A.~Aldweesh, P.~McCorry, and A.~van Moorsel, ``Betrayal,
  distrust, and rationality: Smart counter-collusion contracts for verifiable
  cloud computing,'' in \emph{Proceedings of the 2017 ACM SIGSAC Conference on
  Computer and Communications Security}, 2017, pp. 211--227.

\bibitem{filecoinWhitepaper}
``Filecoin: {A} decentralized storage network, access date: 06-{Oct}-2018.''

\bibitem{Numbero75:online}
Statista, ``Number of social media users worldwide 2010--2021,''
  \url{https://tinyurl.com/statista-worldwide}, (Accessed on 06-{Oct}-2018).

\bibitem{fire2014online}
M.~Fire, R.~Goldschmidt, and Y.~Elovici, ``Online social networks: threats and
  solutions,'' \emph{IEEE Communications Surveys \& Tutorials}, vol.~16, no.~4,
  pp. 2019--2036, 2014.

\bibitem{mitblock86:online}
``mit\_blockchain\_and\_infrastructure\_report.pdf,''
  \url{https://www.getsmarter.com/blog/wp-content/uploads/2017/07/mit_blockchain_and_infrastructure_report.pdf},
  (Accessed on 20-May-2020).

\bibitem{SteemWhi91:online}
``Steemwhitepaper.pdf,'' \url{https://steem.io/SteemWhitePaper.pdf}, (Accessed
  on 06-{Oct}-2018).

\bibitem{dennis2015rep}
R.~Dennis and G.~Owen, ``Rep on the block: A next generation reputation system
  based on the blockchain,'' in \emph{10th International Conference for
  Internet Technology and Secured Transactions (ICITST)}.\hskip 1em plus 0.5em
  minus 0.4em\relax IEEE, 2015, pp. 131--138.

\bibitem{2017Cost74:online}
Accenture, ``2017 cost of cyber crime study,''
  \url{https://tinyurl.com/CostCyberCrimeStudy}, (Accessed on 06-{Oct}-2018).

\bibitem{IECBlock59}
Deloitte, ``Blockchain \& cyber security,'' \url{https://goo.gl/2BXkDb}, 2017,
  (Accessed on 06-{Oct}-2018).

\bibitem{myers2017block}
S.~Myers, ``Block-by-block: Leveraging the power of blockchain technology to
  build trust and promote cyber peace,'' \emph{Yale JL \& Tech.}, vol.~19, pp.
  334--334, 2017.

\bibitem{liu2019survey}
J.~Liu and Z.~Liu, ``A survey on security verification of blockchain smart
  contracts,'' \emph{IEEE Access}, vol.~7, pp. 77\,894--77\,904, 2019.

\bibitem{tezosWhitePaper}
``{Tezos: A Self-Amending Crypto-Ledger, access date: 06-{Oct}-2018},''
  \url{https://tezos.com/static/papers/position_paper.pdf}, 2014.

\bibitem{WhitePap7:online}
``White paper\_v.0.1.pdf - google drive,''
  \url{https://drive.google.com/file/d/0B1jTRGmj_3khUV9RTERnYzNvaE0/view},
  (Accessed on 06-{Oct}-2018).

\bibitem{KSIdatas0:online}
``Ksi data sheet,'' \url{https://tinyurl.com/KSI-data-sheet}, (Accessed on
  09/16/2017).

\bibitem{Obsidian56:online}
``Obsidian platform whitepaper,''
  \url{https://tinyurl.com/obsidian-white-paper}, (Accessed on 06-{Oct}-2018).

\bibitem{polyzosblockchain}
G.~C. Polyzos and N.~Fotiou, ``Blockchain-assisted information distribution for
  the internet of things.''

\bibitem{conti2018survey}
M.~Conti, E.~S. Kumar, C.~Lal, and S.~Ruj, ``A survey on security and privacy
  issues of bitcoin,'' \emph{IEEE Communications Surveys \& Tutorials},
  vol.~20, no.~4, pp. 3416--3452, 2018.

\bibitem{fromknecht2014decentralized}
C.~Fromknecht, D.~Velicanu, and S.~Yakoubov, ``A decentralized public key
  infrastructure with identity retention.'' \emph{IACR Cryptology ePrint
  Archive}, vol. 2014, p. 803, 2014.

\bibitem{prins2011diginotar}
J.~Prins and B.~U. Cybercrime, ``Diginotar certificate authority
  breach’operation black tulip’,'' 2011.

\bibitem{fisher2012final}
D.~Fisher, ``Final report on diginotar hack shows total compromise of {CA}
  servers,'' \emph{ThreatPost, Oct}, vol.~31, 2012.

\bibitem{Trustwav26:online}
``Trustwave sold root certificate for surveillance | zdnet,''
  \url{http://www.zdnet.com/article/trustwave-sold-root-certificate-for-surveillance/},
  (Accessed on 06-{Oct}-2018).

\bibitem{DebianSe99:online}
``Debian -- security information -- dsa-1571-1 openssl,''
  \url{https://www.debian.org/security/2008/dsa-1571}, (Accessed on
  06-{Oct}-2018).

\bibitem{w32stuxn78:online}
N.~Falliere, ``w32\_stuxnet\_dossier.pdf,''
  \url{http://www.symantec.com/content/en/us/enterprise/media/security_response/whitepapers/w32_stuxnet_dossier.pdf},
  February 2011, (Accessed on 06-{Oct}-2018).

\bibitem{seltzer2013securing}
L.~Seltzer, ``Securing your private keys as best practice for code signing
  certificates,'' 2013.

\bibitem{bencsath2011duqu}
B.~Bencs{\'a}th, G.~P{\'e}k, L.~Butty{\'a}n, and M.~F{\'e}legyh{\'a}zi, ``Duqu:
  A stuxnet-like malware found in the wild,'' \emph{CrySyS Lab Technical
  Report}, vol.~14, pp. 1--60, 2011.

\bibitem{bencsath2012cousins}
B.~Bencs{\'a}th, G.~P{\'e}k, L.~Butty{\'a}n, and M.~Felegyhazi, ``{The cousins
  of stuxnet: Duqu, flame, and gauss},'' \emph{Future Internet}, vol.~4, no.~4,
  pp. 971--1003, 2012.

\bibitem{bencsath2012duqu}
B.~Bencs{\'a}th, G.~P{\'e}k, L.~Butty{\'a}n, and M.~F{\'e}legyh{\'a}zi, ``Duqu:
  Analysis, detection, and lessons learned,'' in \emph{ACM European Workshop on
  System Security (EuroSec)}, vol. 2012, 2012.

\bibitem{faisal2012stuxnet}
M.~Faisal and M.~Ibrahim, ``{Stuxnet, duqu and beyond},'' \emph{International
  Journal of Science and Engineering Investigations}, vol.~1, no.~2, pp.
  75--78, 2012.

\bibitem{schmid2015ecdsa}
M.~Schmid, ``{ECDSA}-application and implementation failures,''
  \url{https://tinyurl.com/SchmidProject}, accessed on 6-Oct-2018.

\bibitem{axon2015privacy}
L.~Axon, ``Privacy-awareness in blockchain-based {PKI},'' \emph{Oxford
  University Center for Doctoral Training (CDT) in Cyber Security: CDT
  Technical Paper}, 2015.

\bibitem{qin2017cecoin}
B.~Qin, J.~Huang, Q.~Wang, X.~Luo, B.~Liang, and W.~Shi, ``{Cecoin: A
  decentralized PKI mitigating MitM attacks},'' \emph{Future Generation
  Computer Systems}, 2017.

\bibitem{Patricia69:online}
``Ethereum,'' \url{https://github.com/ethereum/wiki/wiki/Patricia-Tree},
  (Accessed on 06-{Oct}-2018).

\bibitem{axonpb}
L.~Axon and M.~Goldsmith, ``{PB-PKI: a privacy-aware blockchain-based PKI}.''

\bibitem{al2017scpki}
M.~Al-Bassam, ``{SCPKI: A Smart Contract-based PKI and Identity System},'' in
  \emph{Proceedings of the ACM Workshop on Blockchain, Cryptocurrencies and
  Contracts}.\hskip 1em plus 0.5em minus 0.4em\relax ACM, 2017, pp. 35--40.

\bibitem{tewarix509cloud}
H.~Tewari, A.~Hughes, S.~Weber, and T.~Barry, ``{X509Cloud-Framework for a
  Ubiquitous PKI}.''

\bibitem{matsumoto2016ikp}
S.~Matsumoto and R.~M. Reischuk, ``{IKP: Turning a PKI Around with
  Blockchains},'' \emph{IACR Cryptology ePrint Archive}, vol. 2016, p. 1018,
  2016.

\bibitem{underwood2016blockchain}
S.~Underwood, ``Blockchain beyond bitcoin,'' \emph{Communications of the ACM},
  vol.~59, no.~11, pp. 15--17, 2016.

\bibitem{yeoh2017regulatory}
P.~Yeoh and P.~Yeoh, ``Regulatory issues in blockchain technology,''
  \emph{Journal of Financial Regulation and Compliance}, vol.~25, no.~2, pp.
  196--208, 2017.

\bibitem{AbraBuyS84:online}
``Abra,'' \url{https://www.abra.com/}, (Accessed on 06-{Oct}-2018).

\bibitem{cermeno2016blockchain}
J.~S. Cerme{\~n}o, ``Blockchain in financial services: Regulatory landscape and
  future challenges for its commercial application,'' Working Paper, Tech.
  Rep., 2016.

\bibitem{bharadwaj2016blockchain}
K.~Bharadwaj, ``Blockchain 2.0: Smart contracts,'' 2016.

\bibitem{wright2015decentralized}
A.~Wright and P.~De~Filippi, ``Decentralized blockchain technology and the rise
  of lex cryptographia,'' 2015.

\bibitem{sebastian2016blockchain}
J.~Sebastian \emph{et~al.}, ``Blockchain in financial services: Regulatory
  landscape and future challenges,'' Tech. Rep., 2016.

\bibitem{Roadmapf79}
V.~Meguerditchian, ``Roadmap\_for\_blockchain\_standards,''
  \url{https://goo.gl/zbv6p6}, March 2017, (Accessed on 06-{Oct}-2018).

\bibitem{de2017understanding}
J.~de~Kruijff and H.~Weigand, ``Understanding the blockchain using enterprise
  ontology,'' in \emph{International Conference on Advanced Information Systems
  Engineering}.\hskip 1em plus 0.5em minus 0.4em\relax Springer, 2017, pp.
  29--43.

\bibitem{deshpande2017understanding}
A.~Deshpande, K.~Stewart, L.~Lepetit, and S.~Gunashekar, ``Understanding the
  landscape of distributed ledger technologies/blockchain,'' 2017.

\bibitem{tapscott2016blockchain}
D.~Tapscott and A.~Tapscott, \emph{Blockchain Revolution: How the technology
  behind Bitcoin is changing money, business, and the world}.\hskip 1em plus
  0.5em minus 0.4em\relax Penguin, 2016.

\bibitem{Revenuea86}
G.~of~UK~HMRC, ``Revenue and customs brief 9 (2014): Bitcoin and other
  cryptocurrencies - gov.uk,'' \url{https://goo.gl/QSz2GL}, March 2014,
  (Accessed on 06-{Oct}-2018).

\bibitem{network2013application}
F.~C.~E. Network, ``Application of fincen’s regulations to persons
  administering, exchanging, or using virtual currencies,'' \emph{United States
  Department of the Treasury, March}, vol.~18, 2013.

\bibitem{guadamuz2015blockchains}
A.~Guadamuz and C.~Marsden, ``Blockchains and bitcoin: Regulatory responses to
  cryptocurrencies,'' \emph{First Monday}, vol.~20, no.~12, 2015. [Online].
  Available: \url{http://journals.uic.edu/ojs/index.php/fm/article/view/6198}

\bibitem{2016773d76}
E.~Securities and M.~Authority, ``Discussion paper: The distributed ledger
  technology applied to securities markets,'' \url{https://goo.gl/jHncDb}, June
  2016, (Accessed on 06-{Oct}-2018).

\bibitem{walport2016distributed}
M.~Walport, ``Distributed ledger technology: Beyond blockchain. uk government
  office for science,'' Tech. Rep, Tech. Rep., 2016.

\bibitem{GDPRdoc}
``{Official GDPR Document},'' {Official Journal of the European Union},
  \url{https://eur-lex.europa.eu/legal-content/EN/TXT/PDF/?uri=CELEX:32016R0679&from=EN},
  (Accessed on 08-{Nov}-2018).

\bibitem{bacon2017blockchain}
J.~Bacon, J.~D. Michels, C.~Millard, and J.~Singh, ``Blockchain demystified,''
  \emph{SSRN: https://ssrn.com/abstract=3091218}, 2017.

\bibitem{cnil:online}
C.~N. Informatique, ``Blockchain,''
  \url{https://www.cnil.fr/sites/default/files/atoms/files/blockchain_en.pdf},
  (Accessed on 20/05/2020).

\bibitem{sompolinsky2013accelerating}
Y.~Sompolinsky and A.~Zohar, ``Accelerating bitcoin’s transaction processing
  fast money grows on trees,'' \emph{Not Chains}, 2013.

\bibitem{croman2016scaling}
K.~Croman, C.~Decker, I.~Eyal, A.~E. Gencer, A.~Juels, A.~Kosba, A.~Miller,
  P.~Saxena, E.~Shi, E.~G. Sirer \emph{et~al.}, ``On scaling decentralized
  blockchains,'' in \emph{International Conference on Financial Cryptography
  and Data Security}.\hskip 1em plus 0.5em minus 0.4em\relax Springer, 2016,
  pp. 106--125.

\bibitem{BlogTrad16:online}
``Tradeblock blog,'' \url{https://tinyurl.com/tradeblock-blog}, (Accessed on
  6-Oct-2018).

\bibitem{eyal2016bitcoin}
I.~Eyal, A.~E. Gencer, E.~G. Sirer, and R.~Van~Renesse, ``Bitcoin-{NG}: A
  scalable blockchain protocol.'' in \emph{NSDI}, 2016, pp. 45--59.

\bibitem{wattenhoferfast}
C.~D.~R. Wattenhofer, ``A fast and scalable payment network with bitcoin duplex
  micropayment channels.''

\bibitem{yu2020survey}
G.~Yu, X.~Wang, K.~Yu, W.~Ni, J.~A. Zhang, and R.~P. Liu, ``Survey: Sharding in
  blockchains,'' \emph{IEEE Access}, vol.~8, pp. 14\,155--14\,181, 2020.

\bibitem{scal}
Sofia, ``How the bitcoin lightning network could solve the blockchain
  scalability problem, access date: 06-{Oct}-2018,''
  \url{https://goo.gl/SqpMX4}, 2016.

\bibitem{databasesss}
I.~Allison, ``Meet bigchaindb: the scalable blockchain database' hitting one
  million writes per second, access date: 06-{Oct}-2018,''
  \url{https://goo.gl/iBWb0Y}, 2016.

\bibitem{luu2016secure}
L.~Luu, V.~Narayanan, C.~Zheng, K.~Baweja, S.~Gilbert, and P.~Saxena, ``A
  secure sharding protocol for open blockchains,'' in \emph{Proceedings of the
  2016 ACM SIGSAC Conference on Computer and Communications Security}.\hskip
  1em plus 0.5em minus 0.4em\relax ACM, 2016, pp. 17--30.

\bibitem{davids2016research}
C.~Davids, V.~K. Gurbani, G.~Ormazabal, A.~Rollins, K.~Singh, and R.~State,
  ``Research topics related to real-time communications over 5g networks,''
  2016.

\bibitem{duffield2014darkcoin}
E.~Duffield and K.~Hagan, ``Darkcoin: Peertopeer cryptocurrency with anonymous
  blockchain transactions and an improved proofofwork system,'' \emph{Mar-2014
  [Online]. Available: \url{https://cryptopapers.info/assets/pdf/darkcoin.pdf}
  [Accessed: 06-{Oct}-2018]}, 2014.

\bibitem{de2016interplay}
P.~De~Filippi, ``The interplay between decentralization and privacy: the case
  of blockchain technologies,'' 2016.

\bibitem{wang2016maturity}
H.~Wang, K.~Chen, and D.~Xu, ``A maturity model for blockchain adoption,''
  \emph{Financial Innovation}, vol.~2, no.~1, p.~12, 2016.

\bibitem{goldfeder2017cookie}
S.~Goldfeder, H.~Kalodner, D.~Reisman, and A.~Narayanan, ``When the cookie
  meets the blockchain: Privacy risks of web payments via cryptocurrencies,''
  \emph{Proceedings on Privacy Enhancing Technologies}, vol. 2018, no.~4, pp.
  179--199, 2018.

\bibitem{meiklejohn2013fistful}
S.~Meiklejohn, M.~Pomarole, G.~Jordan, K.~Levchenko, D.~McCoy, G.~M. Voelker,
  and S.~Savage, ``A fistful of bitcoins: characterizing payments among men
  with no names,'' in \emph{Proceedings of the 2013 conference on Internet
  measurement conference}.\hskip 1em plus 0.5em minus 0.4em\relax ACM, 2013,
  pp. 127--140.

\bibitem{pass2017fruitchains}
R.~Pass and E.~Shi, ``Fruitchains: A fair blockchain,'' in \emph{Proceedings of
  the ACM Symposium on Principles of Distributed Computing}.\hskip 1em plus
  0.5em minus 0.4em\relax ACM, 2017, pp. 315--324.

\bibitem{eyal2014majority}
I.~Eyal and E.~G. Sirer, ``Majority is not enough: Bitcoin mining is
  vulnerable,'' in \emph{International conference on financial cryptography and
  data security}.\hskip 1em plus 0.5em minus 0.4em\relax Springer, 2014, pp.
  436--454.

\bibitem{bradbury2013problem}
D.~Bradbury, ``The problem with bitcoin,'' \emph{Computer Fraud \& Security},
  vol. 2013, no.~11, pp. 5--8, 2013.

\bibitem{bastiaan2015preventing}
M.~Bastiaan, ``Preventing the 51\%-attack: a stochastic analysis of two phase
  proof of work in bitcoin.''

\bibitem{xu2016blockchains}
J.~J. Xu, ``Are blockchains immune to all malicious attacks?'' \emph{Financial
  Innovation}, vol.~2, no.~1, p.~25, 2016.

\bibitem{koshy2014analysis}
P.~Koshy, D.~Koshy, and P.~McDaniel, ``An analysis of anonymity in bitcoin
  using p2p network traffic,'' in \emph{International Conference on Financial
  Cryptography and Data Security}.\hskip 1em plus 0.5em minus 0.4em\relax
  Springer, 2014, pp. 469--485.

\bibitem{herrera2015research}
J.~Herrera-Joancomart{\'\i}, ``Research and challenges on bitcoin anonymity,''
  in \emph{Data Privacy Management, Autonomous Spontaneous Security, and
  Security Assurance}.\hskip 1em plus 0.5em minus 0.4em\relax Springer, 2015,
  pp. 3--16.

\bibitem{reid2013analysis}
F.~Reid and M.~Harrigan, ``An analysis of anonymity in the bitcoin system,'' in
  \emph{Security and privacy in social networks}.\hskip 1em plus 0.5em minus
  0.4em\relax Springer, 2013, pp. 197--223.

\bibitem{rahouti2018bitcoin}
M.~Rahouti, K.~Xiong, and N.~Ghani, ``Bitcoin concepts, threats, and
  machine-learning security solutions,'' \emph{IEEE Access}, vol.~6, pp.
  67\,189--67\,205, 2018.

\bibitem{bigdata}
``Introduction to blockchains \& what it means to big data,''
  \url{https://www.kdnuggets.com/2017/09/introduction-blockchain-big-data.html},
  (Accessed on 20-May-2020).

\bibitem{di2015bitconeview}
G.~Di~Battista, V.~Di~Donato, M.~Patrignani, M.~Pizzonia, V.~Roselli, and
  R.~Tamassia, ``Bitconeview: visualization of flows in the bitcoin transaction
  graph,'' in \emph{Visualization for Cyber Security (VizSec), 2015 IEEE
  Symposium on}.\hskip 1em plus 0.5em minus 0.4em\relax IEEE, 2015, pp. 1--8.

\bibitem{spagnuolo2014bitiodine}
M.~Spagnuolo, F.~Maggi, and S.~Zanero, ``Bitiodine: Extracting intelligence
  from the bitcoin network,'' in \emph{International Conference on Financial
  Cryptography and Data Security}.\hskip 1em plus 0.5em minus 0.4em\relax
  Springer, 2014, pp. 457--468.

\bibitem{eskandariusability}
S.~Eshkandary, D.~Barrera, E.~Stobert, and J.~Clark, ``A first look at the
  usability of bitcoin key management,'' \emph{NDSS Symposium 2015}, 2015.

\end{thebibliography}


\end{document}